\nonstopmode{}
\documentclass[12pt,
prd, 
aps, 
showpacs, 
superscriptaddress,
groupedaddress,
preprintnumbers,
floatfix,
nofootinbib]{revtex4-1}

\usepackage{epsfig}
\usepackage{graphicx}
\usepackage{amssymb}
\usepackage{amsmath}
\usepackage{longtable}
\usepackage{epstopdf}
\newcommand\bnl{Brookhaven National Laboratory, Upton, NY 11973, USA}
\newcommand\kek{KEK Theory Center, Institute of Particle and Nuclear Studies, High Energy Accelerator Research Organization (KEK), Tsukuba 305-0801, Japan}
\newcommand\guas{School of High Energy Accelerator Science, The Graduate University for Advanced Studies (Sokendai), Tsukuba 305-0801, Japan}
\newcommand\riken{RIKEN-BNL Research Center, Brookhaven National Laboratory, Upton, NY 11973, USA}
\newcommand\tsukuba{Graduate School of Pure and Applied Science, University of Tsukuba, Tennodai 1-1-1, Tsukuba, Ibaraki 305-8571, Japan}
\newcommand\uconn{Physics Department, University of Connecticut,
  Storrs, CT 06269-3046, USA}
\newcommand\nagoya{Department of Physics, Nagoya University,
Nagoya 464-8602, Japan}
\newcommand\nag{Department of Physics, Nagoya University,
Nagoya 464-8602, Japan}
\newcommand\nishina{Theoretical Physics Laboratory, Nishina Center,
RIKEN, Wako, 351-0198, Japan}
\newcommand\wako{RIKEN Nishina Center for Accelerator-Based Science, Wako, Saitama 351-0198, Japan}
\newcommand\Indiana{$^*$Present Address: Department of Physics, Indiana University, Bloomington, Indiana 47405, USA}

\def\non{\nonumber}

\newcommand{\del}{\partial}

\def\non{\nonumber}

\def\simge{
    \mathrel{\rlap{\raise 0.511ex
        \hbox{$>$}}{\lower 0.511ex \hbox{$\sim$}}}}
\def\simle{
    \mathrel{\rlap{\raise 0.511ex 
        \hbox{$<$}}{\lower 0.511ex \hbox{$\sim$}}}}

\begin{document}
\preprint{KEK-CP-235, RBRC-844}
\bibliographystyle{apsrev4-1}

\title{Electromagnetic mass splittings of the low lying hadrons and quark masses from 2+1 flavor lattice QCD+QED}
\author{Tom Blum and Ran Zhou$^*$}\affiliation{\uconn}\affiliation{\riken}\affiliation{\Indiana}

\author{Takumi Doi}\affiliation{\tsukuba}\affiliation{\wako}

\author{Masashi Hayakawa}\affiliation{\nagoya}\affiliation{\nishina}

\author{Taku Izubuchi}\affiliation{\bnl}\affiliation{\riken}

\author{Shunpei Uno}\affiliation{\nag}\affiliation{\riken}

\author{Norikazu Yamada}\affiliation{\kek}\affiliation{\guas}


\begin{abstract}
\vspace*{2cm}
 Results computed in lattice QCD+QED are presented
for the electromagnetic mass splittings of the low lying hadrons.
 These are used to determine the renormalized, non-degenerate,
light quark masses.
 It is found that $m^{\overline{MS}}_{u}=2.24~(10)~(34)$,
$m^{\overline{MS}}_{d}=4.65~(15)~(32)$,
and $m^{\overline{MS}}_{s}=97.6~(2.9)~(5.5)$ MeV
at the renormalization scale 2 GeV,
where the first error is statistical and the second systematic. 
We find the lowest order electromagnetic splitting $(m_{\pi^+}-m_{\pi^0})_{\rm QED}=3.38(23)$ MeV,
the splittings including next-to-leading order, $(m_{\pi^+}-m_{\pi^0})_{\rm QED}=4.50(23)$ MeV,
$(m_{K^+}-m_{K^0})_{\rm QED}=1.87(10)$ MeV, and the $m_u\neq m_d$ contribution to the kaon mass difference,
$(m_{K^+}-m_{K^0})_{(m_u-m_d)}=-5.840(96)$ MeV.
 All errors are statistical only, and the next-to-leading order pion splitting is only approximate in that it does not contain all next-to-leading order contributions. 
We also computed the proton-neutron mass difference, including for the first time, QED interactions in a realistic 2+1 flavor calculation. We find $(m_p-m_n)_{\rm QED}=0.383(68)$ MeV,
$(m_p-m_n)_{(m_u-m_d)}=-2.51(14)$ MeV (statistical errors only),
and the total $m_p-m_n=-2.13(16)(70)$ MeV, where the first error is statistical, and the second, part of the systematic error.
 The calculations are carried out on QCD ensembles generated
by the RBC and UKQCD collaborations,
using domain wall fermions and the Iwasaki gauge action
(gauge coupling $\beta=2.13$
and lattice cutoff $a^{-1}\approx 1.78$ GeV). We use two lattice sizes, $16^{3}$ and $24^{3}$ ( (1.8 fm)$^3$ and (2.7 fm)$^3$ ), to address finite volume effects.
 Non-compact QED is treated in the quenched approximation.
 The valence pseudo-scalar  meson masses
in our study cover a range of about 250 to 700 MeV,
though we use only those up to about 400 MeV to quote final results.
 We present new results for the electromagnetic low energy constants
in SU(3) and SU(2) partially-quenched chiral perturbation theory
to the next-to-leading order, obtained from fits to our data.
 Detailed analysis of systematic errors in our results and methods
for improving them are discussed.
 Finally, new analytic results for ${\rm SU(2)_L\times SU(2)_R}$-plus-kaon
chiral perturbation theory,
including the one-loop logs proportional to $\alpha_{\rm em}\,m$, are given.
\end{abstract}
\date{May 26, 2010, revised: September 2, 2010}
\maketitle

\section{Introduction}

The mass splitting in the meson and baryon systems is an interesting 
topic in hadron spectroscopy. It is related to the quark masses
which are fundamental parameters of the standard model. The mass
splitting in the pseudo-scalar meson octet is a signature of the
breaking of the strong isospin symmetry by the electromagnetic (EM) interaction and
non-degenerate quark masses. 

The hadron spectra are rich in diversity due to two origins: the nonperturbative quantum dynamics of the strong interaction and the presence of flavor symmetry breaking.
In the standard model, the latter originates from non-degenerate quark masses
as well as the difference between up-type quarks and down-type quarks.
These sources of flavor symmetry breaking affect significantly
the hadron spectra less than ($1\sim2$) GeV.
In the baryon octet, for instance, 
the mass difference 
of the proton and neutron is crucial to the
phenomenological model of nuclei because it plays an important role 
in neutron $\beta$-decay, which is related to the stability
of nuclei. If the up and down quark masses were degenerate, the proton would be heavier due to the EM interaction, and our Universe would not exist! 
Even though the mass differences in the baryon octet spectrum have already been measured in experiments to good accuracy,
it is important to confirm that we can {\it predict} these splittings
in the standard model from first-principles computation. 
Parameterizing the calculated 
splittings in terms of low energy constants (LEC) is also useful for effective theories like chiral perturbation theory. 

The mass of a hadron is determined by both quantum chromodynamics (QCD) and quantum electrodynamics (QED), 
though the vast majority of the mass is due to QCD. For the QCD interaction, since the coupling constant is large in the low energy regime ($E \lesssim 1$ GeV), perturbation theory 
is not applicable, and we must turn to the techniques of lattice gauge
theory to solve QCD for the hadron spectrum. On the other hand, the EM
contributions to the masses, which break degeneracies due to flavor
SU(3) (and isospin) symmetry since the up quark has charge $+2/3\,e$ and
the down and strange quarks have charge $-1/3\,e$,  are expected to
depend on the small QED coupling constant, $\alpha_{\rm em}\equiv e^{2}/4 \pi \approx 1/137$. However, since the hadrons are formed from bound-states of the quarks, there is no systematic way to treat the contributions in weak-coupling perturbation theory. Thus, calculations are done nonperturbatively 
in a combined lattice QCD+QED theory (indeed, even if the strong coupling constant were small,
one is forced to a nonperturbative solution for QCD because of confinement). The state-of-the-art in lattice calculations is such that sub-percent errors (statistical and systematic) on low-lying hadron masses and other observable quantities are becoming the norm(for a broad review, see \cite{Scholz:2009yz} and references therein). As the precision of lattice calculations improve, the EM splittings become more and more relevant. Indeed, the splittings themselves can be computed with sub-percent precision, at least for the statistical errors\cite{Blum:2007cy}. 

It is well known that the lowest order EM effect, the so-called Dashen
term\cite{Dashen:1969eg}, which enters at $O(\alpha_{\rm em})$, is the
dominant contribution to the charged-neutral pion mass difference. In
the chiral limit where the quarks are all massless, it is also true for
the kaons. This theorem, known as Dashen's theorem, is broken by terms
of order $O(\alpha_{\rm em} m)$ away from the chiral limit. Using an effective theory of QCD known as chiral perturbation theory, these corrections can be identified in the lattice calculation, and the non-degenerate quark masses determined by matching to the experimentally measured mass splittings\cite{Amsler:2008zzb},
\begin{eqnarray}
m_{\pi^\pm}-m_{\pi^0}&=&4.5936(5) \text{~~MeV}, \label{eq:experiment pi mass}\\
m_{K^\pm}-m_{K^0} &=& -3.937(28) \text{~~MeV},\\
m_n-m_p&=&1.2933321(4) \text{~~MeV}.\label{eq:experiment nuc mass }
\end{eqnarray}
In fact, any three hadron masses are enough to determine the three light
quark masses, and we choose $m_{\pi^{+}}, m_{K^{+}}$ and $m_{K^{0}}$ for
reasons explained later. The determination of the up quark mass, $m_u$, is particularly interesting since
one can check the simplest solution to the strong CP problem, $m_u=0$.
 
 In the study presented here,
we work with lattice domain wall fermions (DWF)~\cite{Kaplan:1992bt,Shamir:1993zy} for the quarks
and the Iwasaki gauge action for the gluons.
 We use an ensemble of gluon configurations
with a single lattice spacing, generated by the RBC and UKQCD collaborations
using (2+1)-flavors of dynamical quarks, a pair of degenerate quarks for
the up and down, and a heavier strange quark~\cite{Allton:2008pn}.
The photons are simulated in non-compact, quenched QED,
as was done in the pioneering quenched QCD+quenched QED calculations~\cite{Duncan:1996xy,Duncan:1996be}.

 There are several differences between this work
and our previous one~\cite{Blum:2007cy}. The most obvious is
that the dynamical strange sea quark has been included here.
 In Ref.~\cite{Blum:2007cy} the QED gauge potential
was fixed to the Coulomb gauge, and here we work in Feynman gauge 
in QED on finite volume, as described in Ref.~\cite{Hayakawa:2008an}.
 Next, we perform fits
to full next-to-leading order (NLO) partially-quenched
chiral perturbation theory (PQ$\chi$PT), including photons,
for both ${\rm SU(3)_L \times SU(3)_R}$ chiral symmetry~\cite{Bijnens:2006mk}
and ${\rm SU(2)_L \times SU(2)_R}$-plus-kaon
\cite{Roessl:1999iu,Ouellette:2001ib,Allton:2008pn},
where the latter treats only the pion-triplet as pseudo-Nambu-Goldstone bosons.
 The NLO PQ$\chi$PT for ${\rm SU(2)_L\times SU(2)_R}$-plus-kaon, including photons, is new,
and is presented here for the first time. 
 Calculations by the RBC and UKQCD collaborations~\cite{Antonio:2007pb,
Allton:2008pn,Kelly:2009fp,Mawhinney:2009jy,RBC:2010},
and more recently PACS-CS\cite{Aoki:2008sm,Kadoh:2008sq}, have shown
that the physical strange quark mass is out of reach
for NLO 
${\rm SU(3)}$ chiral perturbation theory.
 Since we also wish to determine the strange quark mass in our calculation,
we have developed
the chiral perturbation theory
for the ${\rm SU(2)_L \times SU(2)_R}$-plus-kaon system,
including photons.
 In addition, since the photons are not confined,
finite volume effects are expected to be large,
so we work with two lattice sizes, $16^3$ and $24^3$,
with the same lattice spacing to investigate these effects.
 The leading EM finite volume effects have been computed
in PQ$\chi$PT~\cite{Hayakawa:2008an}, which we also use in our analysis.

 This paper is organized as follows.
 In Sec.~\ref{sec:chipt} we summarize the chiral perturbation theories
used to fit our lattice calculations
(details are given in the Appendix \ref{sec:PCHPT}).
 In Sec.~\ref{sec:lattice} the basic framework and details
of the lattice simulations are given.
 Section~\ref{sec:results} contains results
and discussion of the calculation, including the
fitted LEC's and the quark masses.
 Section~\ref{sec:sys errors} discusses systematic errors, and in Secs.~\ref{sec:final quark masses}~and~\ref{sec:final meson} we give final values for the quark masses and meson splittings, respectively. 
We examine the impact of $(m_u-m_d)$ on the decay constant ratio,
$f_K/f_\pi$ in Sec. \ref{sec:fK}.
 The nucleon mass splitting is computed in Sec.~\ref{sec:nucleon}. Finally, this work is summarized in Sec.~\ref{sec:conclusion}.

 We reported the preliminary results 
from SU(3) PQ$\chi$PT study in this work
in Ref.~\cite{Zhou:2008gb,Zhou:2009ku}.
 The MILC collaboration also presented their first results on the EM splittings,
using improved staggered fermions and non-compact,
quenched QED configurations in Ref.~\cite{Basak:2008na}.

\section{Chiral Perturbation Theory}\label{sec:chipt}

 We briefly review the framework and formulas of
partially quenched chiral perturbation theory relevant
for our 2+1 flavor calculation.
 The EM corrections in SU(2) chiral perturbation theory coupled to kaons
are new.
Details are given in Appendix \ref{sec:PCHPT}.

 Recently it has been shown that SU(3) chiral perturbation theory
is poorly convergent for quark masses
near the physical strange quark mass,
and that a straightforward and effective solution is
to treat the strange quark mass $m_{s}$ as large compared
to the light quark masses $m_{l}$ in an expansion in $m_{l}/m_{s}$
\cite{Allton:2008pn,Roessl:1999iu,Aoki:2008sm,Kadoh:2008sq}.
 We carry out fits to the data using both SU(3) and SU(2)
chiral perturbation theory.
 We find the poor convergence extends to the EM sector as well,
and use SU(2) chiral perturbation theory to quote our central results.

 Before proceeding, it is important to discuss the order
in chiral perturbation theory to which we work.
 For the SU(3) case where the kaon is a Nambu-Goldstone boson,
the leading order (LO) includes all terms that are $O(p^{2})$ and $O(e^{2})$,
and the next-to-leading order (NLO) includes
all terms that are $O(p^{4})$, $O( e^{2}p^{2})$, and $O(e^{4})$,
where the conventional power counting is $O(e)\sim O(p)$.
 This counting is the same for the square of the masses
and the mass-squared splittings.
 $O(e^{4})$ contributions
have so far been ignored \cite{Bijnens:2006mk,Blum:2007cy}
since $O(e^4) \ll O(e^2 p^2)$ in practice, and we also follow this here. 

 In the SU(2) theory coupled to kaons the power counting becomes
a bit more complicated for the kaon
(for the pion it is the same as in the SU(3) case).
 Since the kaon is no longer a Nambu-Goldstone boson,
LO for the mass-squared is now $O(p^{0})$,
and NLO is $O(p^{2})$ and $O(e^{2})$.
 The mass-squared EM splitting, however, remains the same
order of magnitude as in (partially-quenched)
SU(3) chiral perturbation theory. 
That is, to obtain the NLO contributions to the mass-squared splittings, we must work to NNLO for the masses.
 Since the aim here is to include all effects up to
and including $O(e^{2} p^2)$ terms in the meson mass-squared splittings
as well as $m_{d}-m_{u}$, we include all $O(e^{2} p^2)$
contributions to the kaon mass.
 Because we compute $m_{d}-m_{u}$
from the neutral-charged kaon mass-squared difference,
the pure QCD effects at $O(p^{4})$,
including the one-loop logarithms~\cite{Roessl:1999iu},
cancel and are not included in our analysis.

{ Finally, to avoid confusion we emphasize that in this paper
we only calculate correlation functions
for ``charged", or ``off-diagonal" mesons.
 However, since we are free to change the charges
and masses of the valence quarks making up these mesons,
the total charge of the (unphysical) meson
may happen to be zero.
 Sometimes we refer to these as ``neutral" mesons,
but it must be kept in mind these never correspond
to the $\pi^0$ meson which requires so-called disconnected quark diagrams
in its correlation function
as well as the full treatment of ``diagonal'' mesons in PQ$\chi$PT.}

\subsection{${\rm SU(3)_L \times SU(3)_R}$}

 The partially quenched chiral perturbation theory
has been worked out in Ref.~\cite{Bijnens:2006mk},
and we adopt their notation. For three non-degenerate sea quarks
and two non-degenerate valence quarks, labeled by ``1'' and ``3'',
the meson mass-squared at NLO is 
\begin{eqnarray}
M_{13}^2 &=&
\chi_{13}+\frac{2 C e^{2}}{F_{0}^{2}}q_{13}^{2}\nonumber\\
&&+\frac{48L^{r}_{6}-24L^{r}_{4}}{F_{0}^{2}}\chi_{13}\bar\chi_{1}
     +\frac{16L^{r}_{8}-8L^{r}_{5}}{F_{0}^{2}}\chi_{13}^{2}\nonumber\\
&& -48 e^{2}\frac{C}{F_{0}^{4}}L^{r}_{4}q_{13}^{2}\bar\chi_{1}
      -16e^{2}\frac{C}{F_{0}^{4}}L^{r}_{5}q_{13}^{2}\chi_{13}\nonumber\\
&& -e^2  \left[ 12 K^{Er}_1 +12K^{Er}_2 -12K^{Er}_7 -12K^{Er}_8\right]  \bar q^2 \chi_{13}\nonumber\\
&& -e^2 \left[4 K_5^{Er}+4 K_6^{Er}\right] q_p^2 \chi_{13}\nonumber\\
&& +e^2 \left[4 K_9^{Er}+4 K_{10}^{Er}\right] q_p^2 \chi_{p}\nonumber\\
&& +12e^2  K_8^{Er} q_{13}^2 \bar\chi_{1}\nonumber\\
&& +8 e^2 \left[K_{10}^{Er}+K_{11}^{Er}\right] q_{13}^2 \chi_{13}\nonumber\\
&& - e^2 \left[8 K_{18}^{Er}+4 K_{19}^{Er}\right] q_{1}q_{3} \chi_{13}\nonumber\\
&&+\frac{1}{3}\frac{1}{16\pi^2 F_{0}^{2}}R^{m}_{n13}\chi_{13}\chi_{m}\log{\frac{\chi_{m}}{\mu^2}}+\frac{1}{3}\frac{1}{16\pi^2 F_{0}^{2}}R^{p}_{q\pi\eta}\chi_{13}\chi_{p}\log{\frac{\chi_{p}}{\mu^2}}\nonumber\\
%
&&- 2 e^2 \frac{C}{F^{4}_{0}} \frac{1}{16\pi^2}\left(\chi_{14}\log{\frac{\chi_{14}}{\mu^2}}q_{14}+\chi_{15}\log{\frac{\chi_{15}}{\mu^2}}q_{15}+\chi_{16}\log{\frac{\chi_{16}}{\mu^2}}q_{16}\right)q_{13}\nonumber\\
&&+ 2 e^2  \frac{C}{F^{4}_{0}}\frac{1}{16\pi^2}\left(\chi_{34}\log{\frac{\chi_{34}}{\mu^2}}q_{34}+\chi_{35}\log{\frac{\chi_{35}}{\mu^2}}q_{35}+\chi_{36}\log{\frac{\chi_{36}}{\mu^2}}q_{36}\right)q_{13}\nonumber\\
&&
-
\frac{\left(q_{13}^2\right) e^2}{16\pi^2}\,
{\chi}_{13}
\left\{
 3\,\log\left(\frac{\chi_{13}}{\mu^2}\right) - 4
\right\}\nonumber\\
&&
+e^{2}\delta_{m_{\rm res}}(q_1^2+q_3^2).
\label{eq:su3 pion}
\end{eqnarray}
Indices $1 - 3$ always refer to valence quarks, $4 - 6$ to sea quarks. 
The coefficients $R^{m}_{n13}$ and $R^{p}_{q\pi\eta}$ 
are the residue functions written
in terms of quark masses and are defined in Ref.~\cite{Bijnens:2006mk}.
The index $p$ implies summation over valence indices 1 and 3, and if $q$ is also present, then the sum is over pairs (1,3) and (3,1). The indices $(m,n)$ signify a sum over pairs ($\pi,\eta$) and ($\eta,\pi$). 
$\chi_{ij}=B_{0}(m_{i}+m_{j})$ is the LO mass-squared for a meson made
of quarks with masses $m_{i}$ and $m_{j}$, $q_{ij}=q_{i}-q_{j}$ where
$q_{i}$ is the electric charge  of the $i$th quark in units of the
fundamental charge $e$. $\bar \chi_{1} = 2 B_{0}(m_{4}+m_{5}+m_{6})/3$ and
$\bar q^2 =(q_{4}^{2}+q_{5}^{2}+q_{6}^{2})/3$. 
$\chi_\pi$ and $\chi_\eta$ are given by the solution of
\begin{align}
&\chi_\pi+\chi_\eta=2\bar{\chi_1},\\
&\chi_\pi \chi_\eta = {4\over 3}\,B_0^2 (m_4 m_5 + m_5 m_6 + m_4 m_6).
\end{align}
The LO Dashen term is
proportional to the low energy constant (LEC) $C$ and the fine structure
constant $\alpha_{\rm em}$. $B_{0}$ and $F_{0}$ \footnote{$F_{0}$ is
normalized such that the physical decay constant is roughly 92 MeV.} are
the LO QCD LEC's, the $L$'s are the Gasser-Leutwyler LEC's at NLO, and
the $K$'s are the EM LEC's at $O(\alpha_{\rm em} m)$. $\delta_{m_{\rm res}}$ is a pure lattice artifact LEC associated with the finite size of the extra dimension for DWF.

 { We note from Eq.~(\ref{eq:su3 pion})
that masses of mesons $\sim \overline{q} q^\prime$ made
from degenerate valence quarks $q,\,q^\prime$
with equal charges
do not have logarithmic corrections at NLO.
 This happens for the SU(2) case as well}.

 Following Ref.~\cite{Bijnens:2006mk},
the EM LEC's can be written in terms of five independent linear combinations of the $K$'s, which is all that can be determined from lattice calculations,
\begin{eqnarray}
Y_{1}&=&  K^{Er}_1 +K^{Er}_2 -K^{Er}_7 -K^{Er}_8, \\
Y_{2}&=&  K_9^{Er}+K_{10}^{Er}, \\
Y_{3}&=&  - K_5^{Er}- K_6^{Er}+2K_{10}^{Er}+2K_{11}^{Er}, \\
Y_{4}&=&  2K_5^{Er}+2K_6^{Er}+2 K_{18}^{Er}+ K_{19}^{Er},\nonumber\\
Y_{5}&=&  K_8^{Er}.
\end{eqnarray}

The EM mass-squared splitting of the pseudo-scalar meson is defined as $\Delta M^{2}= M^2(e\neq0)-M^2(e=0)$.
In terms of the $Y_{i}$'s, it becomes
\begin{eqnarray}
\Delta M^{2} &=& 
\frac{2 C e^{2}}{F_{0}^{2}}q_{13}^{2}\nonumber\\
&& -48 e^{2}\frac{C}{F_{0}^{4}}L^{r}_{4}q_{13}^{2}\bar\chi_{1}
      -16e^{2}\frac{C}{F_{0}^{4}}L^{r}_{5}q_{13}^{2}\chi_{13}\nonumber\\
&&-12e^2  Y_1  \bar q^2 \chi_{13} +4e^2  Y_2 q_p^2 \chi_{p} +4 e^2  Y_3 q_{13}^2 \chi_{13} - 4 e^2  Y_4 q_{1}q_{3} \chi_{13} +12e^2  Y_5 q_{13}^2 \bar\chi_{1}\nonumber\\
&&- 2 e^2  \frac{C}{F^{4}_{0}} \frac{1}{16\pi^2}\left(\chi_{14}\log{\frac{\chi_{14}}{\mu^2}}q_{14}+\chi_{15}\log{\frac{\chi_{15}}{\mu^2}}q_{15}+\chi_{16}\log{\frac{\chi_{16}}{\mu^2}}q_{16}\right)q_{13}\nonumber\\
&&+ 2 e^2  \frac{C}{F^{4}_{0}}\frac{1}{16\pi^2}\left(\chi_{34}\log{\frac{\chi_{34}}{\mu^2}}q_{34}+\chi_{35}\log{\frac{\chi_{35}}{\mu^2}}q_{35}+\chi_{36}\log{\frac{\chi_{36}}{\mu^2}}q_{36}\right)q_{13}\nonumber\\
&&
-
\frac{\left(q_{13}\right)^2 e^2}{16\pi^2}\,
\chi_{13}
\left\{
 3\,\log\left(\frac{\chi_{13}}{\mu^2}\right) - 4
\right\}\nonumber\\
&&
+e^{2}\delta_{m_{\rm res}}(q_1^2+q_3^2).
\label{eq:su3 split}
\end{eqnarray}
Note that $Y_{1}$ is proportional to the sea quark charges.
Since we work with quenched QED,
this LEC can not be obtained from our calculation.

 We carry out the fit in Section \ref{sec:fin vol}
with the finite volume correction to the chiral logarithms
taken into account.
 The finite volume correction to the leading-order chiral logarithms
was computed in Ref.~\cite{Hayakawa:2008an},
\begin{eqnarray}
 \delta M_{13}^2(L)
 &\equiv&
 M_{13}^2(L) - M_{13}^2(\infty)\nonumber\\
 &=&
 \frac{1}{3}\,\frac{1}{16\pi^2 F_0^2}\,
 R^m_{n13}\,\chi_{13}\,\frac{\mathcal{M}(\sqrt{\chi_m}\,L)}{L^2}
 +
 \frac{1}{3}\,\frac{1}{16\pi^2 F_0^2}\,
 R^p_{q\pi\eta}\,\chi_{13}\,\frac{\mathcal{M}(\sqrt{\chi_p}\,L)}{L^2}
  \nonumber\\
 &&
 - 2 e^2\,\frac{C}{F_0^4}\,\frac{1}{16\pi^2}\,q_{13}\nonumber\\
 &&\qquad
 \times
 \left(
  q_{14}\,\frac{\mathcal{M}(\sqrt{\chi_{14}}\,L)}{L^2}
  +
  q_{15}\,\frac{\mathcal{M}(\sqrt{\chi_{15}}\,L)}{L^2}
  +
  q_{16}\,\frac{\mathcal{M}(\sqrt{\chi_{16}}\,L)}{L^2}
 \right)\nonumber\\
 &&
 + 2 e^2\,\frac{C}{F_0^4}\,\frac{1}{16\pi^2}\,q_{13}\nonumber\\
 &&\qquad
 \times
 \left(
  q_{34}\,\frac{\mathcal{M}(\sqrt{\chi_{34}}\,L)}{L^2}
  +
  q_{35}\,\frac{\mathcal{M}(\sqrt{\chi_{35}}\,L)}{L^2}
  +
  q_{36}\,\frac{\mathcal{M}(\sqrt{\chi_{36}}\,L)}{L^2}
 \right)\nonumber\\
 &&
 -
 3\,\frac{\left(q_{13}\right)^2 e^2}{4\pi}\,\frac{\kappa}{L^2}
  \nonumber\\
 &&
 +
 \frac{\left(q_{13}\right)^2 e^2}{\left(4\pi\right)^2}
 \left\{
  \frac{\mathcal{K}\left(\sqrt{\chi_{13}}\,L\right)}{L^2}
  -
  4 \sqrt{\chi_{13}}\,\frac{\mathcal{H}(\sqrt{\chi}_{13}\,L)}{L}
 \right\}\,.\label{eq:fv_su3_pion}
\end{eqnarray}
 $\mathcal{M}(x)$ is the function
appearing in the finite volume correction to the chiral logarithm
induced by the tadpole diagram,
\begin{eqnarray}
 \mathcal{M}(x) &\equiv&
 4\pi
 \int_0^\infty \frac{d\lambda}{\lambda^2}\,
 \exp\left(-\frac{x^2}{4\pi}\,\lambda\right) \mathcal{T}(\lambda)\,,
  \nonumber\\
 \mathcal{T}(\lambda) &\equiv&
 \left(\vartheta_3\left(0,\,i\,\frac{1}{\lambda}\right)\right)^3 - 1\,,
  \label{eq:def_fun_M}
\end{eqnarray}
where $\vartheta_3(v;\,\tau)$ is a Jacobi-theta function, 
\begin{eqnarray}
 \vartheta_3\left(v,\,\tau\right)
 &\equiv&
 \sum_{n=-\infty}^\infty
 \exp
 \left(
  \pi \tau i n^2 + 2 \pi v i n
 \right)\,.\nonumber
\end{eqnarray}
 The other functions and a constant $\kappa$ are given by
\cite{Hayakawa:2008an}
\begin{eqnarray}
 \kappa &\equiv&
 \int_0^\infty
 \frac{d\lambda}{\lambda^2}\,\mathcal{S}(\lambda)
 =
 2.837\cdots\,,\label{eq:def_kappa}\\
 \mathcal{S}(\lambda) 
 &\equiv&   
 - 
 \left\{ 
  \left(\vartheta_3\left(0,\,i \frac{1}{\lambda}\right)\right)^3 
  - 1 - \lambda^{\frac{3}{2}} 
 \right\} \, , \label{eq:fun_S}\\
 \mathcal{H}(x) 
 &\equiv& 
 \pi 
 \int_0^\infty \frac{d\lambda}{\lambda^{\frac{3}{2}}}\, 
 {\rm erf}\left(x \sqrt{\frac{\lambda}{4\pi}}\right) 
 \mathcal{S}(\lambda) \,, \label{eq:fun_H}\\
 \mathcal{K}(x) 
 &\equiv& 
  4\pi \int_0^\infty \frac{d\lambda}{\lambda}\,\frac{1}{\lambda} 
  \left(1 - e^{-\frac{x^2}{4\pi}\,\lambda}\right) 
  \mathcal{S}(\lambda) \, , \label{eq:fun_K}
\end{eqnarray} 
where ${\rm erf}(x)$ is the error function, 
\begin{eqnarray} 
 & 
 \displaystyle{ 
  {\rm erf}(x) = \frac{2}{\sqrt{\pi}} \int_0^x ds\,e^{-s^2} 
  \,. 
 }& \nonumber
\end{eqnarray} 

\subsection{${\rm SU(2)_L \times SU(2)_R}$-plus-kaon}

 Some time ago Roessl~\cite{Roessl:1999iu} worked out
the low energy SU(2) Lagrangian of pions coupled to a kaon.
 Recently, the RBC and UKQCD collaborations
showed that SU(3) chiral perturbation theory
is poorly convergent for quark masses near the strange quark mass
but that SU(2) chiral perturbation theory coupled to a kaon worked well
for pions with masses less than about 400 MeV at NLO~\cite{Allton:2008pn}.
 In Ref.~\cite{Allton:2008pn}, the unitary Lagrangian
was extended to the partially quenched case.
 Here, we extend both works
to include the EM interactions to the order $\alpha_{\rm em} m$
for the kaon mass,
including the one-loop diagrams proportional to $\alpha_{\rm em}$.
 For the pion mass, we begin with the partially-quenched SU(3)
 Lagrangian in Ref.~\cite{Bijnens:2006mk} and expand in $m_l/m_s$.

\subsubsection{pions}

 We derive the ${\rm SU(2)_L\times SU(2)_R}$ result
for the pion mass-squared splitting by expanding
Eq.~(\ref{eq:su3 split})
in $(m_1,m_3,m_4,m_5)/m_6$, where $m_6$ is the strange sea quark mass, $m_1$ and $m_3$ are taken as non-degenerate light valence quark masses, and $m_4$ and $m_5$ the light sea quark masses,
\begin{eqnarray}
\Delta M^2 &=& \frac{2Ce^2}{F_0^2}q_{13}^2\nonumber\\
&& -48 e^{2}\frac{C}{F_0^{4}}L^{r}_{4}q_{13}^{2}\frac{\chi_{4}+\chi_{5}}{3}
      -16e^{2}\frac{C}{F_0^{4}}L^{r}_{5}q_{13}^{2}\chi_{13}\nonumber\\
&&-12e^2  Y_1  \bar q^2 \chi_{13}+4e^2  Y_2 q_p^2 \chi_{p} +4 e^2  Y_3 q_{13}^2 \chi_{13} - 4 e^2  Y_4 q_{1}q_{3} \chi_{13} +12e^2  Y_5 q_{13}^2 \frac{\chi_{4}+\chi_5}{3}\nonumber\\
&& -e^2 \frac{3}{16\pi^2}\chi_{13}\log{\frac{\chi_{13}}{\mu^2}}q_{13}^2+e^2 \frac{1}{4\pi^2}\chi_{13}q_{13}^2\nonumber\\
&&- e^2 \frac{C}{F_0^4} \frac{1}{8\pi^2}q_{13}\left(q_{14}
\chi_{14}\log{\frac{\chi_{14}}{\mu^2}}+q_{15}\chi_{15}\log{\frac{\chi_{15}}{\mu^2}}
-q_{34}\chi_{34}\log{\frac{\chi_{34}}{\mu^2}}-q_{35}\chi_{35}\log{\frac{\chi_{35}}{\mu^2}}
\right)\nonumber\\&&
+e^{2}\delta_{m_{\rm res}}(q_1^2+q_3^2).
\label{eq:su2 split}
\end{eqnarray}
 In Eq.~(\ref{eq:su2 split}) all of the low energy constants
now depend implicitly on the strange sea quark mass which is fixed (we rename them below to distinguish them from their SU(3) counterparts). 
 In addition the Dashen term has absorbed contributions
from the NLO SU(3) LEC's and the logs which do not depend on the charges or masses of
the up and down quarks,
\begin{eqnarray}
\frac{2Ce^2}{F_0^2}q_{13}^2+12e^2 Y_5 q_{13}^2 \frac{\chi_{6}}{3} - \frac{2 e^2 }{16\pi^2}\frac{C}{F_0^4}q_{13}^2
\chi_6\log{\frac{\chi_6}{\mu^2}}
-48 e^{2}\frac{C}{F_0^{4}}L^{r}_{4}q_{13}^{2}\frac{\chi_{6}}{3}.
\label{eq:su2 dashen}
\end{eqnarray}

 Including the contributions in pure QCD~\cite{Allton:2008pn},
the pion mass-squared to NLO becomes
\begin{eqnarray}
M^2 &=& 
\chi_{13}\left\{
1+\frac{24}{F^2}(2 L_6^{(2)}-L_4^{(2)})\frac{\chi_4+\chi_{5}}{3}
+\frac{8}{F^2}(2 L_8^{(2)}-L_5^{(2)})\chi_{13}\right.\non\\
&&\quad\quad
+
\left.
 \frac{1}{2}\,\frac{1}{16\pi^2 F^2}
 \left(
    R^\pi_{13}\, \chi_\pi\,\log \frac{\chi_\pi}{\mu^2}
  + R^1_{\pi 3}\,\chi_1\,\log \frac{\chi_1}{\mu^2}
  + R^3_{\pi 1}\,\chi_3\,\log \frac{\chi_3}{\mu^2}
 \right)
\right\}\nonumber\\
&+&\frac{2C^{(2)}e^2}{F^2}q_{13}^2
\non\\&-&
12e^2  Y^{(2)}_1  \bar q^2 \chi_{13} +4e^2  Y^{(2)}_2 q_p^2 \chi_{p} +4 e^2  Y^{(2)}_3 q_{13}^2 \chi_{13} - 4 e^2  Y^{(2)}_4 q_{1}q_{3} \chi_{13} +12e^2  Y^{(2)}_5 q_{13}^2 \frac{\chi_{4}+\chi_5}{3}\nonumber\\
&-& e^2 \frac{3}{16\pi^2}\chi_{13}\log{\frac{\chi_{13}}{\mu^2}}q_{13}^2+e^2 \frac{1}{4\pi^2}\chi_{13}q_{13}^2\nonumber\\
&-&e^2 \frac{C^{(2)}}{F^4} \frac{1}{8\pi^2}q_{13}\left(q_{14}
\chi_{14}\log{\frac{\chi_{14}}{\mu^2}}+q_{15}\chi_{15}\log{\frac{\chi_{15}}{\mu^2}}
-q_{34}\chi_{34}\log{\frac{\chi_{34}}{\mu^2}}-q_{35}\chi_{35}\log{\frac{\chi_{35}}{\mu^2}}
\right)\nonumber\\&+&
e^{2}\delta_{m_{\rm res}}(q_1^2+q_3^2) \,.
\label{eq:su2 pion mass}
\end{eqnarray}
The LO LEC's $F$ and $B$
are the counterparts of $F_{0}$ and $B_{0}$ from the SU(3) theory, and
the other SU(2) LEC's are denoted by an explicit superscript ``$(2)$''.
 $R^i_{jk}$ is given in $SU(2)$ partially quenched case as
\begin{eqnarray}
 R^i_{jk} &\equiv&
 \frac{\left(\chi_i - \chi_4\right) \left(\chi_i - \chi_5\right)}
      {\left(\chi_i - \chi_j\right) \left(\chi_i - \chi_k\right)}\,.
\end{eqnarray}

 The finite volume correction to Eq.~(\ref{eq:su2 pion mass})
is given by
\begin{eqnarray}
 \delta M^2(L)
 &\equiv&
 M^2(L) - M^2(\infty)\nonumber\\
 &=&
 \frac{1}{2}\,\frac{\chi_{13}}{16\pi^2 F^2}
 \left(
  R^\pi_{13}\,\frac{\mathcal{M}\left(\sqrt{\chi_\pi}\,L\right)}{L^2}
  +
  R^1_{\pi 3}\,\frac{\mathcal{M}\left(\sqrt{\chi_1}\,L\right)}{L^2}
  +
  R^3_{\pi 1}\,\frac{\mathcal{M}\left(\sqrt{\chi_3}\,L\right)}{L^2}
 \right)\nonumber\\
 &&
 - 2 e^2\,\frac{C^{(2)}}{F^4}\,\frac{1}{16\pi^2}\,q_{13}
 \left(
  q_{14}\,\frac{\mathcal{M}(\sqrt{\chi_{14}}\,L)}{L^2}
  +
  q_{15}\,\frac{\mathcal{M}(\sqrt{\chi_{15}}\,L)}{L^2}
 \right.\nonumber\\
 &&\qquad\qquad\qquad\qquad\quad
 \left.
  - q_{34}\,\frac{\mathcal{M}(\sqrt{\chi_{34}}\,L)}{L^2}
  - q_{35}\,\frac{\mathcal{M}(\sqrt{\chi_{35}}\,L)}{L^2}
 \right)\nonumber\\
 &&
 -
 3\,\frac{\left(q_{13}\right)^2 e^2}{4\pi}\,\frac{\kappa}{L^2}
  \nonumber\\
 &&
 +
 \frac{\left(q_{13}\right)^2 e^2}{\left(4\pi\right)^2}
 \left\{
  \frac{\mathcal{K}\left(\sqrt{\chi_{13}}\,L\right)}{L^2}
  -
  4 \sqrt{\chi_{13}}\,\frac{\mathcal{H}(\sqrt{\chi}_{13}\,L)}{L}
 \right\}\,.\label{eq:fv_su2_pion}
\end{eqnarray}
 The constant $\kappa$ and various functions are defined
in Eqs.~(\ref{eq:def_fun_M}), (\ref{eq:def_kappa}),
(\ref{eq:fun_H}) and (\ref{eq:fun_K}).

\subsubsection{kaons}

 The kaon mass can be obtained from  the tree-level Lagrangian,
following Refs.~\cite{Roessl:1999iu,Jenkins:1990jv},
by constructing the kaon from one light and one ``heavy" quark
and writing down all operators with the desired symmetries
in a non-relativistic theory where the power counting is straightforward.
 The needed relativistic Lagrangian is then constructed
such that in the limit that the kaon is heavy, the non-relativistic theory
is recovered.
 This has been done in the case of QCD to NNLO in Ref.~\cite{Roessl:1999iu}
and to NLO in partially quenched QCD in Ref.~\cite{Allton:2008pn}.
 Here we add the order $e^2 p^2$ terms induced by the EM interactions.
 Once the tree-level Lagrangian is known,
the one-loop corrections can be computed.
 The $O(e^2)$ Lagrangian and details of the one-loop calculation
are given in the Appendix \ref{sec:PCHPT}. 

The $O(e^2 p^{2})$ Lagrangian is quite complicated, with many operators appearing. 
While we have listed all possible operators
in the Appendix \ref{sec:PCHPT}
that
contribute, we have not yet reduced them to a linearly independent set
using relativistic invariance and the equations of motion. Still, this
is enough to give the general quark mass and charge dependence. In the
following, this is given by the generic LEC's $x^{(K)}_{3} \sim x_{8}^{(K)}$.

From Eqs.~(\ref{eq:esq kaon mass sq}),~(\ref{eq:chiral_logs_L_quenchedQED})
~and~(\ref{eq:esq-psq kaon mass sq}), 
the mass-squared of the kaon is
\begin{eqnarray}
M_K^2 &=&  M^2 -4B( A_3m_{1}+A_4(m_{4}+m_{5})) \nonumber\\
 &&+ e^{2}\left(2 \left(A_K^{(1,1)}+A_K^{(2,1)}\right)q_{1}^2+
A_K^{(s,1,1)}q^2_{3}+ 2 A_K^{(s,2)}q_{1}q_{3}\right)\nonumber \\&&
-\frac{e^{2}}{(4\pi)^2 F^2}\left(
  (A_K^{(1,1)}+3A_K^{(2,1)})q_{1}^2+A_K^{(s,2)}q_1 q_3 \right)
\sum_{s=4,5}\chi_{1s}\log{\frac{\chi_{1s}}{\mu^2}}\nonumber \\ 
&&+e^{2}m_{1}\left(x_{3}^{(K)}(q_{1}+q_{3})^{2}+x_{4}^{(K)}(q_{1}-q_{3})^{2}+x_{5}^{(K)}(q_{1}^{2}-q_{3}^{2})\right)\nonumber \\
&&+e^{2}\frac{m_{4}+m_{5}}{2}\left(x_{6}^{(K)}(q_{1}+q_{3})^{2}+x_{7}^{(K)}(q_{1}-q_{3})^{2}+x_{8}^{(K)}(q_{1}^{2}-q_{3}^{2})\right)\nonumber \\
&&+e^{2}\delta_{m_{\rm res}}(q_1^2+q_3^2),
\label{eq:kaon mass sq}
\end{eqnarray}
where we have included the explicit chiral symmetry breaking LEC $\delta_{m_{\rm res}}$, the same as for the pion.
Here the subscript ``1" stands for a light valence quark, $u$ or $d$, and ``3"
for the strange valence quark (charge). 
``4'' and ``5'' refer to the $u$ and $d$ sea quarks, respectively.
To avoid confusion we note that 
LEC's without superscripts denote the pure 
QCD LEC's of Refs.~\cite{Roessl:1999iu} and~\cite{Allton:2008pn}
while those with superscripts are EM LEC's 
defined in Appendix \ref{sec:PCHPT}. The finite volume correction to Eq.~(\ref{eq:kaon mass sq}) is  given in Appendix \ref{sec:PCHPT}, Eqs.~(\ref{eq:kaon fv 1})~and~(\ref{eq:kaon fv 2}).

Notice that the LO ``Dashen" term is different
than for the pion: the latter is a single LEC proportional to $q_{13}^2$
while the former consists of three LEC's and depends on the $u$, $d$,
and $s$ charges separately. This is a consequence
of the different chiral symmetries assumed in the two cases.

We remind the reader that we do not keep terms of order $p^{4}$ and $e^{4}$. 

\section{Lattice framework}\label{sec:lattice}

 Following Ref.~\cite{Duncan:1996xy},
the lattice calculation employs combined QCD+QED gauge configurations.
 A combined gluon-photon gauge link is simply the product
of two independent links,
a SU(3) color matrix for the gluons and a U(1) phase for the photons.
\begin{eqnarray}
U_{x,\mu} &=& U^{(3)}_{x,\mu}\times \left(U^{(1)}_{x,\mu}\right)^{Q_i},
\end{eqnarray}
where $Q_i=e q_i$ is the charge of the quark with flavor $i$.
 It is the combined link
that appears in the lattice Dirac operator, in the usual gauge-invariant way.
 The gluon and photon links were generated independently in our calculation, 
so the sea quarks were not electrically charged.
 This quenched QED calculation suffers a systematic error
that is expected to be $O(\alpha_{\rm em}\alpha_s)$
from a simple vacuum polarization argument.
 In chiral perturbation theory,
the charged sea quarks first contribute
at $O(\alpha_{\rm em} m_{\rm val})$ 
for the valence quark mass $m_{\rm val}$,
as we have seen in Sec.~\ref{sec:chipt}.
 This drawback can be eliminated
with the technique of
re-weighting~\cite{Duncan:2004ys,Hasenfratz:2008fg,Jung:2010jt},
which is becoming common in large scale dynamical calculations~\cite{Kelly:2009fp,Mawhinney:2009jy,RBC:2010,Aoki:2009ix}, and is under active investigation by us~\cite{Ishikawa:2010tq,Izubuchi:2010}.
 In a different context, combined dynamical simulations
have also been performed for the first time~\cite{Abramczyk:2009gb},
where the sea quarks are charged from the beginning.

 For the QCD configurations,
we use the $2+1$ flavor QCD configurations generated with DWF
and the Iwasaki gauge action ($\beta=2.13$)
by the RBC and UKQCD collaborations~\cite{Allton:2007hx,Allton:2008pn,RBC:2010}.
 The lattice sizes are $16^3\times32$ and $24^3\times64$.
 The lattice spacing
\footnote{This result is slightly larger
than the published one, 1.729~(28) GeV, in Ref.~\cite{Allton:2008pn}
because it was determined on a larger ensemble of lattices.
 It is also larger than the result of a combined fit, including new
$32^3$, $\beta=2.25$, ensembles~\cite{Kelly:2009fp,Mawhinney:2009jy,RBC:2010}.
 Later, we use the slight difference as a systematic error.}
is $a^{-1}=1.784~(44)$ GeV, as determined
from the $\Omega$ baryon mass on the larger lattice,
and which yields physical volumes
$(1.76~\rm{fm})^3$ and $(2.65~\rm{fm})^3$, respectively.
 The domain wall height $M_5$ and the size of the extra dimension $L_s$
are 1.8 and 16, respectively.
The residual quark mass in the chiral limit for pure QCD is found to be
$m_{\rm res}=0.003148(46)$ and 0.003203(15), for the $16^3$ and 24$^3$
lattice sizes, respectively. The latter is slightly larger than the
value 0.00315(2) determined in Ref.~\cite{Allton:2008pn} on a smaller ensemble of configurations.

 The ensembles and number of measurements on each
are summarized in Table~\ref{tab:ensembles}.
 The stopping criterion in the conjugate-gradient algorithm
used to compute quark propagators was $10^{-8}$,
the same as in Ref.~\cite{Allton:2008pn}.
 To increase our statistics on some of the ensembles,
two or more different locations of the source
are used on each configuration (see Table~\ref{tab:ensembles}).
 The QCD configurations are separated by 20 or 40 Monte Carlo time units 
to suppress the autocorrelations in them.
 Our calculation is for the pseudo-scalar meson at the unitary point
$(m_1=m_3=m_4=m_5)$ 
and the partially quenched point(arbitrary quark mass combination).

 The quenched QED configurations were generated
on the non-compact manifold~\cite{Blum:2007cy,Hayakawa:2008an}.
 Here we employ the Feynman gauge instead of the Coulomb gauge 
which was used previously \cite{Blum:2007cy} in our two-flavor calculation.
 Since the mass is a gauge invariant quantity,
the result should be consistent within the statistical error,
up to the effects of zero-modes.
 Further, removal of the modes also results in the satisfaction
of Gauss' Law on the torus~\cite{Hayakawa:2008an}.
 An advantage of the non-compact QED formalism is that 
the $U(1)$ gauge potential $A_\mu$ can be chosen randomly
with the correct Gaussian distribution in momentum space,
then Fourier transformed to coordinate space,
so there are no autocorrelations in the ensemble.
 Finally, yet another advantage
is that there is no lattice-artifact photon self-interactions
in the action.
 To couple $A_\mu$ to the fermions, the non-compact field
is exponentiated to produce the photon link, 
$U^{(1)}_{x,\mu} = \exp{(i e A_{x,\mu})}$,
where $e=\sqrt{4\pi\alpha_{\rm em}}\approx 0.30286$.

 Since the QED interaction does not confine, it is possible 
that the finite volume may induce a significant systematic error.
 We thus do our simulation on both $16^3$ and $24^3$ lattice configurations
with the same lattice spacing.
 This allows direct investigation of the finite volume effect in the mass spectrum.

\begin{table}[h]
 \centering
 \begin{tabular}{ccccccc}
  \hline  
  lat & $m_{\rm sea}$ & $m_{\rm val}$ & Trajectories & $\Delta$ &
  $N_{\rm meas}$ & $t_{\rm src}$\\
  \hline
   $16^3$ & 0.01 & 0.01, 0.02, 0.03 & 500-4000 & 20  & 352 & 4,20 \\
  $16^3$ & 0.02 & 0.01, 0.02, 0.03 & 500-4000 & 20  & 352 & 4,20 \\
  $16^3$ & 0.02 & 0.01, 0.02, 0.03 & 500-4000 & 20  & 352 & 4,20 \\
  \hline
  $24^3$ & 0.005 & 0.001, 0.005, 0.01, 0.02, 0.03 & 900-8660 & 40  & 195 & 0 \\
  $24^3$ & 0.01 & 0.001, 0.01, 0.02, 0.03 & 1460-5040 & 20 &  180 & 0 \\
  $24^3$ & 0.02 & 0.02 & 1800-3580 & 20 & 360 & 0,16,32,48 \\
   $24^3$ & 0.03 & 0.03 & 1260-3040 & 20 & 360 & 0,16,32,48 \\
    \hline
 \end{tabular}
 \caption{Ensembles of QCD gauge field configurations generated by
   the RBC and UKQCD collaborations~\cite{Allton:2007hx,Allton:2008pn,RBC:2010}
   for $\beta =2.13$ with the Iwasaki gauge action that were used in this work. 
   $\Delta$ is the separation between measurements in molecular dynamics
   time units.
   $N_{\rm meas}$ denotes the total number of measurements, 
   and $t_{\rm src}$ is
   the Euclidean time-slice location of the source.}
 \label{tab:ensembles}
\end{table}

\clearpage
\section{Results}\label{sec:results}

In this section we present our results, focusing on the $24^3$ ensembles, for the electromagnetic pseudo-scalar mass splittings ($\Delta M^2$), EM LEC's in SU(3) and SU(2) chiral perturbation theory describing the pseudo-scalar masses, and the quark masses. Results from the $16^3$ ensemble are used for estimating finite volume effects which are discussed extensively in Sec.~\ref{sec:sys errors}.
Before turning to the results for $\Delta M^2$, we first describe lattice-artifact electromagnetic effects induced by the finite size of the fifth dimension of the DWF used to simulate the four dimensional $u$, $d$, and $s$ quarks.

{ In the following the notation $u\bar u$ ($d\bar d$) denotes a meson whose two-point correlation function is made from just the connected quark diagram using degenerate light quarks with equal charges, $q=2/3\,(-1/3)$. Such a meson is neutral, but should not be confused with the $\pi^0$, which requires disconnected quark diagrams.}

\subsection{Electromagnetic effects in $m_{\rm res}$}\label{subsec:mres}

 We first calculate
the residual mass $m_{\rm res}$ \cite{Furman:1995ky,Blum:1998ud,Antonio:2008zz}
from the pure QCD configurations. 
 Then we consider the residual mass from the combined QCD+QED
configurations so that the QED contribution to $m_{\rm res}$
can be extracted.
 In the lattice DWF, $m_{\rm{res}}$ is determined from the ratio
\begin{eqnarray}
R(t)&=&
 \frac{\displaystyle{\left<\sum_{x}J^a_{5q}({\vec x}, t)\pi^a(0)\right>}}
      {\displaystyle{\left<\sum_{x}J_5^a({\vec x}, t)\pi^a(0)\right>}},
     \label{eq:mres ratio}
\end{eqnarray}
where $t$ is the Euclidean time,
$J^a_{5q}$ is a pseudo-scalar density
evaluated at the mid-point of the extra dimension,
$\pi^a$ denotes the usual $4d$ pseudo-scalar density,
and the superscript $a$ is a non-singlet flavor index.
 The correlation functions in Eq.~(\ref{eq:mres ratio})
are computed from wall source, point sink, quark propagators. 
 
 The residual mass is an ultra-violet, additive shift of the input,
bare quark mass. Because we are interested only in the EM meson mass-squared splittings,
the leading order dependence of $m_{\rm res}$ on the bare quark mass
cancels, and
we use a mass-independent residual mass in our later analysis 
that can be identified
by extrapolating $R(t)$ for the unitary quark masses to $m_f=0$ 
with a suitable $t$-average.

\begin{table}
 \centering
 \begin{tabular}{ccc}
  \hline  
   & $16^3$ & $24^3$\\
  \hline  
  $m_{sea}$ & $m_{res}$ & $m_{res}$\\
  \hline
  chiral limit & 0.003148(46) & 0.003203(15)\\
  $0.005$ & N/A          & 0.003222(16) \\
  $0.01$  & 0.003177(31) & 0.003230(15) \\
  $0.02$  & 0.003262(29) & 0.003261(16) \\
  $0.03$  & 0.003267(28) & 0.003297(15) \\
  \hline
 \end{tabular}
 \caption{The QCD residual mass for
   $16^3$ and $24^3$ lattice sizes. The data correspond to unitary
   mass points. The fit range
   for $R(t)$ (defined in Eq.~(\ref{eq:mres ratio}) is $9 \le t \le N_t/2$.}
 \label{tab:mres}
\end{table}

\begin{figure}[ht]
\includegraphics[scale=1.]{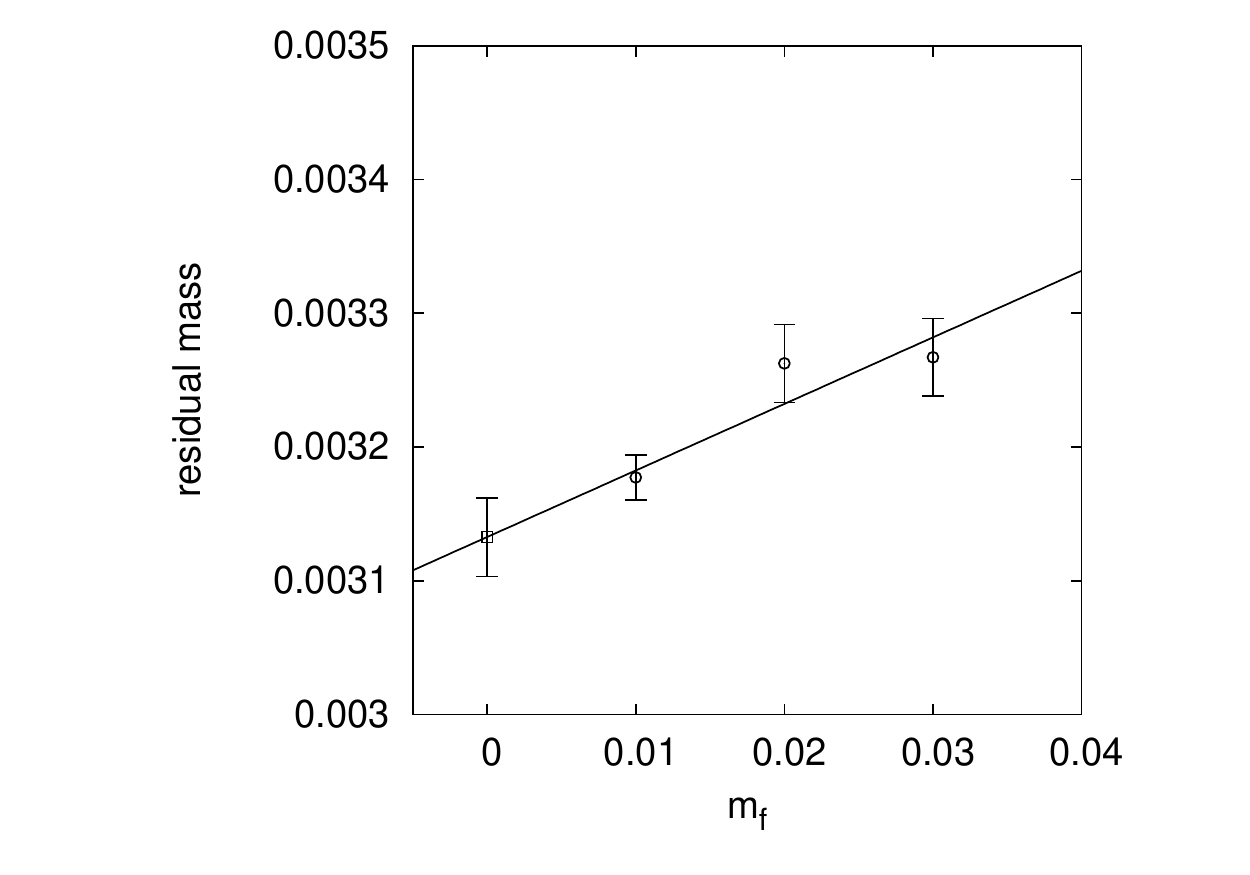}
\includegraphics[scale=1.]{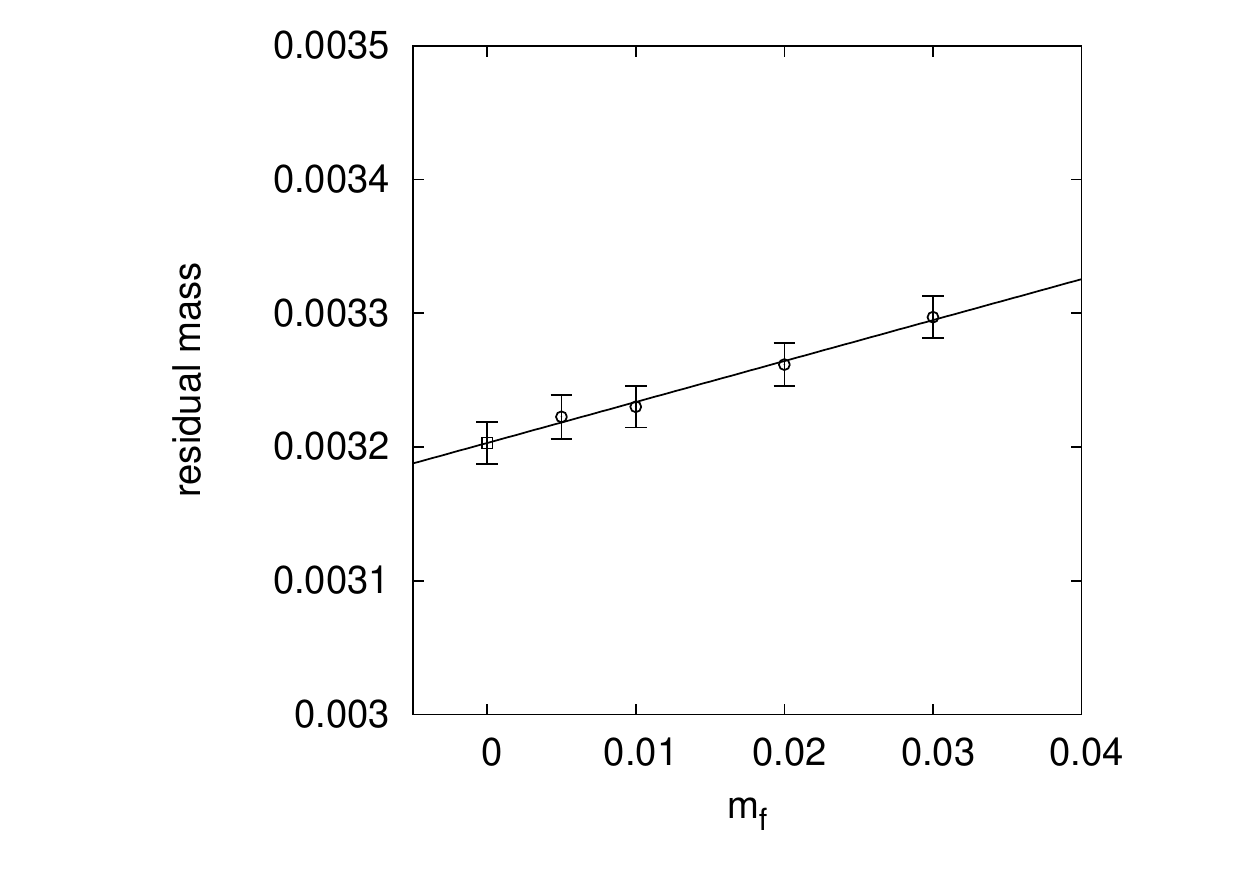}
\caption{The QCD residual mass for
   $16^3$ (upper) and $24^3$ (lower) lattice sizes. The data correspond to unitary
   mass points. The linear chiral extrapolation to the $m_{f}=0$ limit is 
   also shown on the plot.}
\label{Fig:mres}
\end{figure}

 Table~\ref{tab:mres} shows the numerical result of the residual mass 
computed from the QCD configurations alone.
 In each case, $R(t)$ was averaged over the range
$9\le t\le N_t/2$ 
for the size $N_t$ of the lattice in the time direction, 
after folding the correlation function about $N_t/2$.
 Figure \ref{Fig:mres} shows the chiral extrapolation of $m_{\rm res}$.
 The residual mass at the chiral limit is very close between
the $16^3$ and $24^3$ lattices. $m_{\rm{res}}$ is around 0.003,
which is comparable to the lightest input sea quark mass, $m_l=0.005$,
and larger than the smallest valence quark mass $m_{f}=0.001$,
so the effect of the explicit violation of chiral symmetry 
from finite $L_s$ is not negligible in our calculation.
 Our measured values are roughly consistent with those found
by the RBC and UKQCD collaborations~\cite{Allton:2007hx,Allton:2008pn}.

Next we consider the QED contribution to the residual quark mass. The
QED contribution from quark flavor $i$ can be expressed as
\begin{eqnarray}
m_{\text{res},i}({\rm QCD+QED})-m_{\text{res}}({\rm QCD})
= e^2 C_2\,q_i^2,\label{eq:defOfC2}
\end{eqnarray}
where $m_{\text{res},i}({\rm QCD+QED})$ means the residual mass computed on 
the combined QCD+QED configurations and $m_{\text{res}}({\rm QCD})$,
the residual mass computed on the pure QCD configurations.
 Both are evaluated at $m_f=0$ and the former with physical quark charge $q_i$.
$C_2$, which is of order $O(m_{\rm res})$, parametrizes the QED contribution
to the additive shift of the quark mass.
 Although we compute this correction
via the Ward-Takahashi identity for DWF~\cite{Furman:1995ky},
using a neutral meson made with degenerate, equally charged quarks,
the form of Eq.~(\ref{eq:defOfC2}) is completely consistent
with a calculation in weak-coupling perturbation theory,
say from the one-loop self-energy Feynman diagram for a quark with charge $q_i$.
 In our chiral perturbation theory power counting,
the QED contribution to the residual mass is
$O(\alpha_{\rm em} m_{\rm res})$
and must therefore be included in our NLO analysis discussed in the next section.

 To compute the residual mass and extract the EM contribution
via Eq.~(\ref{eq:defOfC2}),
we use $u\bar u$ or $d\bar d$ correlation functions in Eq.~(\ref{eq:mres ratio})
\footnote{
 In an earlier paper~\cite{Blum:2007cy} we mistakenly
included an independent contribution, proportional to $q_i q_j$, to the residual mass
for the charged mesons made of quarks with charges $q_i$ and $q_j$.
 This is clearly inconsistent with flavor conservation
and the definition of a renormalized quark mass defined
in perturbation theory.}
.
 The total contribution to the meson mass-squared
due to explicit chiral symmetry breaking is,
as in the case of pure DWF QCD,
just the sum of contributions from each quark in the meson, modulo higher order than $O(\alpha_{\rm em} m_{\rm res})$-corrections. 
 Table~\ref{tab:mresC1C2} shows the results for $C_2$ from $u\bar u$ and
$d\bar d$ correlation functions. 
 They agree well up to two digits, which implies that the
${O}(\alpha_{\rm em}^2 m_{\rm res})$ contribution is quite small.
 These differences are higher order in chiral perturbation theory
relative to the one we work to in this paper,
so we neglect them.
 We note that $C_2\,e^2\,q_i^2$ is the expected size,
$O(\alpha_{\rm em} m_{\rm res})$.
 The attained statistical precision on $C_2$,
which is impressive, of course stems from the fact that
$m_{{\rm res},i}({\rm QCD+QED})$
and $m_{\rm{res}}({\rm QCD})$ are computed on {\it exactly}
the same set of gluon configurations,
so they are highly correlated,
and the QCD fluctuations cancel between them.
 In addition, $C_2$ appears to be insensitive to the volume (see Tab.~\ref{tab:mresC1C2}),
presumably because the residual mass arises from the UV, short distance, regime.

\begin{table}
 \centering
 \begin{tabular}{ccc}
  \hline
    $L_s$ & $u\bar{u}$ & $d\bar{d}$ \\
  \hline
  \multicolumn{3}{c}{$16^3$ lattice size}\\
  \hline
  16  & 2.597(23) & 2.532(22) \\
  32  & 0.309(16)& 0.301(16) \\
   \hline
  \multicolumn{3}{c}{$24^3$ lattice size}\\
  \hline 
  16 & 2.585(7) & 2.519(7)  \\
  \hline\hline
 \end{tabular}
 \caption{$C_2$ ($\times10^3$) in Eq.~(\ref{eq:defOfC2}), representing
   the electromagnetic contribution to the residual mass.
   $u\bar u$ and $d\bar d$ denote the type of correlation function
   used to extract $C_2$.}
 \label{tab:mresC1C2}
\end{table}

\subsection{Meson mass splittings}

The electromagnetic mass splittings are determined
from the pseudo-scalar masses computed with $e\neq0$ and $e=0$,
using the same gluon configurations.
 We use the additional trick of averaging correlation functions
over $\pm e$,
configuration-by-configuration~\cite{Doi:2006xh,Blum:2007cy}.

\begin{figure}[h]
  \includegraphics[scale=1]{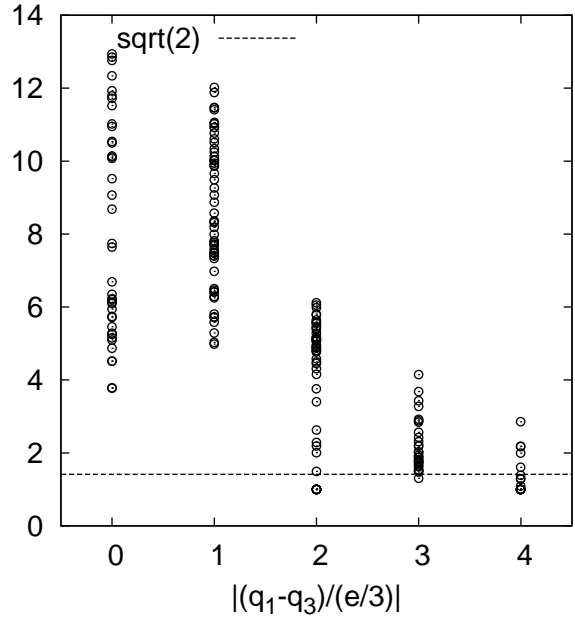}
\caption{A comparison of the statistical errors for the meson mass-squared splitting
with and without the $\pm e$ averaging trick~\cite{Doi:2006xh,Blum:2007cy}.
The vertical axis shows the ratio of the error {\it without} the average
to that with the average, so that larger values indicate smaller
statistical error from the $\pm e$ averaging trick.
In most cases there is a large decrease ($\sim$ 1/10) in the
error over the naive factor of $\sqrt{2}$ that would result
simply from doubling of the measurements
(dashed line). The few points with ratio exactly equal to one correspond to combinations that are trivially invariant under the change $e\to-e$, {\it i.e.}, $m_1=m_3$ and $q_1=-q_3$.
}
\label{fig:epm trick}
\end{figure}

In Fig.~\ref{fig:epm trick},  the improvement due to the $\pm e$ averaging is demonstrated
for the meson mass-squared splitting.
The vertical axis shows the ratio of the statistical error
{\it without} the trick to that with the trick,
so that larger values indicate smaller statistical error
for the $\pm e$ averaging trick.
In most cases there is a large decrease ($\sim$ 1/10)
in the error over the naive factor of
$\sqrt{2}$ that would result simply from doubling of the measurements
(dashed line), while the few points with ratio exactly equal to one correspond to combinations that are trivially invariant under the change $e\to-e$.
This procedure corresponds to including the QED configuration
$-A_{x,\mu}$ for each $A_{x,\mu}$ in the path integral and
can exactly cancel unphysical $O(e)$ noise with finite statistics
which would have obscured the physical $O(e^2)$ signal of interest,
only the latter of which survives in the infinite statistics limit.
Together, the complete procedure yields mass-squared splittings
with sub-percent statistical precision.

\begin{figure}[h]
  \includegraphics[scale=1]{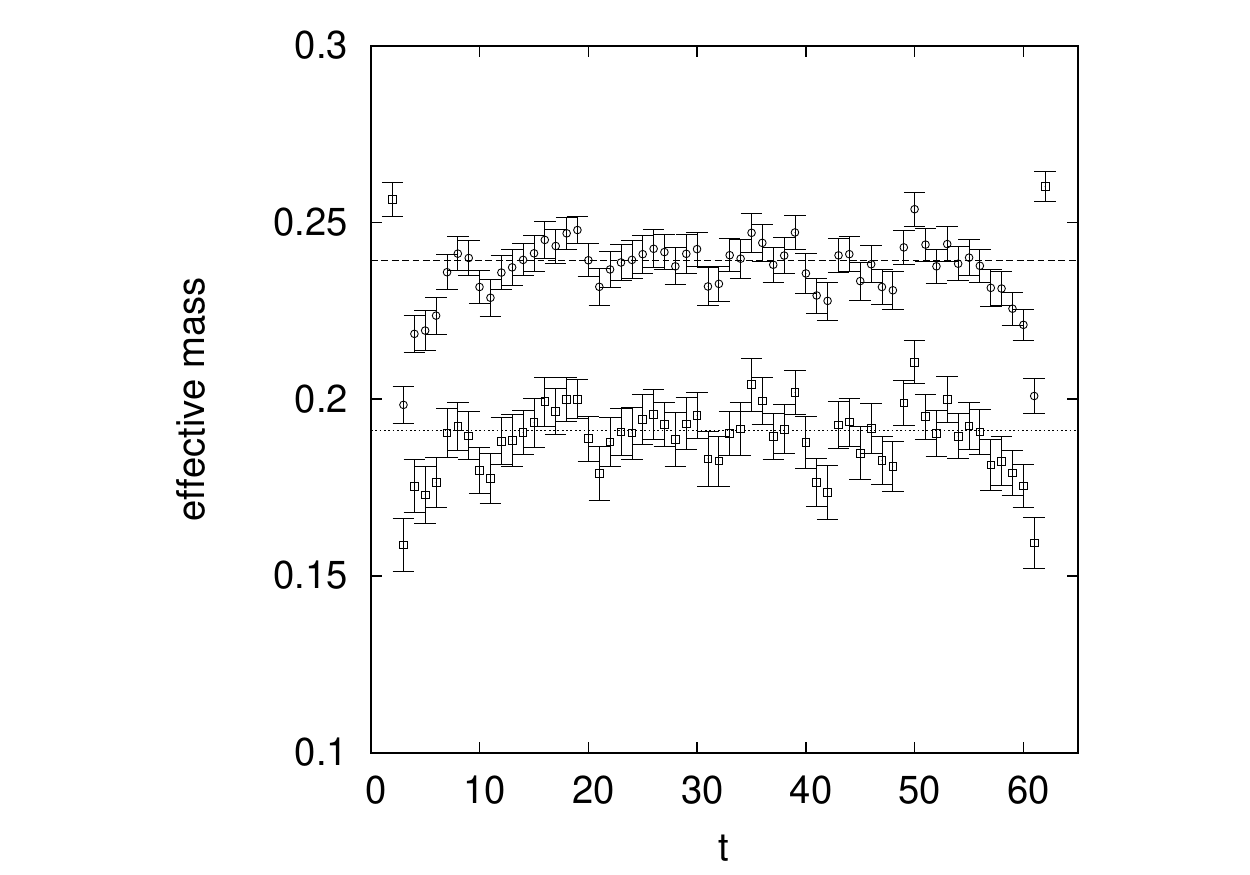}
\caption{Representative effective masses. Lattice size $24^3$. 
  $m_{sea}$=0.005, $m_1$=$m_3$=0.01, $q_1=1/3$ and 
  $q_3=0$ (upper points) and $m_{sea}$=$m_1$=$m_3$=0.005, $q_1=1/3$ and 
  $q_3=-1/3$ (lower points). The horizontal lines represent the fit result.}
\label{fig:eff mass}
\end{figure}

 The pseudo-scalar meson masses are obtained
from single state fits to wall source, point sink correlation functions
with periodic boundary conditions in time with use of the fit function
\begin{eqnarray}
C_{\rm fit}(t-t_{\rm src})&=&A[e^{-M(t-t_{\rm src}+N_t)\%N_t}+e^{-M(N_t-t+t_{\rm src})\%N_t}],
\end{eqnarray}
where $M$ is the ground state meson mass,
and $t_{\rm src}$ is the time slice where the source is placed.
 To improve statistics in some cases,
we average results from two sources (see Table~\ref{tab:ensembles}).
 The fitting procedure is done with the standard $\chi^2$ minimization
(maximum likelihood),
and the error on the mass is obtained by the standard jackknife method.
 Since the meson correlation function is symmetric about the midpoint
(from the source) in the time direction,
we fold the data about this point and fit
with a time range smaller than $N_t/2$.
 Based on the obtained effective masses (a representative example is shown in Fig.~\ref{fig:eff mass}),
for all correlation functions we chose a fit range of $9\le t-t_s\le N_t/2$.

The pseudo-scalar meson masses are tabulated in Tabs.~\ref{tab:mps16}~-~\ref{tab:mps24}.
We have extracted the masses in two ways, one being from the fits to the correlation functions using the full covariance matrix and the other being uncorrelated fits following~\cite{Allton:2008pn,RBC:2010}.
The values of $\chi^2$/dof for the covariant fits are roughly one for
the $24^3$ ensembles, but somewhat higher in some cases for the $16^3$
ones and for the heavier quark masses on both ensembles. Such behavior
for the $16^3$ ensembles was seen in the earlier, pure QCD, analysis
using these configurations, and was attributed to an inferior gauge
field evolution algorithm~\cite{Allton:2007hx}. An improved algorithm was used to generate the $24^3$ ensemble. From Tabs.~\ref{tab:mps16}~-~\ref{tab:mps24} the masses and errors determined with either fit method agree quite well. Our final analysis is based on the masses from the uncorrelated fits in order to be consistent with the analysis in~Refs.~\cite{Allton:2008pn,RBC:2010} from which we take the pure QCD LEC's. 
The typical statistical error on the mass is at the half of a percent level and smaller.

\begin{figure}
  \includegraphics[scale=1]{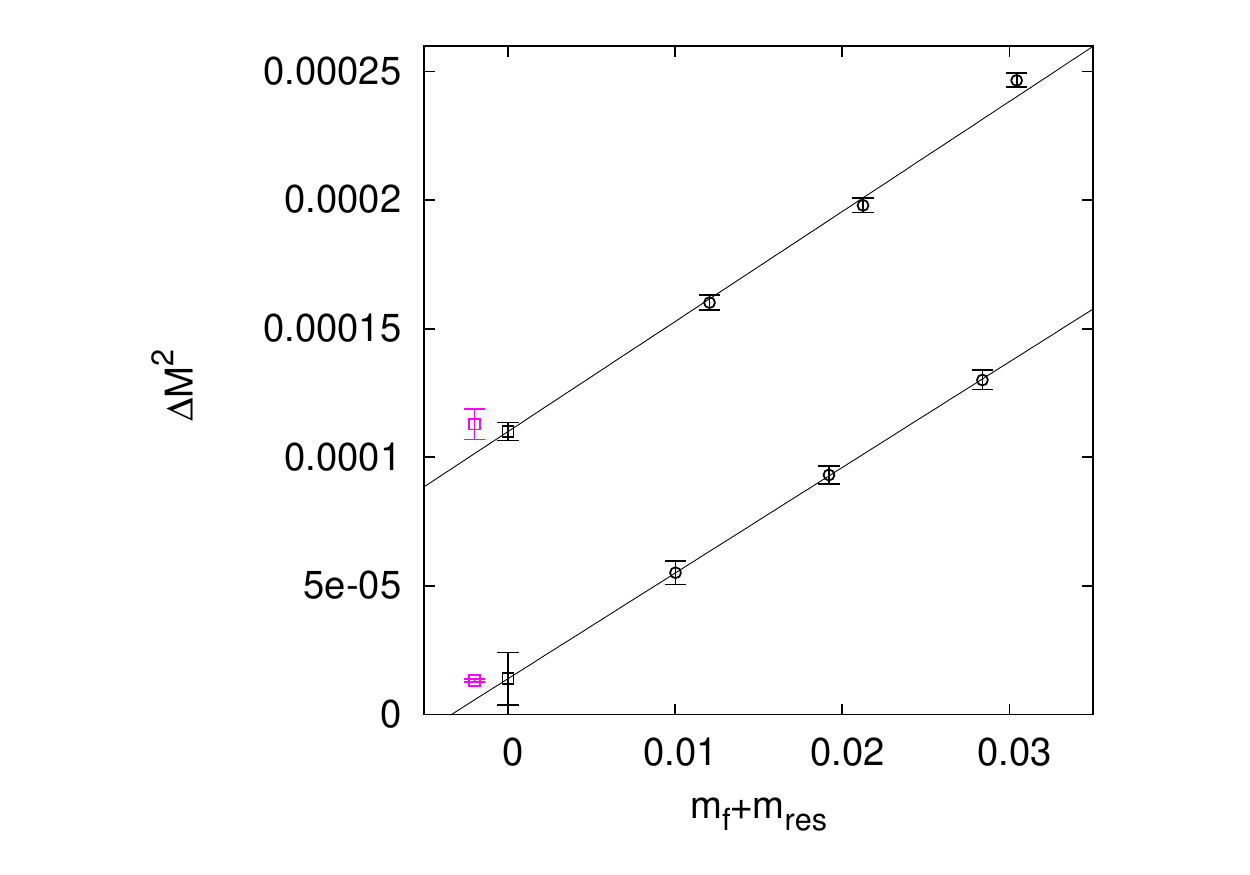}
\caption{$\Delta M^2$ for the  $d\bar d$ meson,
  with $L_s=16$ (upper set of line and plots) and $32$, $16^3$
  lattice size from the SU(3) fit. The extrapolated values (box) are 
  $e^2\delta_{m_{\rm res}}(q_1^2+q_3^2)$. For comparison, we also show the values of 
  $B_0 C_2 e^2 (q_1^2 + q_3^2)$ obtained from the Ward-Takahashi Identity
  (values are slightly shifted horizontally to the left for clarity). 
  $\delta_{m_{\rm res}}$ is obtained from the fit range of 0.01-0.02
  for $L_s=16$ and 0.01-0.03 for $L_s=32$. 
  The error on $B_0 C_2 e^2 (q_1^2 + q_3^2)$ comes mostly from the error on $B_0$.}
\label{fig:dmres 16}
\end{figure}

\begin{figure}
  \includegraphics[scale=1]{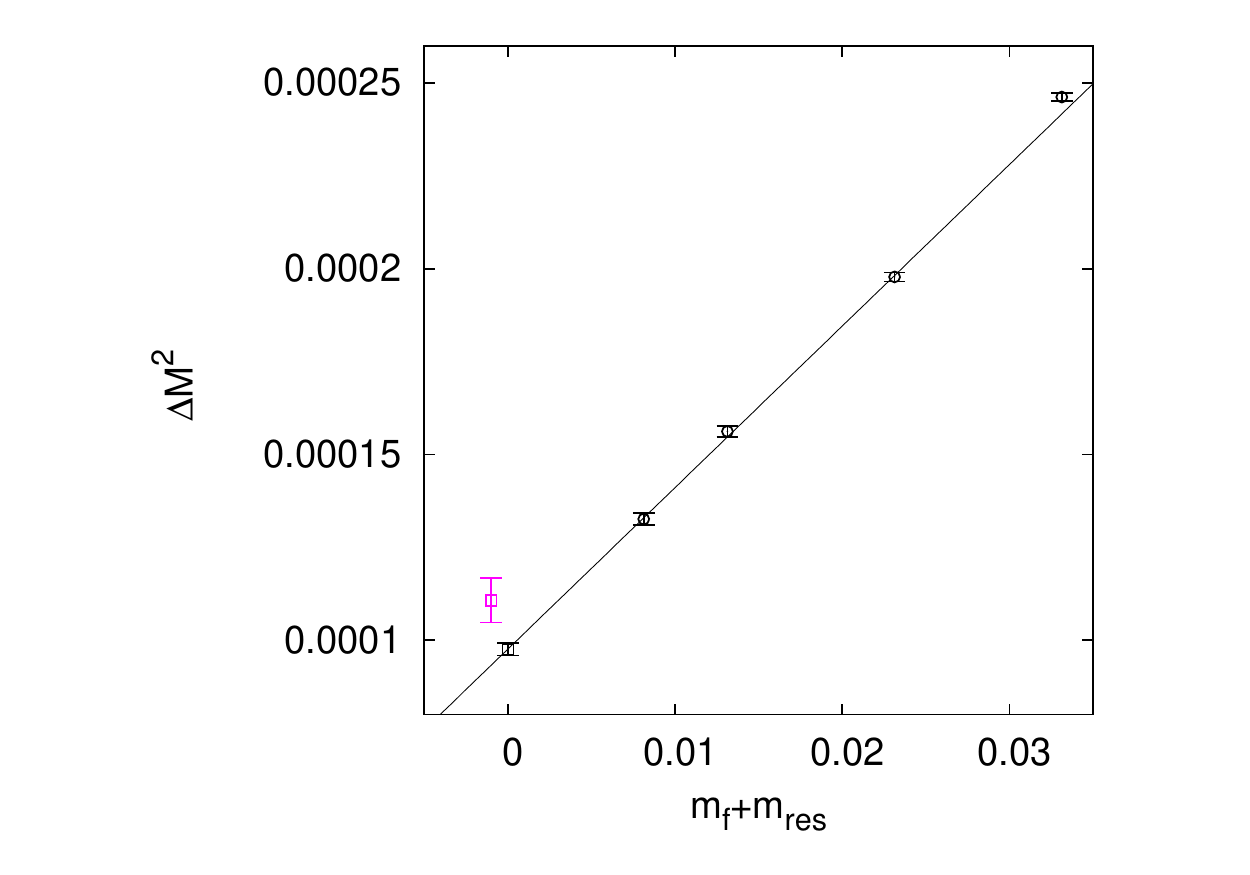}
\caption{Same as Fig.~\ref{fig:dmres 16} but for lattice size $24^3$ and $L_s=16$.}
\label{fig:dmres 24}
\end{figure}

The meson mass-squared splittings are given by $\Delta M^2=
M^2(e\neq0)-M^2(e=0)$, and the errors are again computed using a
jackknife procedure. As an example, in Figs.~\ref{fig:dmres 16} and~\ref{fig:dmres 24},
$\Delta M^2$ for the $\bar d d$ meson is shown. Only the unitary points appear in the figure. 
A full summary of the mass-squared splittings is given in Tabs.~\ref{tab:dm2_16}~and~\ref{tab:dm2_24}. 
The promised statistical precision is observed. 
Even though the errors on the masses themselves are of the same order as the mass difference, the splitting is statistically well resolved under the jackknife analysis thanks to the strong
statistical correlation between $e=0$ and $e\neq 0$. 

We pause to compare the observed explicit chiral symmetry breaking
effects to those expected from the discussion of the residual mass in
the previous section. In the chiral limit, $m_f=-m_{\rm res}(\rm QCD)$,
and in the absence of EM induced explicit chiral symmetry breaking ($L_s\to\infty$), the
neutral meson mass-squared should vanish (up to $\alpha_{\rm em}^2$
corrections which we ignore), and so too should the splittings. But it
is clear from Figs.~\ref{fig:dmres 16}~and~\ref{fig:dmres 24} that the
$\bar d d$ mass-squared splitting does not (the same is true for the $u\bar u$ meson). 
Following the discussion in
Sec.~\ref{subsec:mres} and from the result of the pseudo-scalar mass-squared
at the lowest-order in chiral perturbation theory, the shift in the
splitting in the chiral limit should be 
$2 B_0 C_2 e^2 q^2_d$ or $2 B C_2 e^2 q^2_d$, depending on whether we choose SU(3) or SU(2) chiral perturbation theory. A simple linear fit, also shown in Fig.~\ref{fig:dmres 16}, suggests this is true. Note that at NLO there are no logs in the splitting of neutral mesons made from only connected quark propagators, that is, a ``charged" meson whose net charge happens to be zero. Further, by making $L_s$ larger, this lattice artifact should be (exponentially) reduced, which is also clear from the Fig.~\ref{fig:dmres 16} where for $L_s=32$ the shift has been reduced by roughly a factor of ten, and the splitting nearly vanishes. Similar results hold for the $\bar u u$ mesons. The result based on the Ward-Takahashi Identity depends also on the value of $B_0$ or $B$, depending on whether we choose SU(3) or SU(2) chiral perturbation theory, which introduces some uncertainty. On the other hand, Figs.~\ref{fig:dmres 16}~and~\ref{fig:dmres 24} clearly show this effect is due to finite $L_s$ chiral symmetry breaking, and that it can be precisely subtracted from the physical splitting by introducing a new lattice-artifact LEC to the fit, $e^2\delta_{m_{\rm res}}(q_i^2+q_j^2)$. 
We conclude that the explicit chiral symmetry breaking artifacts induced by finite $L_s$ and QED interactions are precisely quantifiable at NLO in chiral perturbation theory and that higher order terms can be safely neglected, so these artifacts can be robustly eliminated, just as in the case for pure (DWF) QCD.

\subsubsection{infinite volume fits}\label{sec:inf vol}

\begin{figure}[h]
  \includegraphics[scale=0.9]{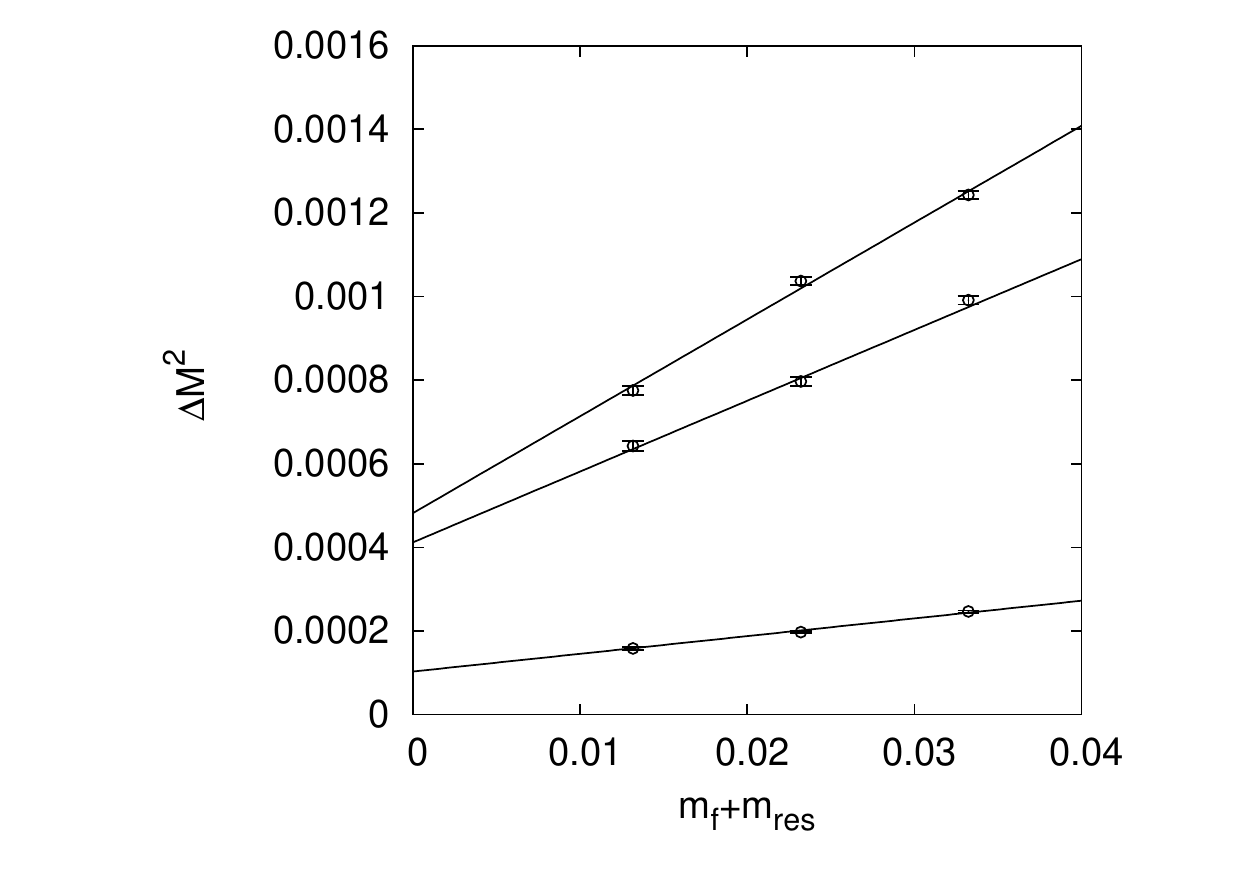}
  \includegraphics[scale=0.9]{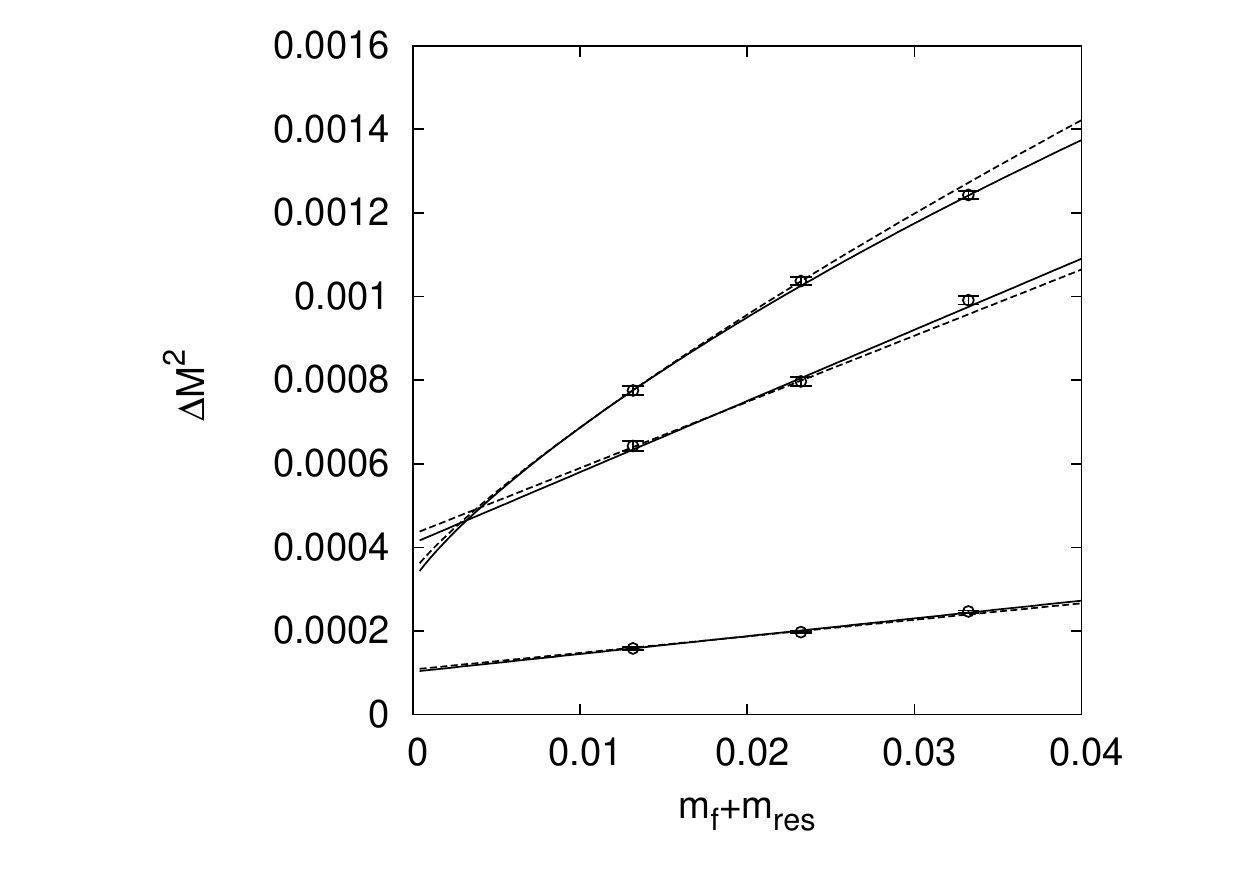}
  \caption{Meson mass-squared splittings. $16^3$ lattice size. Infinite volume linear fit (upper panel) and 
    infinite volume SU(3) chiral log 
    fit (lower panel). The fit range of the linear fit is 0.01-0.03.
    Fit ranges of chiral log fits, 0.01-0.03 (solid line) 
    and 0.01-0.02 (dashed line). Data points 
    correspond to $u\bar d, u\bar u$ and $d\bar d$ mesons, 
    respectively, from top to bottom. Only unitary points are 
    shown, although all of the partially quenched points
    were used in the fit.}
\label{fig:16split}
\end{figure}

\begin{figure}[h]
  \includegraphics[scale=0.9]{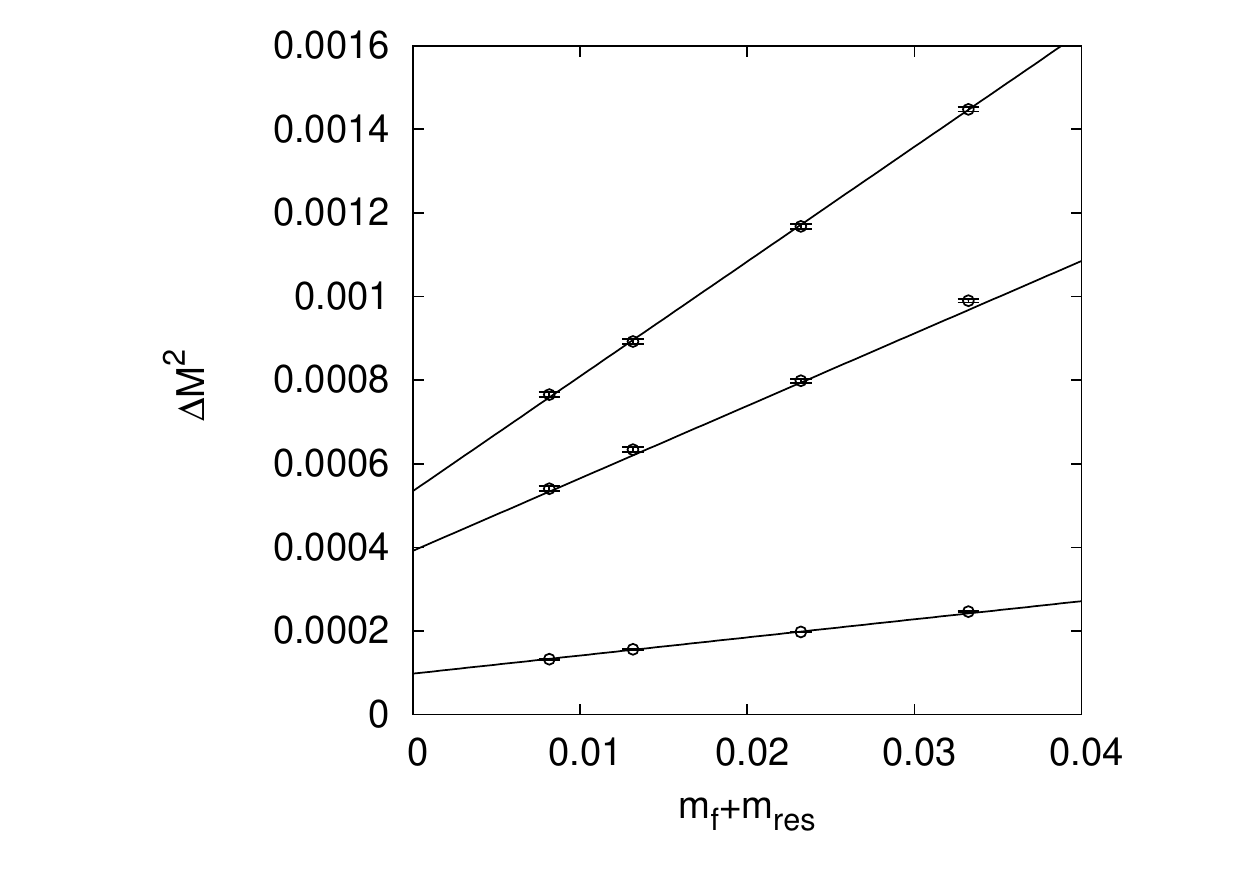}
  \includegraphics[scale=0.9]{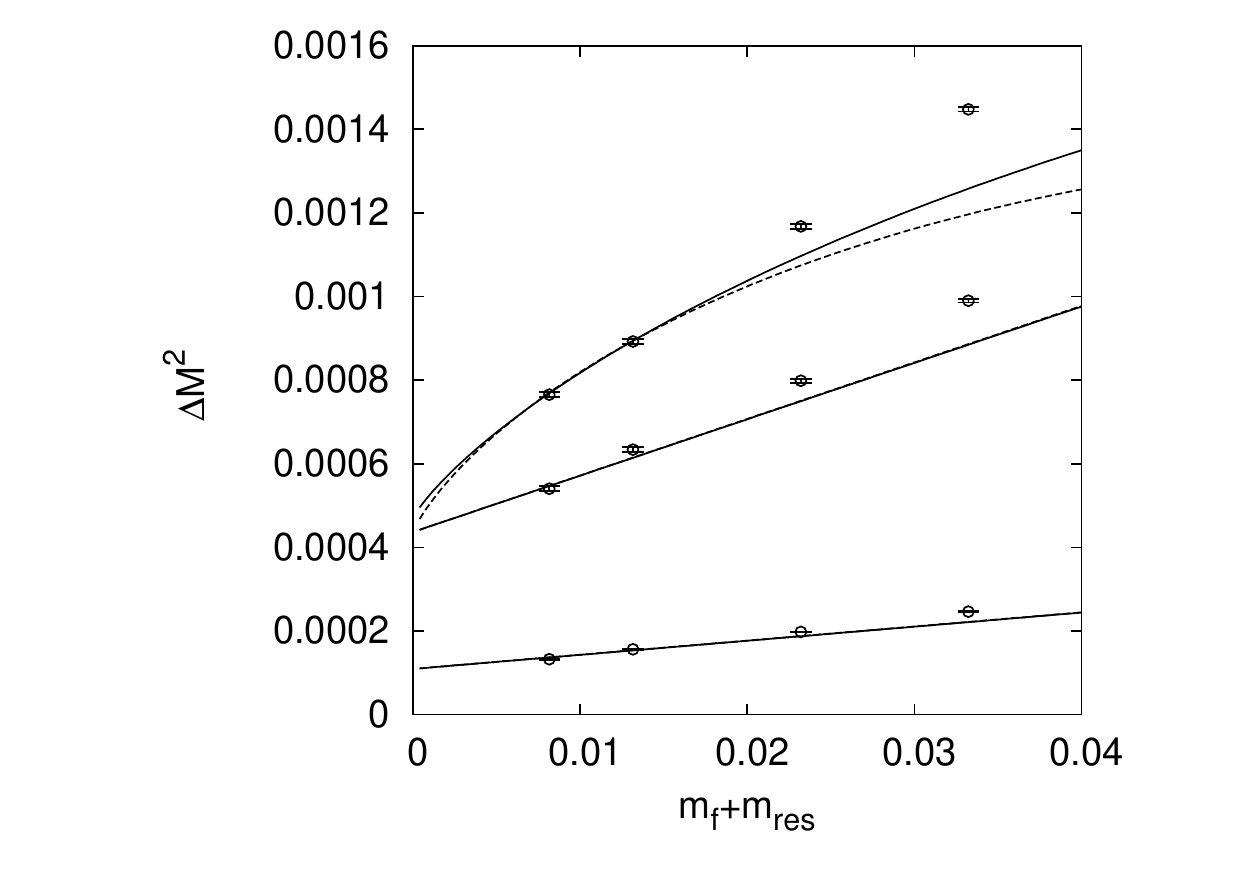}
  \caption{Meson mass-squared splittings. $24^3$ lattice size. Infinite volume linear fit (upper panel), and 
    infinite volume SU(3) and SU(2) chiral log 
    fits (lower panel). The fit range of the linear fit is 0.005-0.03.
    Fit range of chiral log fits is 0.005-0.01. 
    The solid (dashed) line in the lower panel represents the SU(3) 
    (SU(2)) fit. Data points correspond to 
    $u\bar d, u\bar u$ and $d\bar d$ mesons, 
    respectively, from top to bottom. Only unitary points are 
    shown, although all of the partially quenched points
    were used in the fit.}
\label{fig:24split}
\end{figure}

 In Figs.~\ref{fig:16split} and~\ref{fig:24split},
the meson mass-squared splittings are shown
for the unitary quark mass points,
for both $16^3$ and $24^3$ ensembles, respectively.
 For now, we concentrate on the $24^3$ ensemble,
and fit the mass-squared splittings to the infinite volume,
NLO, chiral perturbation theory formulas described
in Sec.~\ref{sec:chipt}.
 The formulas require the values for the pure QCD LEC's,
some of which we have not computed.
 The pure QCD LEC's, including $F_0$, $F$, $B_0$, $B$,
and the Gasser-Leutwyler $L$'s,
have been calculated already by the RBC and UKQCD collaborations
from a larger ensemble of which the present one is a subset.
 We use these values in our fits,
in a combined super-jackknife analysis 
so that the statistical errors on the QCD parameters are fed into our analysis.

 Figure~\ref{fig:24split} shows
the fit to the full ${\rm SU(3)_L \times SU(3)_R}$ NLO formula,
which is summarized in Tables~\ref{tab:qcd lecs}~and~\ref{tab:qed lecs}.
 The quark mass range in the fit is $m_1,m_3\le 0.01$,
and the $\chi^2$/dof for these uncorrelated fits is about two.
 $\chi^2$ degrades significantly if larger quark mass points
are used in the fits.
 Only unitary points are shown in the figure for clarity
while all of the (allowed) quark mass and charge combinations
for the mesons have been used in the analysis.
 For the $24^3$ ensemble,
this amounts to 52 data points for $m_1,\, m_3\le 0.01$.
 The charged meson splittings should not vanish in the chiral limit,
$m_f=-m_{\rm res}$;
this is just the LO Dashen term proportional to $\alpha_{\rm em}$
and the lattice-artifact chiral symmetry breaking.
 The neutral meson splittings do not vanish either due to the latter.
 The chiral logarithms reduce the LO Dashen term relative
to the value given by a simple linear ans\"atz.
{ Recall that the splittings of ``neutral" mesons made from connected quark diagrams only do not contain logs at NLO, so their chiral behavior is particularly simple.}

\begin{figure}[h]
  \includegraphics[scale=0.9]{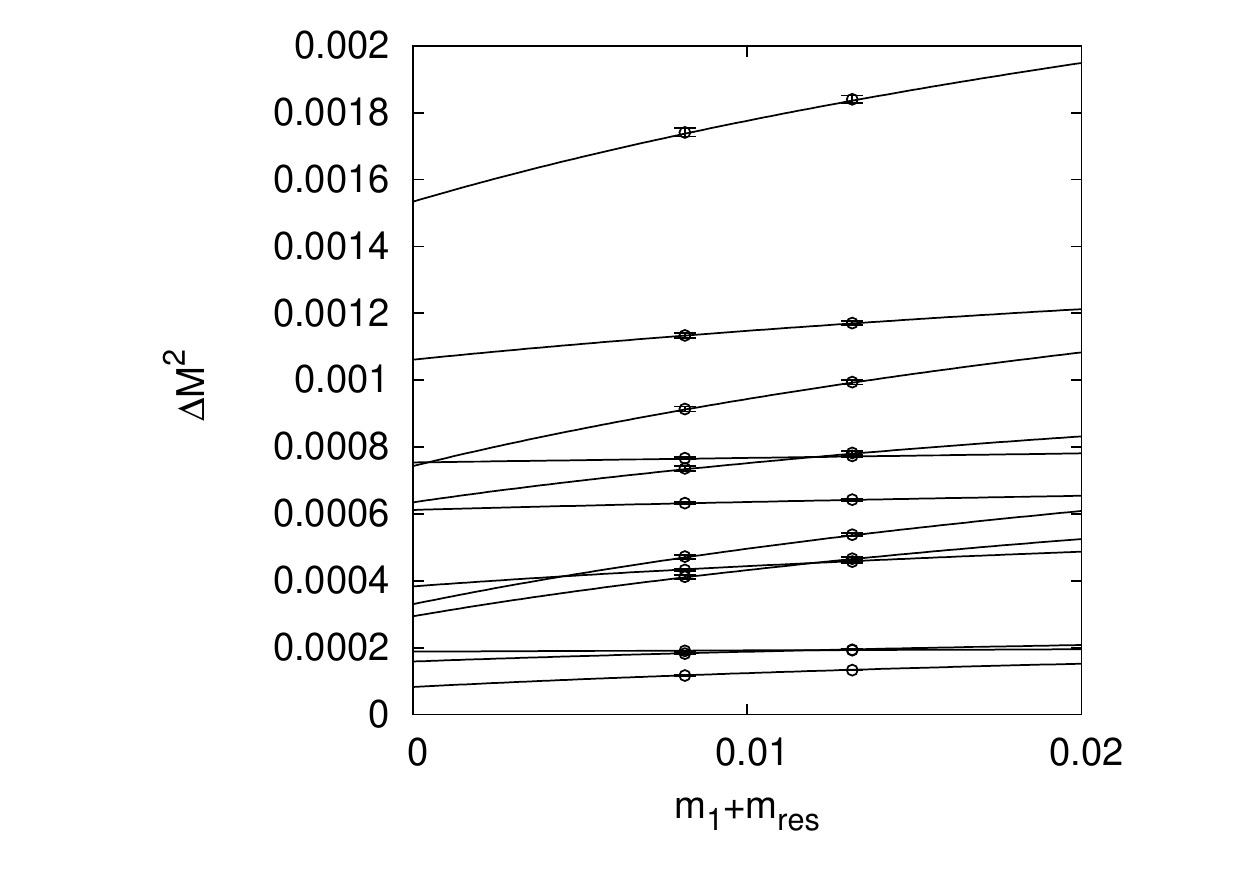}
  \caption{Kaon mass-squared splitting and infinite volume SU(2) kaon fit. 
    The mass of the strange valence quark is fixed
    at 0.03, and $m_{\rm sea}=0.005$. Different lines in the plot correspond to different
    charge combinations of the valence quarks.}
\label{fig:kaon}
\end{figure}

 Figures~\ref{fig:24split}~and~\ref{fig:kaon}
show similar fits for ${\rm SU(2)_L\times SU(2)_R}$-plus-kaon chiral perturbation theory
for the pions and kaons, respectively.
 Here we use the same range for the light quark masses,
and for the kaons
the valence strange quark is fixed to either 0.02 or 0.03. $\chi^2$/dof is similar to the SU(3) case for the pion and also significantly degrades when the quark mass range is extended upwards. For the kaon fits $\chi^{2}$ is small. The total number of data points in the fits are 52 for the pions and 36 for the kaons.
The SU(2) LEC's are also summarized
in Tables~\ref{tab:qcd lecs}~-~\ref{tab:kaonqedlec}. 

{ Before proceeding, we address a subtly in the kaon fits that was not recognized until after the correlation functions had already been computed. Our original plan was to use an SU(3) chiral perturbation theory analysis only, for quark masses in the range $0.005-0.03$, and non-degenerate meson correlation functions were computed for these masses in all possible combinations. However, learning first from the pure QCD analysis~\cite{Allton:2008pn,Kelly:2009fp,Mawhinney:2009jy}, and later from our own, it became clear that 0.02 and 0.03 were too heavy, and that SU(2) chiral perturbation theory would be needed to access the physical strange quark mass regime. We decided to include a lighter valence mass point, 0.001, to augment our fits, but since this was a new, separate calculation, only the mass-degenerate mesons could be computed. Thus, in our kaon fits, we have only two valence and two sea quark mass combinations available for the region $m_{u,d}\le 0.01$. Now the subtly: it turns out these combinations of quark masses and
charges are not enough to constrain all 10 LEC's appearing in Eq.~(\ref{eq:kaon mass sq}). There is one direction in the multi-dimensional parameter space that is not linearly-independent from the rest. Fixing any one of the LEC's to zero, except $A_{5}^{(s,1,1)}$ or $A_{5}^{(s,2)}$, results in a stable fit with the same $\chi^{2}$, but with different values of the LEC's. While these fits all agree exactly when evaluated at the data points used in the fits, they differ elsewhere. 
There are two ways to fix this problem of
an accidental flat direction in the $\chi^2$ function at our disposal. First, keeping the same quark mass range,  use the technique of singular-value-decomposition~\cite{press} (SVD) to determine all 10 LEC's. Second, increase the number of sea or valence quark mass points in the fit, so the parameter directions are all linearly independent. While treating the (next available) mass 0.02 quark simultaneously as light and strange contradicts our assumption that $m_{l}/m_{s}\ll 1$, nevertheless it allows the LEC's to be linearly independent, and only slightly increases $\chi^{2}$ which is still small. 
In practice, we only added the 0.02 valence quark mass to the kaon fit, keeping the light sea quark mass $\le 0.01$. As it happens, the quark masses determined from these two methods agree well, giving confidence that the SVD fit procedure, which we use for our central values, is reliable. 
Further, in the case where 0.02 data points were used, setting each of the LEC's to zero in turn resulted in much bigger $\chi^2$ values except for $x_{6}^{(K)}-  x_{8}^{(K)}$, the ones related to sea quark masses which are not constrained as well. $x_{8}^{(K)}=0$ gave the smallest $\chi^2$. In each of these cases the quark masses agreed within statistical errors to the full SVD fit.
We use the difference in the central values of the quark masses from the two procedures as an estimate of one of the systematic errors due to fitting.
}

 From Table~\ref{tab:qed lecs},
we can see a large effect on $C$ going from SU(3),
where it is almost zero,
to SU(2) where it is almost ten times larger.
 Recall that in the SU(2) theory the contributions
of the { strange quark terms in the SU(3) theory
are absorbed into $C^{(2)}$ (see Eq.~(\ref{eq:su2 dashen})).}
 This situation is reminiscent of the pion decay constant
in pure QCD computed on these lattices
and for the same range of quark masses;
the logs in that case also tend to significantly
reduce the LO contribution over a simple analytic function,
and the physical value~\cite{Allton:2008pn,Kelly:2009fp,Mawhinney:2009jy}.
 Here, especially in the SU(3) case, the effect is even more dramatic.
 The other pion electromagnetic LEC's
are roughly the same in both theories.
In the SU(2) case, the size of the NLO EM correction turns out to
be smaller than the LO one,
showing compatibility with the chiral expansion.
 Finally, in Table~\ref{tab:qed lecs}, we show LEC's corresponding to the phenomenological
parameter set presented in Ref.~\cite{Bijnens:2006mk}. { The fact that the SU(3) NLO LEC's computed here (left-most column) do not agree is not surprising since the LO LEC, $C$, is clearly underestimated by a large degree. Note that to compare values of $C$, a factor of $a^{-4}$ needs to be introduced, as well.}
 We discuss the Dashen term further in Sec.~\ref{sec:sys errors},
after presenting the finite volume fits.


\begin{table}[ht]
  \centering
  \begin{tabular}{ccc}
    \hline\hline
    & SU(3) inf.v & SU(2) inf.v \\
    \hline
    $10^0B_0$ & 2.15(11) & 2.348(44) \\
    $10^2F_0$ & 3.43(19) & 4.55(10) \\
    $10^6(2L_6-L_4)$ & $-$2.6(29.6) & 2.9(45.3) \\
    $10^4(2L_8-L_5)$ & 5.42(29) & 4.36(31)\\
    $10^5L_4$ & 1.7(5.5) & 2.48(89) \\
    $10^4L_5$ & 2.02(63) & 5.49(47) \\
    $10^3m_{\rm res}$ & 3.131(27) & 3.131(27)\\
    $a^{-1}({\rm GeV})$ & 1.784(44) & 1.784(44)\\
    \hline\hline
  \end{tabular}
  \caption{The QCD LEC's from RBC/UKQCD collaboration's infinite volume fits 
    on $24^3$ lattices with SU(3) and SU(2) PQ$\chi$PT~\cite{RBC:2010}. They were computed from a larger ensemble of lattices than used in~\cite{Allton:2008pn}. All of the QCD LEC's are 
    defined at the chiral scale $\Lambda_\chi=1$ GeV. The labels in the first column correspond to SU(3) definitions; the analogous LEC for SU(2) is given in the third column.}
    \label{tab:qcd lecs}
\end{table}

\begin{table}[ht]
  \centering
  \begin{tabular}{cccccc}
    \hline\hline
    & \multicolumn{2}{c}{SU(3)} & \multicolumn{2}{c}{SU(2)} \\ \hline
    & inf.v. & f.v. & inf.v. & f.v. & SU(3)+phenom.\\
    \hline
    $10^7C$ & 2.2(2.0) & 9.3(2.4) & 18.3(1.8) & 32.9(2.3) & 41 \\
    $10^2Y_2$ & 1.63(10) & 1.451(92) & 1.416(50) & 1.301(49) & 0.19\\
    $10^3Y_3$ & $-$11.85(74) & $-$5.37(70) & $-$10.61(62) & $-$4.00(62) & 1.25\\
    $10^3Y_4$ & 13.4(1.7) & 9.7(1.7) & 10.6(1.1) & 8.3(1.1) & 2.17\\
    $10^3Y_5$ & 2.06(72) & 1.12(74) & 1.95(50) & 1.61(54) & $-$1.17\\
    $10^3\delta_{m_{\rm res}}$ & 5.356(98) & 5.357(98) & 5.355(98) &
    5.355(98) & - \\
    \hline
    Dashen's term(MeV) & 0.40(37) & 1.68(39) & 1.88(18) & 3.38(23) & 3.7\\
    \hline
    $\chi^2$/dof & 2.11(73) & 2.12(78) & 2.27(78) & 2.19(82) & - \\
 \hline\hline
  \end{tabular}
  \caption{The SU(3) PQ$\chi$PT and SU(2) pion PQ$\chi$PT QED LEC's from fits 
    of the mass-squared splittings measured on the $24^3$ lattices. 
    All of the LEC's are defined at chiral scale $\Lambda_\chi=1$
    GeV and are given in lattice units. The quark mass range in the fits is $m_{1,3}<=0.01$. ``inf.v.'' and ``f.v.''
    means infinite and finite volume
    fits, respectively. ``SU(3)+phenom.''  refers to a parameter set presented based on phenomenology and using SU(3) $\chi$PT~\cite{Bijnens:2006mk}. Labels in the first column correspond to SU(3) definitions. ``Dashen's term" is the LO result for the mass splitting in the chiral limit.}
    \label{tab:qed lecs}
\end{table}

\begin{table}[ht]
  \centering
  \begin{tabular}{cccccccc}
    \hline
    \hline
    & \multicolumn{3}{c}{inf.v} & \multicolumn{3}{c}{f.v} \\ \hline
    $m_s^{\rm val}$ & 0.02 & 0.03 & $m_s^{\rm phys}$ & 0.02 & 0.03 & $m_s^{\rm phys}$\\
    \hline
   \hline  
   $10^2M^2$ & 4.804(88) & 6.89(10)    &  7.37(36) & - & - & -\\
   $10^1A_3$ & -2.199(44) & -2.198(45) &  -2.198(46) &  - & - & - \\
   $10^2A_4$ & -1.89(45) & -2.15(52)   &  -2.21(56) & - & - & -\\
   \hline
   $10^3A_K^{(1,1)}$ & -9.1(1.1) & -8.9(1.3)   &  -8.8(1.4)  & -6.4(1.0) & -5.8(1.2) & -5.7(1.3) \\
   $10^3A_K^{(2,1)}$ & 8.29(86) & 8.15(99)     &  8.1(1.0)   & 7.16(81) & 6.92(93)   & 6.87(99)  \\
   $10^2A_K^{(s,1,1)}$ & 0.958(26) & 1.254(30) &  1.321(64)  & 1.241(26) & 1.577(31) & 1.653(70) \\
   $10^3A_K^{(s,2)}$ & -4.22(20) & -4.68(22)   &  -4.79(25)  & -6.74(20) & -7.56(23) & -7.75(28) \\
   $10^2x^{(K)}_3$ & 1.41(32) & 1.93(39)       &  2.05(46)   & 2.34(35) & 3.00(42)   & 3.15(52)  \\
   $10^2x^{(K)}_4$ & 4.60(36) & 5.06(47)       &  5.16(50)   & 3.52(38) & 3.83(49)   & 3.90(51)  \\
   $10^1x^{(K)}_5$ & 0.376(42) & 0.366(51)     &  0.364(53)  & 0.361(41) & 0.350(50) & 0.348(53)  \\
   $10^2x^{(K)}_6$ & -0.83(0.94) & -0.99(1.01) &  -1.0(1.0)  & -0.086(0.959) & -0.14(1.02)& -0.16(1.05)\\
   $10^2x^{(K)}_7$ & -0.11(1.82) & -0.27(2.00) &  -0.30(2.05)& -0.81(1.82) & -1.0(2.0) & -1.1(2.0)\\
   $10^2x^{(K)}_8$ & -8.28(47) & -8.65(78)     &  -8.73(86)  & -8.23(47) & -8.60(78)   & -8.69(86)\\
   \hline

    $\chi^2$/dof & 0.4578(52) & 0.2869(40) & - & 0.4578(52) & 0.2869(40) & -\\
    \hline\hline
  \end{tabular}
  \caption{Kaon QCD and QED LEC's extracted from $24^3$ lattice size data. LEC's are in lattice units. The kaon 
    is composed by one light- ($m_1$) and one strange- ($m_3$) quark. 
    We choose $m_1\leq 0.01$ and $m_3$=0.02 or 0.03. The light sea quark 
    is chosen as $m^{sea}\leq 0.01$. The mass of the strange 
    sea quark is fixed at 0.04. The kaon QCD LEC's are quoted from
    RBC/UKQCD's work~\cite{RBC:2010}. $\chi^2$/dof refers to the fit using the SVD method \cite{press}.}
  \label{tab:kaonqedlec}
\end{table}

\subsubsection{finite volume fits}\label{sec:fin vol}

\begin{figure}[h]
  \includegraphics[scale=0.9]{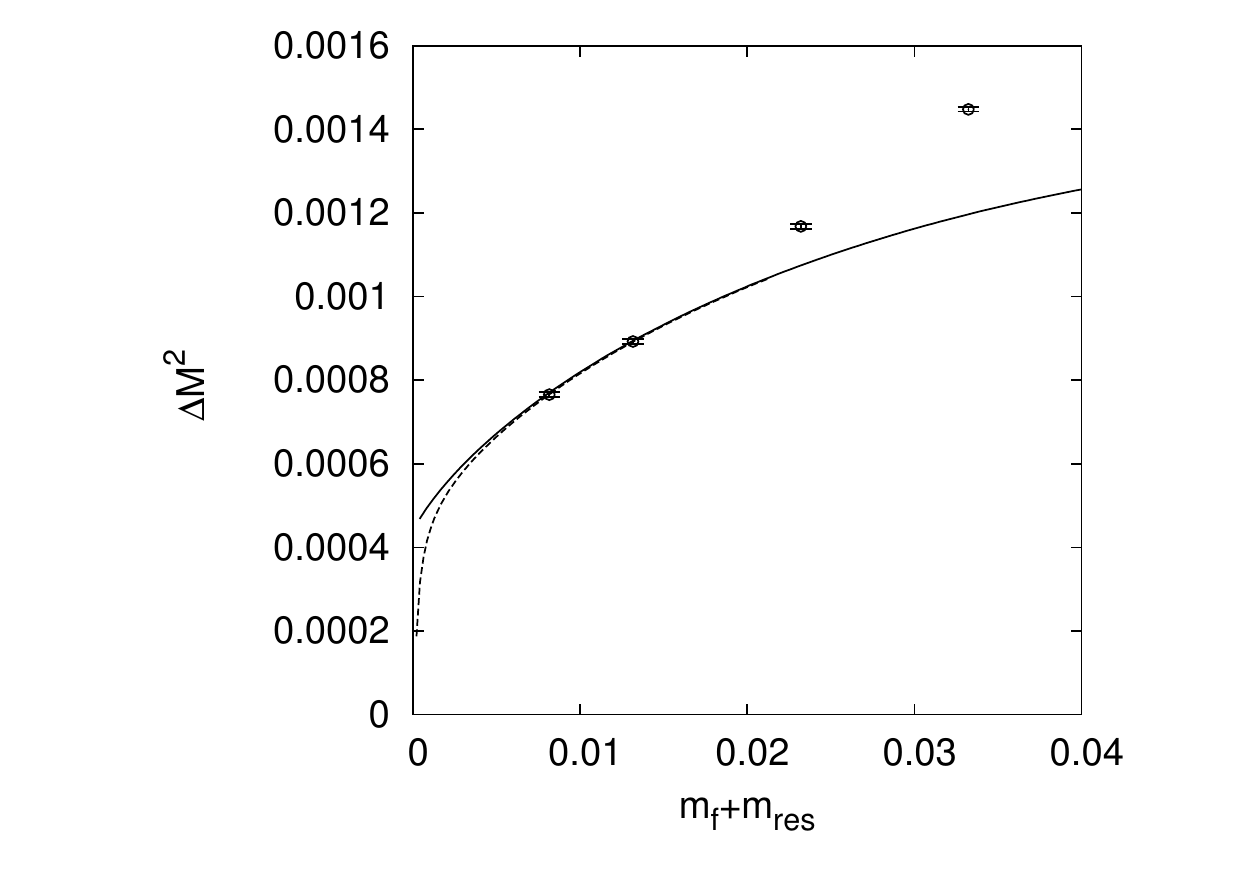}
  \includegraphics[scale=0.9]{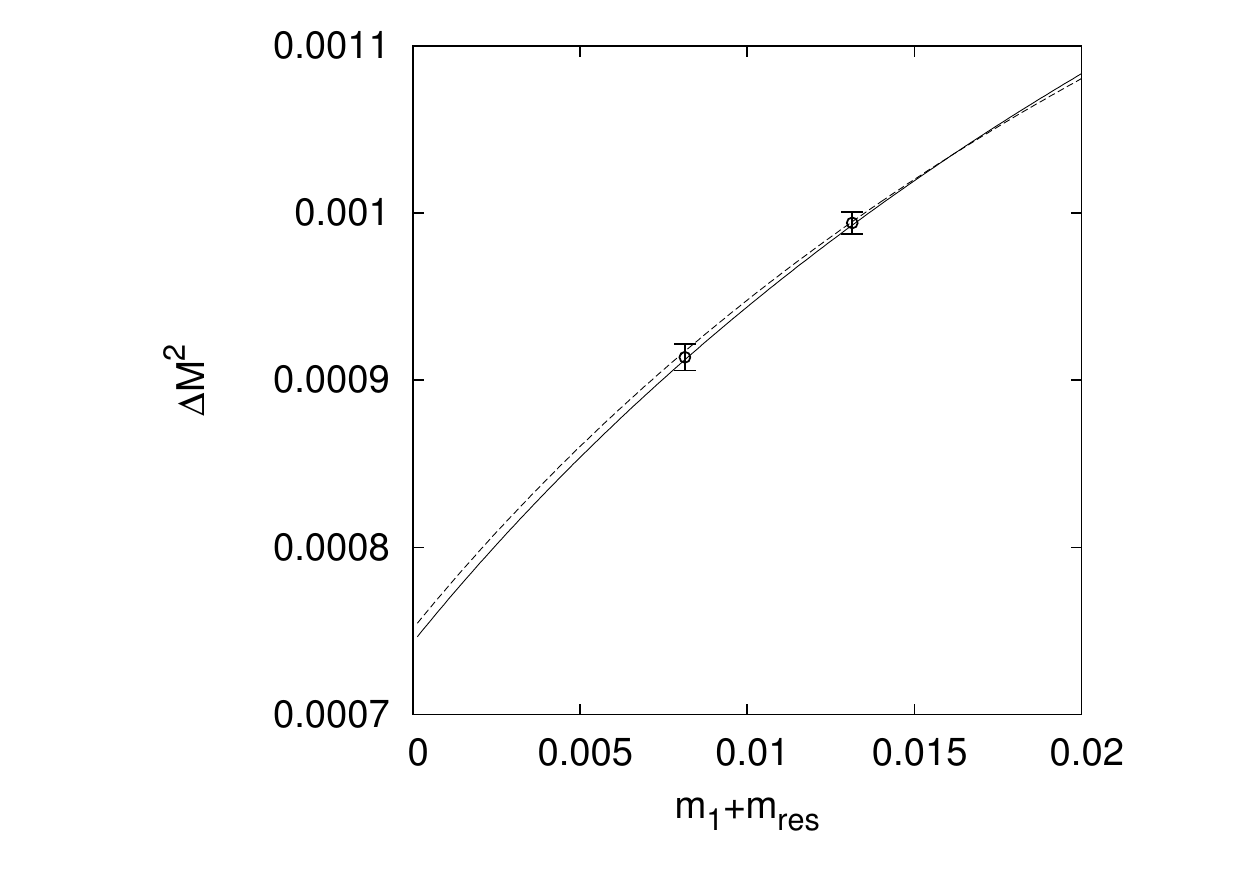}
  \caption{$24^3$ SU(2) chiral log infinite volume and finite volume
    fits for pion (upper) and kaon (lower) mass-squared splittings. 
    Lines correspond to fit results. The fit 
    range is 0.005-0.01. The solid (dashed) line
    represents the infinite (finite) volume fit. In the upper panel, the fit curves are evaluated for degenerate unitary light quarks. For the lower panel, the curves are evaluated for $m_{sea}=0.005$ and  $m_3=0.03$. Data points in the plot correspond to $q_1=2/3e$ and 
$q_3=-1/3e$, but all partially quenched points allowed by the fit range 
were used in the fit.}
  \label{fig:fin vol fit}
\end{figure}

Next we include in our fits the finite volume corrections to the chiral
logarithms
using Eq.~(\ref{eq:fv_su3_pion}) for the SU(3) fit,
and Eq.~(\ref{eq:fv_su2_pion})
for the pion and the results in Appendix \ref{subsec:kaon_su2}
for the kaon in the SU(2) fit.
 We continue to use the pure QCD, infinite volume, LEC's from~\cite{RBC:2010}. Since the finite volume effects in QCD are very small compared to the QED ones, we ignore the former.
Figure~\ref{fig:fin vol fit} shows the modified fits for the pions and kaons on the $24^3$ lattice. The LEC's are given in Table~\ref{tab:qed lecs} and~\ref{tab:kaonqedlec}. The largest change by far is in $C$, the LO Dashen term, which roughly doubles { in the SU(2) case and increases by a factor of four in the SU(3) case. Note, it is still much larger for the SU(2) fit}. This is consistent with the observed large effect in the charged meson splitting compared to the neutral. Fortunately, this huge change does not greatly affect the values of the quark masses, as we shall see. 

\subsection{quark masses}\label{sec:quark masses}

 Having determined the LEC's to NLO describing the pseudo-scalar masses
in chiral perturbation theory,
we now turn to fixing the physical quark masses
at the (arbitrary) low energy scale of 2 GeV.
 First, the bare quark masses are determined
by solving Eq.~(\ref{eq:su3 pion})
or Eqs.~(\ref{eq:su2 pion mass}) and~(\ref{eq:kaon mass sq})
evaluated at the physical meson masses~\cite{Amsler:2008zzb} (in MeV)
\begin{eqnarray}
 M_{\pi^\pm}&=&139.57018\pm0.00035\\
 M_{K^0}&=&497.614\pm0.024\\
 M_{K^\pm}&=&493.667\pm0.016,
\end{eqnarray}
where only the central values are used in our analysis since the errors are negligible compared to the lattice results.
 Using $a^{-1}=1.784(44)$ GeV
and the pure QCD nonperturbative renormalization constant
$Z_m=1.546~(2)~(43)$~\cite{RBC:2010}
computed by the RBC and UKQCD collaborations,
$\overline{MS}$ light quark masses are given
in Table~\ref{tab:quark masses}, for infinite volume, finite volume,
SU(3), and SU(2) fits.
 We have not included the $O(\alpha_{\rm em})$ renormalization
of the quark mass from QED interactions.
 These are similar to those found
in our earlier two flavor work~\cite{Blum:2007cy}, also using DWF,
but which used a more crude chiral perturbation theory analysis
that did not include logarithms.
 The strange quark mass is somewhat lower here,
which may be a real flavor-dependent effect~\cite{Scholz:2009yz,Blum:2007cy}.
 We also note that in the combined continuum limit analysis
mentioned earlier, the RBC and UKQCD collaborations
find that the strange quark mass is even smaller~\cite{Kelly:2009fp,Mawhinney:2009jy,RBC:2010}.
 The average light quark mass is close to the value determined
in pure QCD~\cite{Allton:2008pn,Kelly:2009fp,Mawhinney:2009jy}.

\begin{table}[ht]
  \centering
  \begin{tabular}{ccccc}
    \hline\hline
    & \multicolumn{2}{c}{SU(3)} & \multicolumn{2}{c}{SU(2)} \\ \hline
    & inf.v & f.v & inf.v. & f.v. \\
    \hline
    $m_u$ & 2.606(89) & 2.318(91) & 2.54(10) & 2.24(10) \\
$m_d$ & 4.50(16) & 4.60(16) & 4.53(15) & 4.65(15) \\
$m_s$ & 89.1(3.6) & 89.1(3.6) & 97.7(2.9) & 97.6(2.9) \\
$m_d-m_u$ & 1.900(99) & 2.28(11) & 1.993(67) & 2.411(65) \\
$m_{ud}$ & 3.55(12) & 3.46(12) & 3.54(12) & 3.44(12) \\
$m_u/m_d$ & 0.578(11) & 0.503(12) & 0.5608(87) & 0.4818(96) \\
$m_s/m_{ud}$ & 25.07(36) & 25.73(36) & 27.58(27) & 28.31(29) \\
    \hline\hline
  \end{tabular}
  \caption{The $u$, $d$ and $s$ quark masses
    determined from QCD+QED interaction on $24^3$ lattices.
    The values are given in MeV and the $\overline{\rm MS}$ 
    scheme at renormalization scale $\mu=2$ GeV. SU(3) or SU(2) mean 
    quark masses from SU(3) PQ$\chi$PT or SU(2) PQ$\chi$PT + kaon 
    theory.}\label{tab:quark masses}
\end{table}

\clearpage

\section{Systematic errors}\label{sec:sys errors}

In this section we examine the important systematic errors
in our calculation: the chiral extrapolations, finite volume,
non-zero lattice spacing, and QED quenching.
 In each case we estimate the size of the effect on the values of the quark masses
and investigate the effect on the LO electromagnetic LEC's.
 Similar systematic uncertainties have been given
for the pure QCD sector~\cite{Allton:2008pn,Kelly:2009fp,Mawhinney:2009jy}. 

 To estimate the systematic errors,
the change in a quantity is computed
under the influence of a change in how that quantity is computed,
for example, by using a different fit formula.
Since the data is the same, or there is significant overlap,
in each case, we compute
the change under the (super-)jackknife procedure
in order to assess its significance. Central values of all quantities are quoted for the finite volume, SU(2) chiral perturbation theory fits which we believe give the most accurate results. The systematic errors computed in the following come from comparison to these central values. 

\subsection{chiral extrapolations}

 Previous studies have used the difference
in analytic and chiral perturbation theory fits
to estimate the chiral extrapolation error
that stems from using unphysical heavy
quarks~\cite{Allton:2008pn,Antonio:2007pb,Kelly:2009fp,RBC:2010,Lellouch:2009fg,Durr:2008zz}.
 One can also estimate the error in chiral perturbation theory alone
by comparing the relative sizes of LO, NLO, or even NNLO corrections
to a given quantity. For the latter to work, 
the estimates of the higher order contributions must be accurate. 

 It is perhaps not surprising to find that the meson mass-squared splittings
show little trace of the chiral logarithms. 
 For the mass range of pions in this study,
it is well known that low energy observables
like the meson mass-squared or decay constant exhibit more or less
linear dependence on the quark mass.
 In Fig.~\ref{fig:24split},
the charged pseudo-scalar splitting appears linear
over the range of unitary points shown in the figure.
 Nevertheless, the fits to our data do show
that NLO chiral perturbation theory (chiral logs)
is {consistent} with the data.
 A similar conclusion was reached
in the pure QCD case~\cite{Allton:2008pn,Kelly:2009fp,Mawhinney:2009jy,RBC:2010}.
To NLO in chiral perturbation theory,
there are no logs for the neutral mesons { made from connected quark propagators like those studied here},
and indeed the neutral splittings, too, appear to be quite linear. 

{ We do point out one aspect of the EM logs that leads one to expect a noticeable affect. They behave like $\alpha m \log m$, not $m^2 \log m$ as the pure QCD logs do. A factor of $\alpha$ has replaced a factor of the quark mass. In fact they are like the quenched logs in pure QCD in this respect. 
}

 The first step in estimating the systematic error
is to determine the fit range, or range of quark masses
included in the fit.
 The available ranges are summarized
in Table~\ref{tab:ensembles}. In
 Refs.~\cite{Allton:2008pn,Antonio:2007pb,Kelly:2009fp,Mawhinney:2009jy,RBC:2010}
 it was shown that for the same ensembles used in this work,
SU(3) and SU(2) chiral perturbation theories give sensible fit results for pion masses
less than about 400 MeV, or for bare quark masses satisfying $m_{f}\le 0.01$.
 It is possible that the range is different, perhaps larger,
for the EM splittings.
 After all, most of the pure QCD contributions at LO and NLO completely
cancel in the EM splittings
(some of the pure QCD LEC's survive at $O(\alpha_{\rm em} m)$).
 We work with uncorrelated fits, though our data are highly correlated,
because there are too many mass and charge combinations
to accurately determine the correlations on this finite statistical ensemble.
 The uncorrelated fits have been shown to agree with correlated ones
when the covariance matrix is well determined,
and when it is not, the correlated fits break down~\cite{Dawson:2009,RBC:2010}.
 As already mentioned, when the quark mass range is extended upwards,
for both SU(3) and SU(2) (pion) fits, $\chi^2$/dof increases noticeably,
by more than a factor of two.
 Since we use uncorrelated fits, this $\chi^2$ is not
an absolute test of goodness of fit,
though we expect changes do indicate relative goodness of fit.
 Thus we stick with the range $m_1,m_3\le 0.01$ for the light quarks
to quote central values and to estimate systematic errors. One of the systematic errors is the difference in the central values for the quark masses determined from this restricted range and those values computed from the range $m_1,m_3\le 0.02$ for the light quarks.
$m_1,m_3\le 0.01$ corresponds to valence pions in the range 250-420 MeV. 

For the mass range $m_1,m_3\le 0.01$, the most important change, which is anticipated in Fig.~\ref{fig:24split}, is that the Dashen term increases significantly when the logs are omitted from the SU(3) fit. 
$C$ increases by about a factor of five, although it is still small compared to the  value one would obtain from the physical splitting.  
For SU(2), the situation is much different; $C$ changes very little, about two percent. Presumably, the large logs containing the strange quark mass contribute to the LO term in this case, and the remaining effect of the light logs is not as important.
The higher order terms change more, but without logs to affect  their running, there is not much sense in comparing the changes.

\begin{figure}[h]
  \includegraphics[scale=1.0]{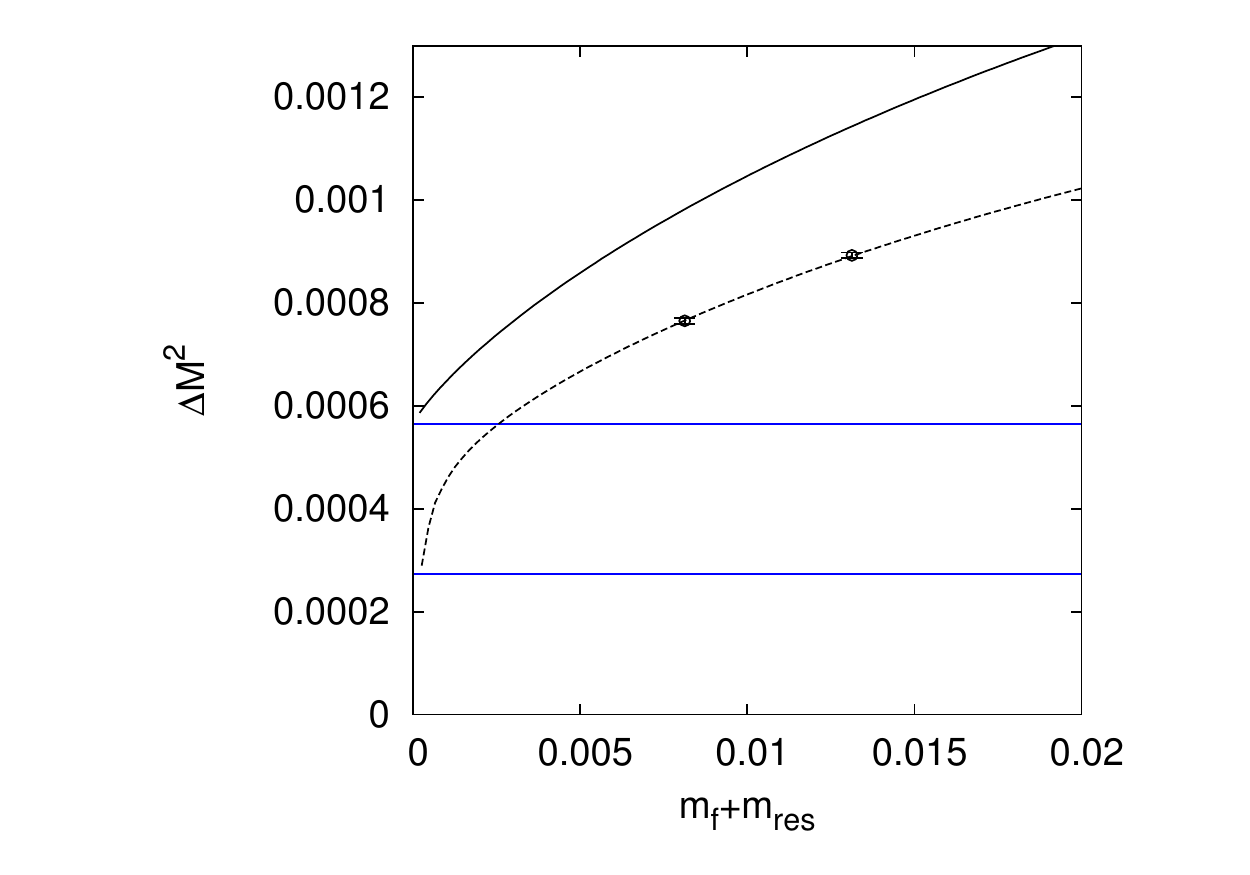}
  \caption{The LO and NLO in finite volume SU(2) chiral perturbation theory
    contributions to the EM meson mass splitting.  
    The dashed line corresponds to the finite volume fit
    result. The data points shown are for charged mesons with $q_1=2/3$ and $q_3=-1/3$. The lower horizontal line gives the contribution of the lattice
    artifact $e^2 \delta_{m_{\rm res}}(q_1^2+q_3^2)$ while the upper horizontal 
    line gives the sum of this contribution and Dashen's Term (in other words, their difference is just the LO contribution). The solid line
    corresponds to the total LO+NLO+$e^2 \delta_{m_{\rm res}}(q_1^2+q_3^2)$ contributions based on the fitted, finite volume LEC's, but evaluated with the infinite volume logarithms. The fit curves are evaluated for degenerate unitary light quarks. }
\label{fig:LO and NLO}
\end{figure}

 In Fig.~\ref{fig:LO and NLO} the LO and NLO contributions
in finite-volume ${\rm SU(2)_L\times SU(2)_R}$ chiral perturbation theory
for the charged pion mass splittings are shown.
 At the values of quark masses used in our calculation,
after accounting for the $\delta_{m_{\rm res}}$ contribution,
the NLO contributions are about 50-100 \% of the LO contribution. 
 It is interesting to see how $m_d-m_u$ is affected
at the various orders in chiral perturbation theory.
 Using the LEC's determined
in the full ${\rm SU(2)_L\times SU(2)_R}$-plus-kaon fits,
we find that the NLO contributions increase $m_d-m_u$
by a bit less than 2\%.

Taking all the above uncertainties due to fitting into account,
we estimate systematic errors of about four and zero percent
for the up and down, and strange quark masses, respectively.
These are collected in Tab.~\ref{tab:quark mass sys error}.

\clearpage

\subsection{finite volume}

\begin{figure}[h]
 \includegraphics[scale=1.0]{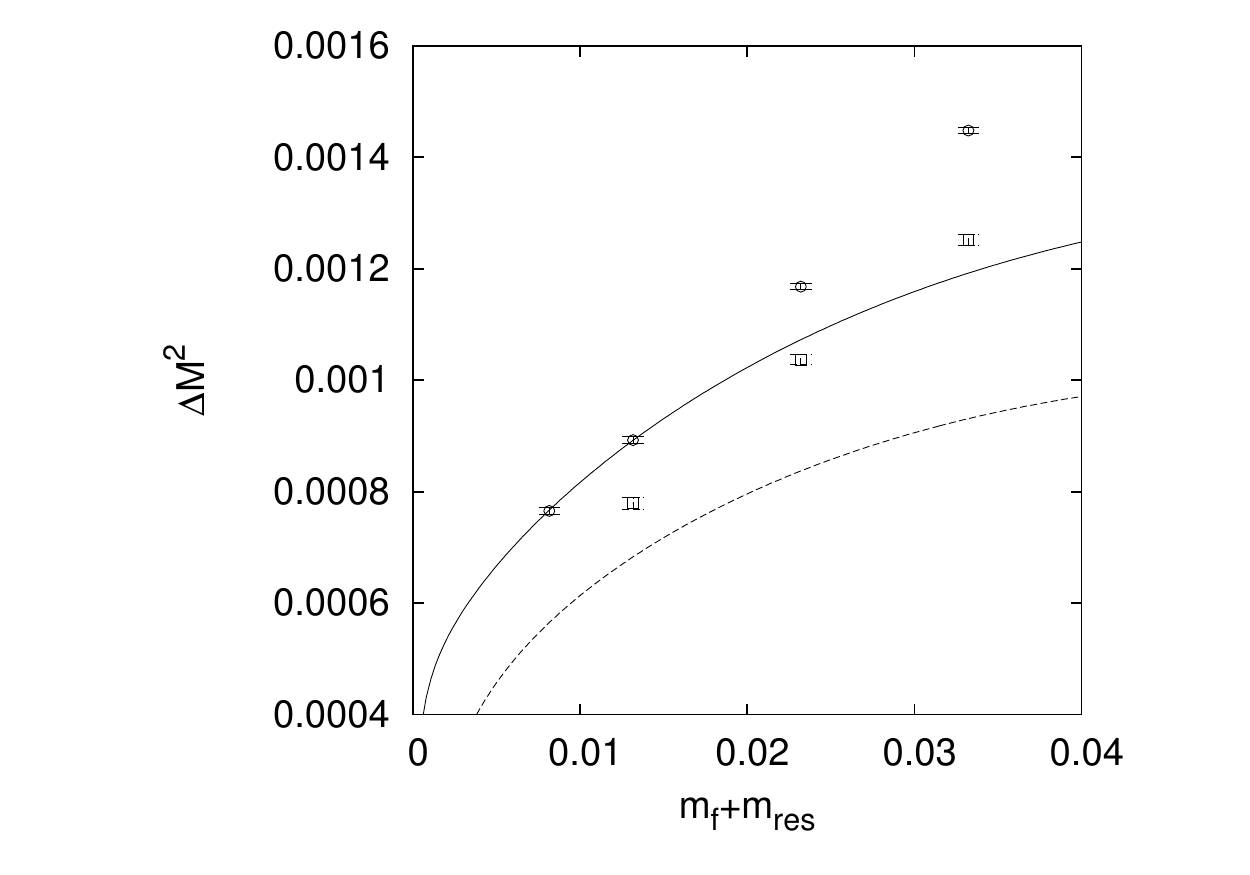}
 \caption{Finite volume effect in the measured EM splittings. All
of the data points have $q_1 = 2/3$ and $q_3 = -1/3$. Circles
and squares correspond to $24^3$ and $16^3$ lattice sizes,
respectively. The solid line is from the finite volume fit
on $24^3$ ensembles. The dashed line is the theoretical prediction
for $16^3$ lattices based on the LEC's extracted from $24^3$
finite volume fit. The fit curves are evaluated for degenerate unitary light quarks.}
\label{fig:finite vol}
\end{figure}

The effect of finite volume on the measured charged-meson splittings is large, as we have seen. In Fig.~\ref{fig:finite vol} the difference between the measured $16^3$ and $24^3$ EM splittings is about 15-20\%.  The LO LEC $C$ changes dramatically, by about a factor of two, when the finite volume corrections at NLO are included in the chiral perturbation theory fits (see Table~\ref{tab:qed lecs} and Fig.~\ref{fig:LO and NLO}). Besides the usual special functions that replace the infinite volume logs, a large, negative constant appears in the finite volume formula, $-3 \kappa q_{13}^2/4 \pi L^2$~\cite{Hayakawa:2008an} with $\kappa\approx 2.837$, which cancels against an enhanced value of $C$. 

To estimate how reliable the NLO finite volume corrections are, one can use the LEC's from the $24^3$ fits to predict the finite volume shift in the $16^3$ splitting. The fit and prediction are shown in Fig.~\ref{fig:finite vol}. First, the SU(2) fit agrees well with the $24^3$ results for $m_f\le 0.01$ which is the quark mass range used in the fit. For larger masses the fit deviates significantly from the data and suggests that NLO chiral perturbation theory is not reliable for these masses. Even for $m_{f}=0.01$, where we may trust NLO chiral perturbation theory, the theory over-predicts the shift on the $16^{3}$ lattice by about a factor of two. The NLO LEC's $Y_3$ and $Y_5$ also have large finite volume shifts. 

From Table~\ref{tab:kaonqedlec} the shifts in the kaon mass-squared LEC's are much smaller, especially the ones representing the LO Dashen term ($A_K^{(1,1)}$, $A_K^{(2,1)}$, $A_K^{(s,1,1)}$, and $A_K^{(s,2)}$).

Even though some of the LEC's show large finite volume shifts, the ultimate shifts in the quark masses are smaller.  The largest, about 14\%, occurs for the up quark mass.  The down quark mass is affected much less, about 3\%, and the shift in the strange quark mass is negligible.

From the pure QCD calculations, we know the finite volume effects
in the $24^{3}$ meson masses are
at about the one percent level~\cite{Kelly:2009fp,Mawhinney:2009jy,RBC:2010}, and therefore the QED finite volume corrections dominate. 

The finite volume errors on the quark masses are summarized in Table~\ref{tab:quark mass sys error}.

\subsection{non-zero lattice spacing}

Since our calculation has only been done at a single lattice spacing, we can not estimate the non-zero lattice spacing errors directly. However, by now there is much evidence that these ${O}(a^2 + m_{res} a)$ discretization errors are small in pure DWF QCD, and they should largely cancel in the splittings. Even assuming they do not cancel, there is  no reason to expect they are enhanced over the pure QCD case. In the first QCD calculation using the $24^3$ ensemble, it was estimated that scaling errors were at about the four percent level for low energy quantities like the pion decay constant and the kaon \cite{Allton:2008pn}. Since then, a new calculation at the same physical volume but smaller lattice spacing has shown this estimate was about right, or perhaps a bit conservative~\cite{Kelly:2009fp,Mawhinney:2009jy,RBC:2010}. 
Of course, here we are interested only in the mass splittings. The pion
and kaon masses are fixed to their continuum values, so they have no
scaling errors. Instead, the lattice spacing errors enter in the LEC's
and the physical quark masses. Therefore we assign a robust four percent
scaling error to the quark masses, which will be eliminated in up-coming
calculations on the finer lattice spacing
ensemble~\cite{Kelly:2009fp,Mawhinney:2009jy,RBC:2010}. This error also
encompasses the uncertainty in setting the lattice scale itself, which
as mentioned earlier differs by about $2 \sim 3$ percent from the scale given
in Ref.~\cite{Allton:2008pn}.

The non-zero lattice spacing errors on the quark masses are summarized in Table~\ref{tab:quark mass sys error}.

\subsection{QED quenching}

As mentioned our calculation is done in quenched QED where the sea quarks are neutral. In chiral perturbation theory, we have neglected terms of order $O(\alpha_{\rm em} m_{\rm sea})$, including logs. From a weak-coupling perturbation theory perspective in QCD+QED, we have neglected vacuum polarization effects at order $O(\alpha_{\rm em} \alpha_s)$. For the pions, the consequence is that the single (linear combination) LEC $Y_{1}$ can not be determined. For the kaons there are several LEC's that can not be determined (see Eq.~(\ref{eq:kaon mass sq})).
However, we do note that sea quark charge effects from the logs can be included {\it a posteriori} in our determination of the quark masses.

Since the LEC's absorb changes of scale in the logs, one way to estimate the effect of the missing LEC's, or counter-terms, is to mark the change in the quark masses when these logs are included, or not. This leads to a negligible change in the quark masses.
From Table~\ref{tab:qed lecs}, the other EM LEC's have magnitudes
roughly in the range 0.01 to 0.001. Setting $Y_1$ at the high end, $Y_{1}=\pm 0.01$, the quark masses again change very little.
Of course, the LEC's calculated with $q_{\rm sea}=0$ will differ from those with $q_{\rm sea}\neq 0$, by $O(\alpha_{\rm em})$. This is higher order for all the LEC's determined here except $C$ for the pions and $A_K^{(1,1)}$, $A_K^{(2,1)}$, $A_K^{(s,1,1)}$, and $A_K^{(s,2)}$ for the kaons. 
Taking all of the above into account, we quote a conservative two percent systematic error in our quark mass determination, stemming from the quenched approximation to QED. 

Of course the above is only a rough estimate, so presently we are investigating the use of so-called re-weighting techniques to eliminate the quenching effects~\cite{Duncan:2004ys,Hasenfratz:2008fg,Jung:2010jt,Ishikawa:2010tq,Izubuchi:2010,Luscher:2008tw}. Re-weighting is simply the use of ratio(s) of fermion determinants in observable averages in order to include the desired dynamical-quark effects. The calculation of a determinant which is non-local in the fields is quite expensive, so stochastic estimators must be used to make the calculation tractable. Re-weighting in the strange quark mass to the {\it a posteriori} determined physical value
has proved quite useful and efficient in recent 2+1 flavor simulations~\cite{Kelly:2009fp,Mawhinney:2009jy,RBC:2010,Aoki:2009ix}.

\section{quark masses}\label{sec:final quark masses}

 We use finite volume SU(2) chiral perturbation theory and the light
quark mass range $m\le 0.01$ to obtain our central values of the physical quark
masses.

{
 The physical strange quark mass is determined from the kaon mass-squared which is an implicit function of the bare sea and valence strange quark masses, through its LEC's which are calculated for fixed valence strange quark masses 0.02 and 0.03 and fixed sea strange quark mass 0.04. Assuming that the $m_s$ dependence is modest in this region, the physical kaon mass-squared is determined from a linear extrapolation in the valence strange quark mass. A similar procedure was carried out in~\cite{Allton:2008pn} where three data points in the range 0.02-0.04 showed the kaon mass-squared is  well approximated by a linear function
(it turns out that
the physical strange quark corresponds to about 0.035). Because we have only carried out calculations at a single strange sea quark mass value of 0.04, the kaon mass-squared can not be evaluated at the physical strange sea quark mass. This partial quenching leads to a small systematic error that was conservatively estimated in~\cite{Allton:2008pn} to be 2\% for $m_s$ which we adopt here. It is added in quadrature to the total systematic error for $m_{s}$ which appears below. The systematic error on the light quark masses is about 0.7\% which is negligible compared to the other systematic errors, so we ignore it.  }

 The statistical errors come from fits underneath a standard jackknife analysis. The QCD LEC's come from an analysis of the extended RBC/UKQCD $24^{3}$ ensembles; the results are consistent with
those in Ref.~\cite{Allton:2008pn}. All of the fits and corresponding
LEC's are analyzed under a super-jackknife analysis so that statistical
errors on all quantities, from all ensembles, are included. The
systematic errors assigned have been discussed in this section. The
mass-independent quark mass renormalization factor is
\begin{align}
Z^{\overline{\rm MS}}_m ( \mu = 2 \text{GeV} ) = 1.546(2)(43),
\end{align}
which is obtained via nonperturbative renormalization using the
RI/SMOM$_{\gamma_\mu}$ scheme \cite{RBC:2010,Sturm:2009kb,Gorbahn:2010bf,Aoki:2010yq,Almeida:2010ns,Arthur:2010ht}.
The second error is systematic, including ${O}((\mu a)^2)$,
which will be removed when we take the continuum limit in future work
 (the ${O}(\alpha_\text{em})$ QED correction to $Z_m$ is omitted). The final values are
\begin{eqnarray}
m_{u}      &=& 2.24  \pm 0.10   \pm 0.34~~\rm MeV\label{eq:mu}\\
m_{d}      &=& 4.65  \pm 0.15   \pm 0.32~~\rm MeV\label{eq:md}\\
m_{s}      &=& 97.6  \pm 2.9    \pm 5.5~~\rm MeV\label{eq:ms}\\
m_d-m_u    &=& 2.411 \pm 0.065  \pm 0.476~~\rm MeV\label{eq:md-mu}\\
m_{ud}     &=& 3.44  \pm 0.12   \pm 0.22~\rm MeV\label{eq:mud} \\
m_u/m_d    &=& 0.4818\pm 0.0096 \pm 0.0860 \label{eq:mu/md}\\
m_s/m_{ud} &=& 28.31 \pm 0.29   \pm 1.77,\label{eq:ms/mud}
\end{eqnarray}
where the first error is statistical, and the second is a total systematic error, derived by adding the individual errors summarized in Table~\ref{tab:quark mass sys error} in quadrature. We remind the reader that these central values are obtained from our SU(2), finite volume fits on the 24$^3$ ensembles.

We note that the up quark mass obtained here is different from zero
by more than six standard deviations, which
seems to rule out the $m_u=0$ solution to the strong CP problem.
However, there is an extensive literature concerning this scenario to which we refer the interested reader.
For a discussion of extracting the up quark mass by using chiral perturbation theory, and its consequences, see~\cite{Kaplan:1986ru,Leutwyler:1996qg,Nelson:2003tb}. The possibility of instanton effects additively shifting the up quark mass is discussed in many places~\cite{'tHooft:1986nc,Choi:1988sy,Banks:1994yg,Davoudiasl:2007zx}.
In \cite{Creutz:2003xc}, renormalization scheme dependence
of the renormalized quark mass was discussed in the context the
isospin breaking. The effect vanishes in the (perturbative) $\overline{MS}$
scheme. At this point, there seem to be no common consensus on
if there is any non-perturbative contribution, which is related to the
aforementioned instanton effects, and how large it might be.
Our results are potentially susceptible to this
uncertainty, as are all other quark masses renormalized in a perturbative scheme.

\begin{table}[htdp]
\caption{Summary of quark mass systematic errors. Central values quoted from the finite volume, SU(2), chiral perturbation theory fit. Masses given in MeV. The quark mass renormalization error comes from the nonperturbative QCD result~\cite{RBC:2010} plus a one percent error from QED, added in quadrature. Systematic errors are given as a percent (\%). The algebraic sign of each change comes from the difference (quantity under change) $-$ (central value).
}
\begin{center}
\begin{tabular}{cccccccc}
\hline\hline
 & value (stat. error) & fit  & fv & lat. spacing & QED quenching & $m_{s}$ quenching & renorm\\
\hline
$m_{u}$      & 2.24(10)   & +4.02 & +13.50  & 4 & 2 &  -& 2.8 \\
$m_{d}$      & 4.65(15)   & +3.55 & -2.48  & 4 & 2 & -& 2.8 \\
$m_{s}$      & 97.6(2.9)  & +0.23 & +0.07   & 4 & 2 & 2 & 2.8 \\
$m_d-m_u$    & 2.411(65)  & +7.77 & -17.35 & 4 & 2&- & 2.8 \\
$m_{ud}$     & 3.44(12)   & +2.75 & +2.71   & 4 & 2 &-& 2.8 \\
$m_u/m_d$    & 0.4818(96) & +5.45 & +16.40  & 4 & -& & -   \\
$m_s/m_{ud}$ & 28.31(29)  & +2.91 & -2.56  & 4 & 2 &2 & -   \\
\hline\hline
\end{tabular}
\end{center}
\label{tab:quark mass sys error}
\end{table}%

\clearpage

\section{Meson Mass Splittings}\label{sec:final meson}

In Tab.~\ref{tab:qed lecs} we give the contribution to the charged pion mass splitting in the chiral limit, or Dashen's term. The physical splitting, given in Eq.~(\ref{eq:experiment pi mass}), is 4.5936(5) MeV. The SU(3) fit gives a very small value, about half an MeV. The finite volume fit dramatically increases the value, but it is less than half the physical value. The SU(2) fit gives a bigger value still, and after including finite volume corrections, it gives the LO EM correction to the pion mass difference $(m_{\pi^\pm}-m_{\pi^0})_{\rm QED}= 3.38(23)$ MeV. Coincidentally, this is about the same value obtained from the linear fit, {
3.22(25)MeV. The value of $m_{\pi^\pm}^2$ in the chiral limit, which comes from the LO EM correction is 929(64) MeV$^2$ and is similar to the values, using a sum rule and lattice-computed vector and axial-vector correlation functions in pure QCD, reported in~\cite{Shintani:2008qe,Boyle:2009xi}.
Our value for $(m_{\pi^\pm}-m_{\pi^0})_{\rm QED}$ is roughly consistent with, but two statistical standard deviations smaller than, the value from phenomenology and SU(3) chiral perturbation theory reported in~\cite{Bijnens:2006mk}, 3.7 MeV.

The above suggests that NLO contributions at the physical quark masses may be as large as 25\% of the total pion mass difference, $m_{\pi^{+}}-m_{\pi^{0}}$. Away from the chiral limit, there are corrections to $m_{\pi^{0}}$ that we have not computed in the lattice calculation (disconnected diagrams), nor in chiral perturbation theory (logs). However, we can estimate some of them by 
evaluating Eq.~(\ref{eq:su2 pion mass}) for $m_{u}=m_{d}=m_{ud}$,
$q_{1}=q_{3}=q_{u}$ and averaging it with the case for
$q_{1}=q_{3}=q_{d}$
\begin{eqnarray}
 \overline{M}^2(q_1,\,q_3;\,m_1)
 &\equiv&
 \frac{1}{2}
 \left\{
  M^2(m_1,\,q_1,\,m_1,\,q_1) + M^2(m_1,\,q_3,\,m_1,\,q_3)
 \right\}\,.\label{eq:avr_pi^0}
\end{eqnarray}
 This form can be inferred 
for the $\pi^0$ made with degenerate light valence quarks
($m_3 = m_1$) in our current study
in which
only the connected valence quark diagram is computed and QED is quenched.
 We focus on the one-particle irreducible two-point function
$\Sigma_{\pi^0}(p^2)$ of $\pi^0$, 
and pick out the part depending on the valence EM charges
induced from the connected diagram.
 $\Sigma_{\pi^0}(p^2)$ can be divided into a pure QCD part
$\Sigma_{\pi^0}^{{\rm QCD}}(p^2)$
and a QED correction $\Sigma_{\pi^0}^{\rm QED}(p^2)$
at order $e^2$,
\begin{eqnarray}
 \Sigma_{\pi^0}(p^2)
 &=&
 \Sigma_{\pi}^{\rm QCD}(p^2)
 +
 \Sigma_{\pi^0}^{\rm QED}(p^2)\,,\label{eq:pi^0_se}\\
 \Sigma_{\pi}^{\rm QCD}(p^2)
 &=&
 \left(\frac{1}{\sqrt{2}}\right)^2
 {\rm tr}\left(\left(\tau_3\right)^2\right)
 A_{\rm QCD}(p^2)
 = A_{\rm QCD}(p^2)\,,\label{eq:pi_QCD}\\
 \Sigma_{\pi^0}^{\rm QED}(p^2)
 &=&
 \left(\frac{1}{\sqrt{2}}\right)^2
 {\rm tr}\left(
  \tau_3 Q \tau_3 Q
 \right)\,D_1(p^2)
 +
 2 \times \left(\frac{1}{\sqrt{2}}\right)^2
 {\rm tr}\left(
  \left(\tau_3\right)^2 Q^2
 \right)\,D_2(p^2) \nonumber\\
 &=&
 \frac{1}{2}\left(q_1^2 + q_3^2\right) D_1(p^2)
 +
 \left(q_1^2 + q_3^2\right) D_2(p^2)\,.\label{eq:pi^0_QED}
\end{eqnarray}
where $\tau_a$ ($a = 1,\,2,\,3$) denote the Pauli matrices
and $Q = {\rm diag}\left(q_1,\,q_3\right)$.
 In Eq.~(\ref{eq:pi^0_QED}), the first term originates from
the Feynman diagram in which a virtual photon 
is exchanged between two valence quark lines,
while a photon propagates on the same valence quark lines
and induces the second term.
 Because the functions $D_{1,\,2}(p^2)$ are given by 
QCD dynamics weighted by the photon propagator,
the self-energy $\Sigma_{\pi^+}(p^2)$ of the charged pion
is also expressed in terms of these functions
\begin{eqnarray}
 \Sigma_{\pi^+}(p^2)
 &=&
 \Sigma_{\pi}^{\rm QCD}(p^2)
 +
 \Sigma_{\pi^+}^{\rm QED}(p^2)\,,\label{eq:pi^+_se}\\
 \Sigma_{\pi^+}^{\rm QED}(p^2)
 &=&
 {\rm tr}\left(
  \tau_+ Q\,\tau_- Q
 \right)\,D_1(p^2)
 +
 {\rm tr}\left(\left(\tau_+ \tau_- + \tau_- \tau_+\right) Q^2\right)
 D_2(p^2) \nonumber\\
 &=&
 q_1 q_3\,D_1(p^2)
 +
 \left(q_1^2 + q_3^2\right) D_2(p^2)\,.\label{eq:pi^+_QED}
\end{eqnarray}
 From Eqs.~(\ref{eq:pi^0_se})-(\ref{eq:pi^+_QED}),
the charge dependence of $m_{\pi^0}^2$ and $m_{\pi^+}^2
= M^2(m_1,\,q_1,\,m_1,\,q_3)$,
to the order relevant to us, is found as
\begin{eqnarray}
 M^2(m_1,\,q_1,\,m_1,\,q_3) &=&
 K + q_1 q_3 F_1 + \left(q_1^2 + q_3^2\right) F_2\,,\nonumber\\
 m_{\pi^0}^2 &=&
 K + \frac{1}{2}\left(q_1^2 + q_3^2\right) F_1
 + \left(q_1^2 + q_3^2\right) F_2\nonumber\\
 &=&
 \overline{M}^2\left(q_1,\,q_3;\,m_1\right)\,,
\end{eqnarray}
where $K$ denotes the QCD part to NLO of chiral perturbation, 
and $F_{1,\,2}$ the $O(e^2)$ and $O(e^2 m_l)$ part.
 The chiral symmetry as well as QED gauge invariance
should give
$\left.F_1\right|_{m_1 = 0}
= - 2\left.F_2\right|_{m_1 = 0}$
to reproduce the EM charge dependence $(q_1 - q_3)^2$
of the LO EM correction to $m_{\pi^+}^2$.
 Using Eq.~(\ref{eq:avr_pi^0}) for $m_{\pi^{0}}^2$, 
we find the ${\rm LO+NLO}$ EM pion mass difference
at the physical point to be $m_{\pi^{+}}-m_{\pi^{0}}=4.50(23)$ MeV.
 Phenomenology predicts that a small part of the total NLO correction
is due to $m_{u}-m_{d}\neq0$, 0.17(3) MeV~\cite{Gasser:1984gg}
and 0.32(20) MeV~\cite{Amoros:2001cp}. 

For the kaons, the pure EM mass difference is
$m_{K^+}-m_{K^0}=1.87(10)$ MeV, while the contribution from
$m_{u}-m_{d}\neq0$ is $-5.840(96)$ MeV. Here, of course, the result includes all NLO corrections, and LEC's from the finite volume fit are used.
These values are obtained by taking the SU(2) formula for the kaon mass-squared
$M_K^2(m_1, q_1, m_3, q_3)$, Eq.~(\ref{eq:kaon mass sq})
\begin{align}
M_K^2(m_u, 2/3, m_s, -1/3) - M_K^2(m_u, -1/3, m_s, -1/3)
&=& \Delta^{\text{(EM)}} M_K^2 + \Delta^{(m_u-m_d)} M_K^2 \nonumber\\ &&+ {\cal O}(e^2 (m_u-m_d))
\end{align}
where the contributions to the mass-squared splitting are
defined as
\begin{align}
&\Delta^{\text{(EM)}} M_K^2 =
M_K^2(m_{ud},2/3,m_{ud},-1/3) - M_K^2(m_{ud},-1/3,m_{ud},-1/3),\\
&\Delta^{(m_u-m_d)} M_K^2 = M_K^2(m_u,0,m_s,0) - M_K^2(m_d,0,m_s,0).
\end{align}
$\Delta^{\text{(EM)}/(m_u-m_d)} M_K^2/(M_{K^0}+M_{K^\pm})$ are quoted above. 
So out of a physical mass-squared splitting $(M_{K^0})^2 - (M_{K^\pm})^2=3902.7$ MeV$^2$,
 about $-47 (2)\%$ is $\Delta^{\text{(EM)}} M_K^2$ and $+148 (2)\%$ is 
$\Delta^{(m_u-m_d)} M_K^2$.
}

{The breaking of Dashen's theorem can also be parametrized by $\Delta E$
\cite{Aubin:2004fs},
\begin{eqnarray}
\Delta E &=&
\frac{M^2_K(m_1,q_1,m_3,q_3) - M^2_K(m_1,q_3,m_3,q_3)}
     {M^2(m_1,q_1,m_1,q_3) - M^2(m_1,q_3,m_1,q_3)} -1,
\end{eqnarray}
where $m_1$ is the light quark mass and $m_3$ is the strange.
 $M^2(m_1,q_3,m_1,q_3)$ is used here to represent $m_{\pi^{0}}^2$;
no significant change of $\Delta E$ is observed in our numerical study
even when the average (\ref{eq:avr_pi^0}) is adopted for $m_{\pi^{0}}^2$
in place of $M^2(m_1,q_3,m_1,q_3)$.
In the SU(3) chiral limit $\Delta E=0$ since the LO Dashen terms are the same in the numerator and denominator. If the strange quark mass is fixed to its physical value, then it does not vanish, and can be much larger than zero, even in the light quark chiral limit.
Notice that $\Delta E$ vanishes trivially in both SU(2) and SU(3) theories when $m_1\to m_3$.

We show $\Delta E$ for our data in Fig.~\ref{fig:deltaE} where the artifact $\delta_{m_{\rm res}}(q_i^2+q_j^2)$ has been subtracted for each value of the meson mass-squared. In the upper panel, fit results are shown for SU(3). The fit evaluated at the simulated mass points does a reasonable job of reproducing the data, though as $m_3$ increases differences emerge. This is not surprising since only $m_1,m_3\le 0.01$ points were used in the fit, and including larger values yielded significantly poorer fits. More troublesome is the light quark extrapolation which yields a large value of $\Delta E$ at the physical point, which can be understood from two primary causes. First, the numerator is quite large since $m_3$ is evaluated at the physical strange quark mass, leading to a large $O(\alpha m)$ correction to the charged kaon mass-squared. Second, the denominator becomes quite small because the LO Dashen term is quite small in the SU(3) fit (compared to NLO terms). Both facts, of course, signal a breakdown in SU(3) chiral perturbation theory which renders the SU(3) $\Delta E$ unreliable. As noted in~\cite{Bijnens:2006mk}, the sea quark charge LEC's drop out of $\Delta E$, and only known logarithms remain. Adding these to the (cyan) physical curve in Fig.~\ref{fig:deltaE} changes it only slightly.

In the lower panel of Fig.~\ref{fig:deltaE} we show analogous results for the SU(2) fits. While the SU(2) fits are more reliable since the LO contribution is larger compared to NLO, the latter corrections are still large (recall Fig.~\ref{fig:LO and NLO}). As expected, the fit agrees better with the data points for larger values of $m_{1}$, but the extrapolated value at the physical point and infinite volume is still much larger than the data points. We find in quenched QED that $\Delta E=0.628(59)$ where the error is statistical only. This is much larger than the value reported in our previous two flavor paper~\cite{Blum:2007cy} and not much smaller than phenomenology and SU(3) chiral perturbation theory~\cite{Bijnens:2006mk}. The main difference is that here we use full NLO chiral perturbation theory with finite volume corrections while in~\cite{Blum:2007cy} only simple analytic fits were used.
To properly address these large corrections, one needs to simulate with larger volumes and smaller quark masses, a project that is now underway.

\begin{figure}[htbp]
\begin{center}
    \includegraphics[scale=1.]{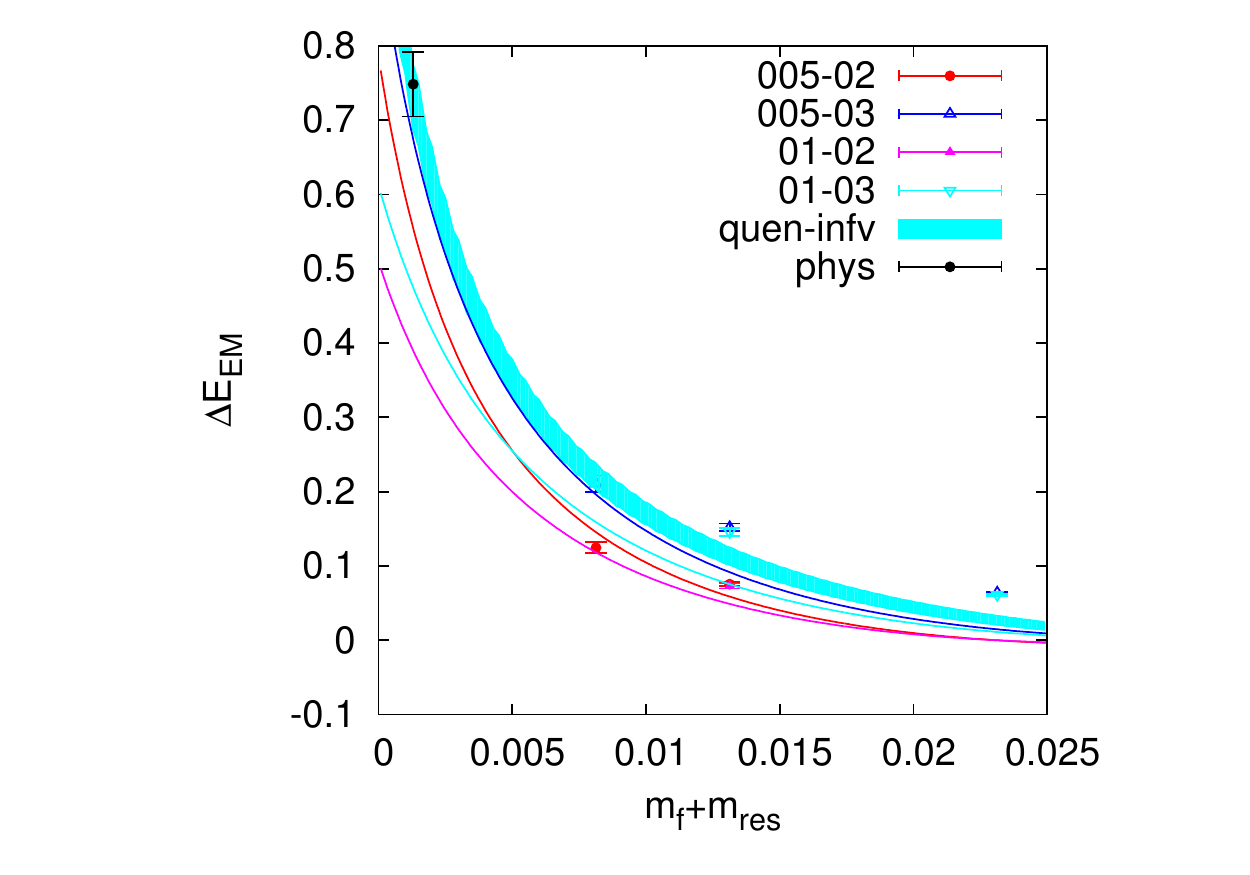}
    \includegraphics[scale=1.]{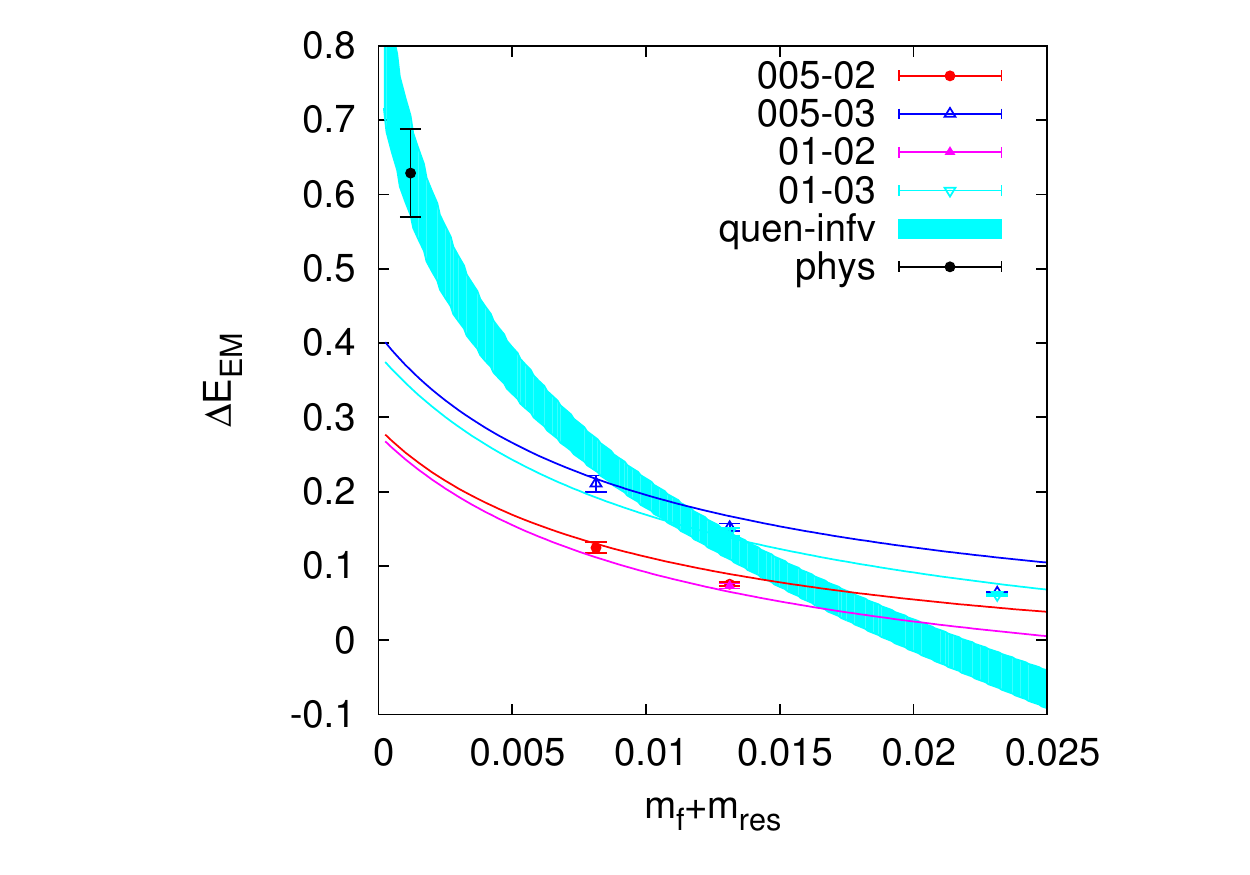}
\caption{Breaking of Dashen's theorem for the quenched QED case. The unphysical contribution $\delta_{m_{\rm res}}(q_1^2+q_3^2)$ has been subtracted from the data. Data for two values of the strange quark, 0.02 and 0.03, are shown. The curves correspond to the SU(3) fits (upper panel) and SU(2) fits (lower panel). The cyan bands denote the infinite volume extrapolations with one standard deviation statistical errors, using the LEC's extracted from the finite volume fits; the sea and strange quark masses are fixed at their physical values.} 
\label{fig:deltaE}
\end{center}
\end{figure}

Perhaps more useful for other pure QCD simulations are the ``physical" values of $m_\pi$ and $m_K$ in pure QCD deduced from our SU(2) fits with $m_u=m_d=m_{ud}$:
\begin{eqnarray}
m_\pi^{\rm (QCD)} &=& 134.98(23)~{\rm MeV} ,\\
m_K^{\rm (QCD)} &=& 494.521(58)~{\rm MeV}.
\end{eqnarray}
The small statistical errors result because the physical pion and kaon meson masses were used to determine the physical quark masses from our fit.

}

Finally, based on the quark masses in Eqs.~(\ref{eq:mu})-(\ref{eq:ms}) and~(\ref{eq:mud}), we examine the ratio introduced in Ref.~\cite{Gasser:1984pr},
\begin{equation}
\kappa_\text{quark mass} \equiv  {m_d - m_u \over m_s - m_{ud} }  {2
m_{ud} \over m_s+ m_{ud} }~,
\end{equation}
which is equal to
\begin{eqnarray}
\kappa_\text{meson} &\equiv&
{ (M^2_{K^0} - M^2_{K^\pm})_\text{QCD}  \over  M_K^2 - M_\pi^2}
{M_\pi^2 \over M_K^2}\\
&=& { M^2(m_d, 0, m_s,0) - M^2_K(m_u, 0, m_s,0) \over
   M_K^2(m_s, 0, m_{ud}, 0) - M^2(m_{ud,} 0, m_{ud},0) }
{ M^2(m_{ud}, 0, m_{ud},0)  \over M_K^2(m_s,0, m_{ud},0) }
~~,
\end{eqnarray}
up to NNLO in SU(3) ChPT \cite{Gasser:1984gg}.
For SU(3) we obtain
\begin{eqnarray}
\kappa_\text{quark mass} &=& 0.00201 (3),\\
\kappa_\text{meson} &=& 0.00201 (3),
\end{eqnarray}
while for SU(2),
\begin{eqnarray}
\kappa_\text{quark mass} &=& 0.00176 (4),\\
\kappa_\text{meson} &=& 0.00191 (3),
\end{eqnarray}
where the errors are statistical only.
For SU(3) the values are quite consistent with each other, while for SU(2) there is a small difference. In~\cite{Gasser:1984pr},  $\kappa$ extracted from $\eta\to
\pi^0\pi^+\pi^-$ decays is
$0.0019(3)$  while  the $ O(p^6)$ analysis  in \cite{Amoros:2001cp} gives
$\kappa=0.00260$ at $m_s/m_{ud}=24$.

\clearpage

\section{Isospin breaking effects on the kaon decay constant}
\label{sec:fK}

In our results the up quark mass is about 35\% 
smaller than average of the up and down quark masses, $m_{ud}$. 
In principle, this isospin breaking effect may
cause visible effects on phenomenologically important
quantities when they are measured with sufficient accuracy.
As we saw in the previous section, 
a major part of the Kaon mass splitting comes from
the quark mass difference, $m_u-m_d$.

Here we examine isospin breaking effects on 
the Kaon decay constant, $f_K$.
By combining the experimental decay widths,
$\Gamma(K\to \nu \mu(\gamma))$ and $\Gamma(\pi\to \nu \mu(\gamma))$, and
$f_\pi$ and $f_{K}$,
one can extract the corresponding ratio of CKM matrix elements~\cite{Marciano:2004uf}.
In the latest global analysis by the FlaviaNet Working group on Kaon Decays
\cite{Antonelli:2010yf},
${f_K\over f_\pi} \left| {V_{us}\over V_{ud}}\right|$
is obtained from experimental results with an accuracy of 0.2\%.
The ratio of the decay constants used are from their world average of
lattice QCD simulations, and is
\begin{equation}
{f_K\over f_\pi} = 1.193(5)~~~~~[0.4 \%].
\end{equation}
We address a question: how far does the value of $f_K$ 
shift when the light quark mass in the Kaon is changed from
from $m_{ud}$ to $m_u$?
Some lattice determinations of $f_K$ use $m_{ud}$ 
as the light quark mass while the experiments measure
decays of the {\em charged} Kaon to obtain $f_{K^\pm}$, 
which is made of an up (and strange) quark. So it is relevant
to know if the shift $f_K(m_{ud})-f_K(m_u)$ is comparable in size to the total error on the ratio, 0.4\%.
We note the analyses of $V_{us}/V_{ud}$ 
in \cite{Marciano:2004uf,Antonelli:2010yf} 
(see also \cite{Cirigliano:2007ga}) correct 
for the QED effects of the decay constants,
and we only consider the decay constant for $e=0$ 
but $m_u\neq m_d$ in this section.

\begin{figure}
\includegraphics[width=4in]{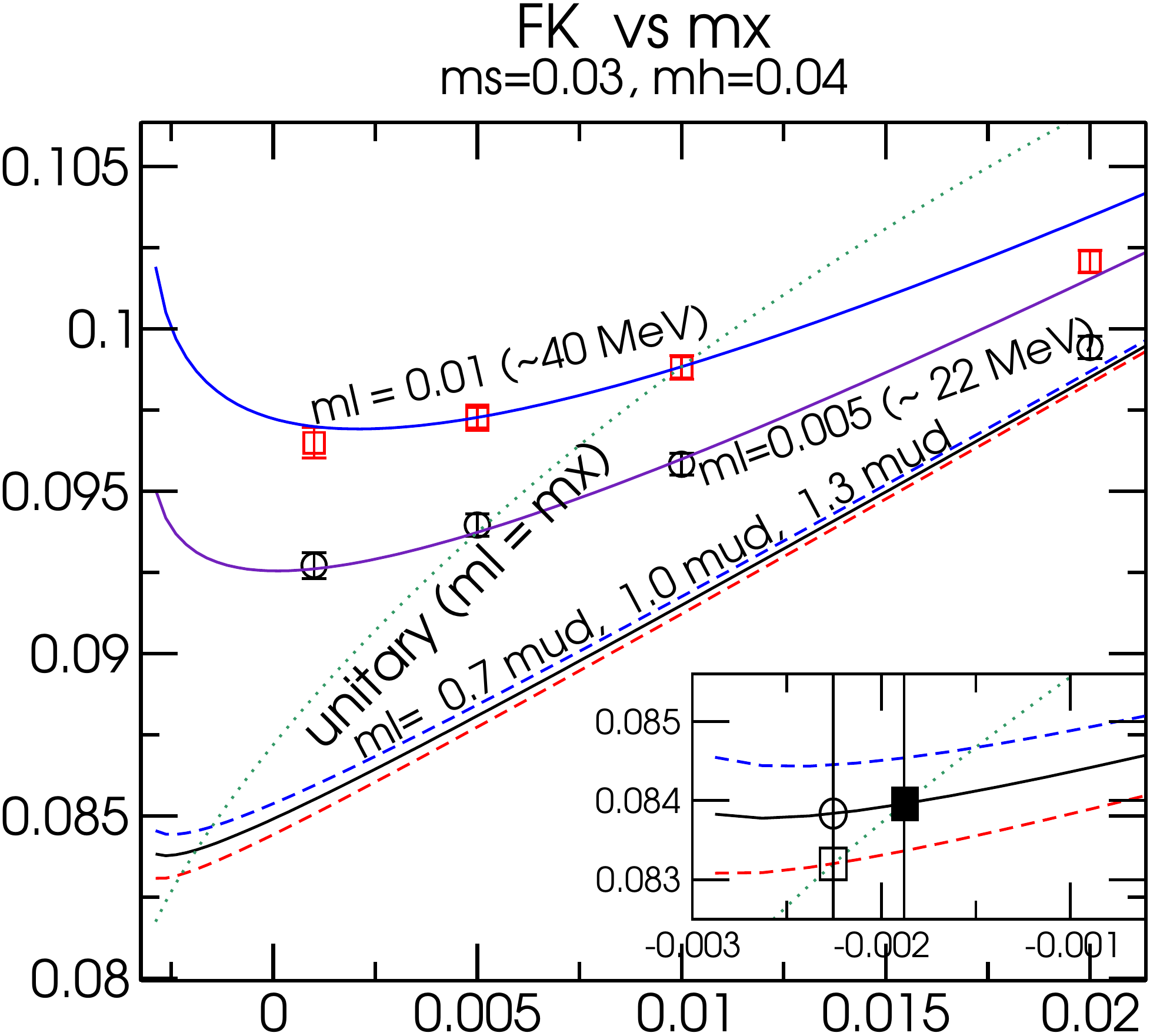}
\caption{
Kaon decay constant in pure QCD \cite{Allton:2008pn} computed from the same ensembles as used in this work.  The valence strange quark mass is fixed to 0.03 and the sea strange quark mass is 0.04.
}
\label{fig:f_K}
\end{figure}

In Fig.\ref{fig:f_K}, $f_K(m_x)$ obtained by the RBC/UKQCD collaboration 
~\cite{Allton:2008pn} is plotted as a function
of valence light quark mass $m_x$. The sea and valence 
strange quark masses are fixed. The square points are
from light sea quark mass $m_l=0.01 (\sim 40 \text{MeV})$
while the circle data are for $m_l=0.005 (\sim 22~\text{MeV})$.
The curves are from the partially quenched SU(2) ChPT fits.
The upper two curves denote
$f_K(m_x)$ at fixed degenerate sea quark 
masses $m_l=0.01$ (upper) and $0.005$ (lower),
while the doted curve is evaluated for unitary quark mass, $m_x=m_l$.
The lower three, almost degenerate, curves 
are for $m_l=0.7\,m_{ud}$, $m_{ud}$, and $1.3 \,m_{ud}$.

The inset magnifies the region close to the physical point.
The filled square is  $f_K$ for equal  up and down
quark masses, $m_l=m_x=m_{ud}$. 
When the valence quark mass is decreased to a 30\% smaller value, 
$0.7\times m_{ud}$, $f_K$ decreases by about 1\%, if we 
simultaneously decrease the light sea quark masses 
to $0.7\times m_{ud}$ (empty square). This setting of quark masses 
(empty square) underestimates the value of $f_K$ in Nature, 
since the down sea quark mass is also decreased 
to $0.7\times m_{ud}$ for the 
empty square\footnote{We thank Chris Sachrajda for pointing this out.}. 

The more accurate estimation of $f_K$ for non-degenerate valence up and down 
quark masses is the empty circle, where the degenerate sea quark mass 
is fixed to $m_l=m_{ud}$,
and only the valence quark mass is 
set to the lighter mass, $m_x=0.7 m_{ud}$.
We note that the non-degenerate quark mass effect in the sea sector is
suppressed by $(m_u-m_d)^2$, and setting degenerate sea quark mass
to $m_l=m_{ud}$ is a good approximation to estimate the $f_K$ shift due to
the isospin breaking in the up and down quark masses.
Because of the (accidental) decrease in the slope of $f_K(m_x)$ 
around the physical sea quark mass $m_l=m_{ud}$, the difference between
$f_K(m_{ud})$ and $f_K(0.7\times m_{ud})$ is only about $-0.2 \%$, which is nevertheless
sizable compared to the total error of 0.4\% in the current 
world average of $f_K/f_\pi$. 

A similar analysis was done in\cite{Aubin:2004fs}, 
where $f_{K^+}$ was properly estimated at $m_x=m_u$. An indirect error on $f_{K^+}$ induced from 
(their) EM uncertainty in $m_u/m_d$ ($\sim 19\%$) was 
estimated to be $\sim 0.07\%$. So their shift of
$f_K$ due to the quark mass difference between $m_{ud}$ and $m_u$ 
would be roughly  $0.07/0.19 \times (m_{ud} / m_u - 1)  \sim $ 0.25 \%
from their value of $m_u/m_{ud} \approx 0.6$.
This shift is slightly larger than our estimation, 0.2\%,
in part because $m_u/m_{ud}$ in [58] is smaller than ours by about 10\%.

\clearpage

\section {Nucleon Mass Splittings}\label{sec:nucleon}
Isospin breaking also occurs in the nucleon system. The proton is
slightly lighter than the neutron, which makes the proton
a stable particle. In conjunction 
with baryon PQ$\chi$PT, the lattice simulation helps us understand
the relation between the baryon masses and their quark 
content~\cite{Beane:2006fk}.

In Nature, $m_p-m_n$=$-1.293321(4)$ MeV as determined by experiments, and 
it is explained  by two mechanisms. One is the EM interaction.
The proton is a charged particle, but the neutron is neutral, so the
QED interaction makes the proton heavier. The other is 
due to non-degenerate $u,d$ quark masses. The valence quark content
in the proton and neutron is $uud$ and $udd$, respectively. So the proton is lighter than 
the neutron due to the fact that the $d$ quark is heavier than the $u$. Combining these 
two effects in our lattice calculations, we can compute the $p$-$n$ mass
splitting.

For non-degenerate quark masses, we study the splitting using the pure QCD
configurations. The nucleon mass in two flavor QCD 
is given by baryon PQ$\chi$PT, to NLO~\cite{Beane:2006fk}, 
\begin{eqnarray}
  m_p &=& M_0+\frac{1}{3}(5\alpha+2\beta)m_u+\frac{1}{3}(\alpha+4\beta)m_d
  +\frac{1}{2}\sigma(m_j+m_l) \label{eq:baryonchipt}\\
  m_n &=& M_0+\frac{1}{3}(\alpha+4\beta)m_u+\frac{1}{3}(5\alpha+2\beta)m_d
  +\frac{1}{2}\sigma(m_j+m_l)
  \label{eq:baryonchipt neutron}
\end{eqnarray}
where $m_u$, $m_d$ are the masses of the valence quarks and $m_j$, 
$m_l$ are the masses of the sea quarks. 
The mass difference 
between the proton and the neutron is
\begin{eqnarray}
  (m_p-m_n)_{(m_d-m_u)} &=& -\frac{1}{3}(4\alpha-2\beta)(m_d-m_u).
  \label{eq:dif-np}
\end{eqnarray}
We note only the sum of sea quark masses, $m_j+m_l$, appears in Eqs.~(\ref{eq:baryonchipt})~and~(\ref{eq:baryonchipt neutron})
and the difference $m_j-m_l$ appears first at NNLO
in any observable due to the symmetry under switching
sea up quark to sea down quark. So our degenerate sea up and down quark mass
is enough to extract the isospin breaking to NLO
(We will ignore possible contributions of ${O}(e^2 (m_u-m_d))$).

Next, we test the EM induced mass splitting on QCD+QED 
configurations with unitary (and therefore degenerate) mass points.  The lowest order mass 
difference is parametrized as:
\begin{eqnarray}
  (m_p-m_n)_\text{QED} &=& \alpha_{\rm em}(A_0 +A_1 m_{ud})
\end{eqnarray}
where $m_{ud}=(m_u+m_d)/2$, and the dependence on $\alpha_{\rm em}$ is made explicit to remind ourselves that the splitting vanishes in the absence of QED. 

All of the above LEC's here can be extracted from fits to lattice data.

We first extract the nucleon masses from the two-point correlation
function. The correlation function measured on the lattice with
anti-periodic boundary condition in time has the form~\cite{Sasaki:2001nf}:
\begin{eqnarray}
  G(t)&=&(1+\gamma_4)A_{B^+}e^{-M_{B^+}t}-
  (1-\gamma_4)A_{B^+}e^{-M_{B^+}(N_{t}-t)} \nonumber \\
  &&+(1+\gamma_4)A_{B^-}e^{-M_{B^-}(N_{t}-t)}
  - (1-\gamma_4)A_{B^-}e^{-M_{B^-}t},
\end{eqnarray}
where $B^+$ represents the nucleon state which has positive parity 
and $B^-$ represents the excited state of the nucleon which has 
negative parity. $N_{t}$ is the time-size of the lattice. 
Since the mass of the excited state is much 
heavier than the ground state, we neglect its contribution. The nucleon and anti-nucleon terms left in the correlation function are 
picked up by multiplying $G(t)$ by the projection operator $1\pm\gamma_4$ 
and taking the trace. Then we average these two terms by taking $t\to N_{t}-t$ for the anti-nucleon 
to improve the
statistics of our measurements. The $\pm e$ trick is also
used when QED configurations are included. Finally the nucleon 
masses are extracted from single state fits to 
point-sink correlation functions as
\begin{eqnarray}
  G(t)&=&Ae^{-Mt},
  \label{eq:nucleon-mass}
\end{eqnarray}
where $M$ is the ground state nucleon mass, and $A$ measures 
the overlap between the nucleon state and the nucleon interpolation 
operator.  

\begin{figure}[h]
  \includegraphics[scale=1.1]{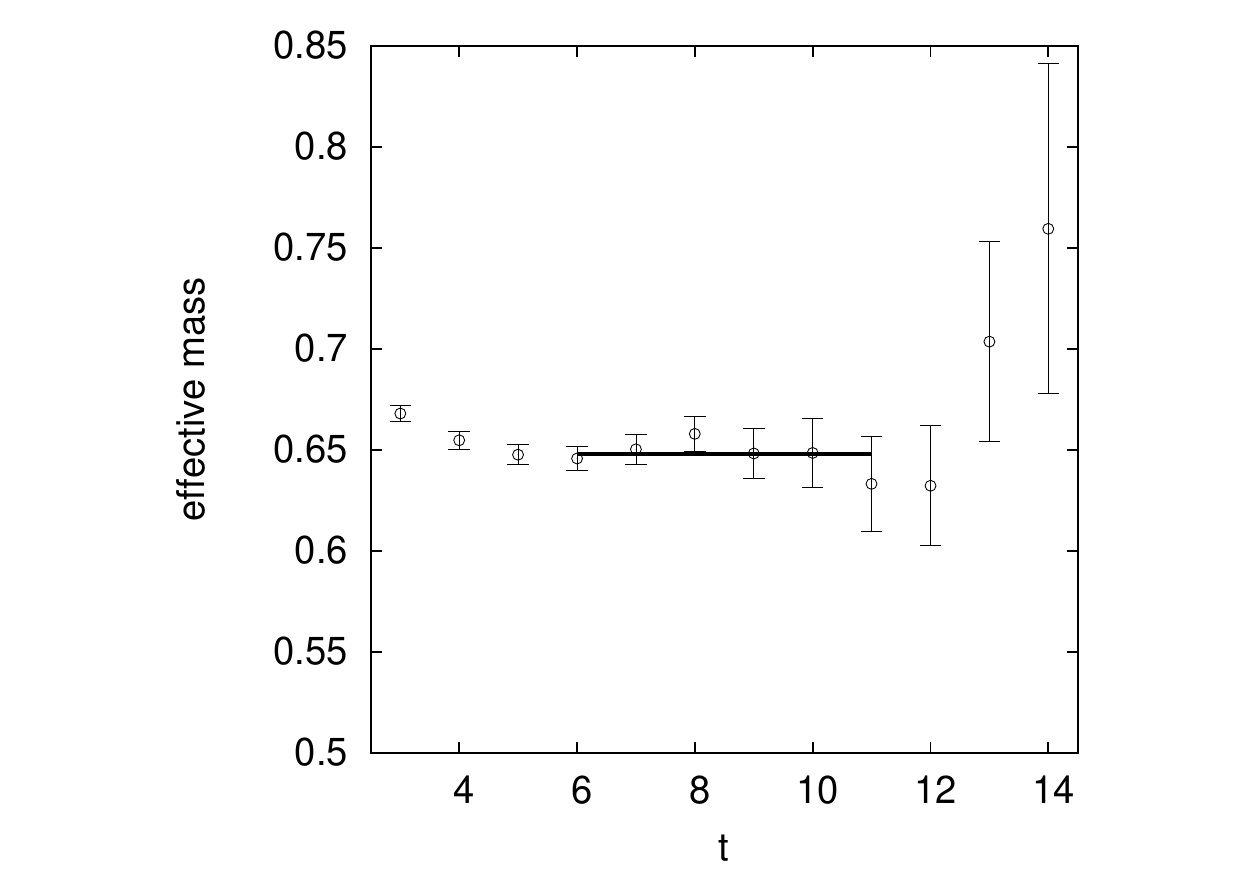}
  \includegraphics[scale=1.1]{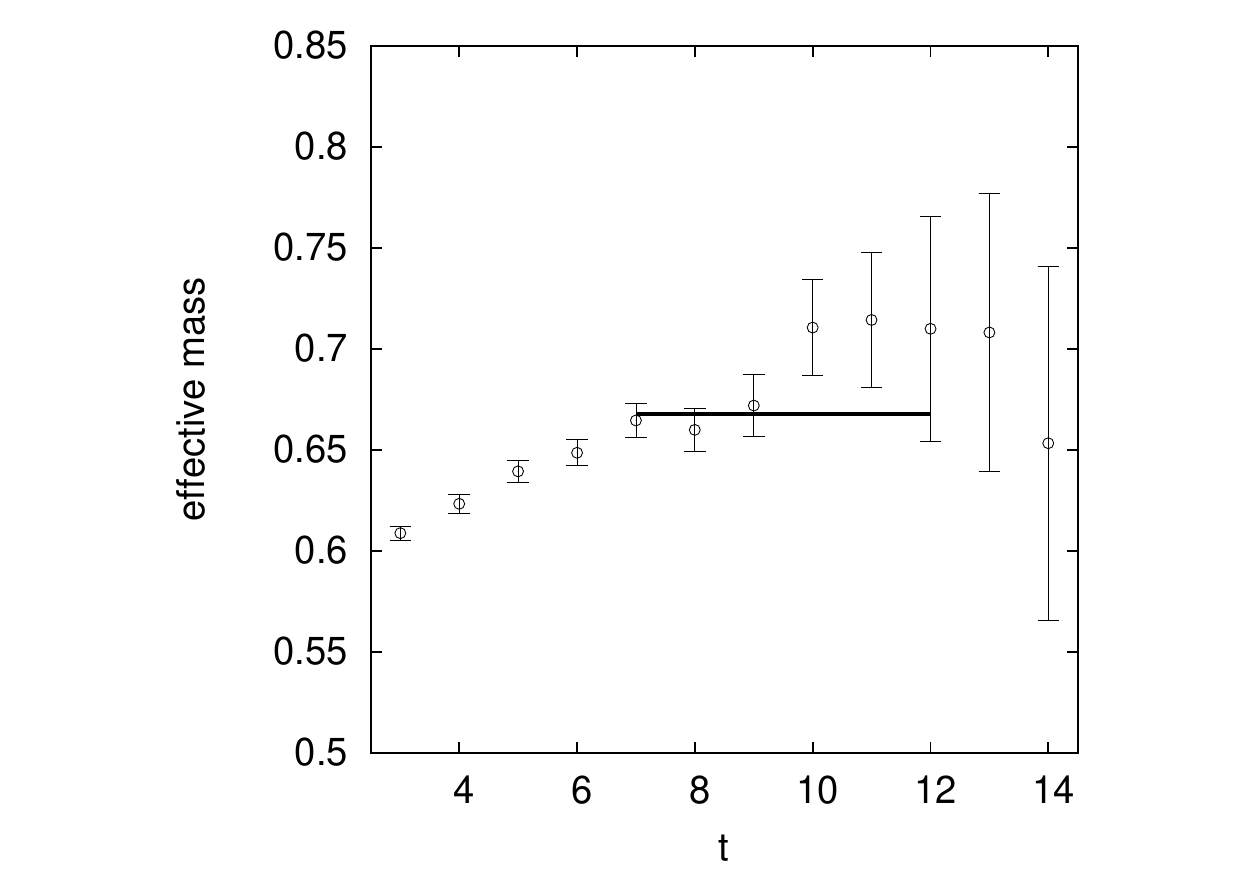}
  \caption{Proton effective masses, $24^{3}$ lattice size, $m_l=0.005$. The upper panel is for the unitary point and box source. The lower panel is for a non-degenerate case and wall source.}
\label{fig:nuc plateaus}
\end{figure}

Initially, nucleon correlation functions were computed from the same wall source propagators used for the meson splitting analysis. However, on the $24^{3}$ ensembles these exhibited poor plateaus and had poor signals for the EM neutron-proton mass difference. We then switched to box sources (of size $16^3$), which gave much better plateaus and signals, but only on the unitary points because of the additional computational cost.
Thus, for the $16^3$ and $24^3$ QCD 
configurations the masses come from wall source 
correlation functions while for the $24^3$ QCD+QED configurations, the masses 
are from box source correlation functions. The configuration information of the additional measurements 
is listed in Tab.~\ref{tab:np-ensembles}.
Figure~\ref{fig:nuc plateaus} shows representative plateaus for the sea quark mass 0.005 ensemble.

\begin{table}[h]
  \centering
  \begin{tabular}{ccccccc}
    \hline  
    \hline  
    lat & $m_{\rm sea}$ & $m_{\rm val}$ & Trajectories & $\Delta$ &
    $N_{meas}$ & $t_{src}$\\
    \hline
    $24^3$ & 0.005 & 0.005 & 900-8000 & 20 & 355 & 0 \\
    $24^3$ & 0.01 & 0.01 & 1460-8540 & 40 & 534 & 0,16,32 \\
    $24^3$ & 0.02 & 0.02 & 1800-3560 & 20 & 534 & 0,8,16,24,32,48 \\
    $24^3$ & 0.03 & 0.03 & 1260-3020 & 20 & 534 & 0,8,16,24,32,48 \\
    \hline  
    \hline
  \end{tabular}
  \caption{Summary of additional configurations used for the box source nucleon 
    calculation on the $24^3$ lattices. QCD gauge
    configurations  generated by
    the RBC and UKQCD collaborations~\cite{Allton:2007hx,Allton:2008pn,RBC:2010}. $\Delta$ is 
    the separation between measurements in molecular dynamics 
    time units. The Iwasaki gauge coupling is $\beta =2.13$.}
  \label{tab:np-ensembles}
\end{table}

The nucleon masses are listed in Tabs.~\ref{tab:mass-em} 
and~\ref{tab:mass-isospin}. They come from a standard $\chi^2$ 
minimization with correlated fit, and the error on the mass is 
from the standard jackknife method. The results in
Tab.~\ref{tab:mass-em} for the unitary masses and non-zero $\alpha_{\rm
em}$ on the $24^{3}$ ensembles are consistent with the pure QCD results
obtained on the same ensembles reported
in Ref.~\cite{Allton:2008pn,RBC:2010},
except for the $m_{l}=0.005$ case.
 {The proton and neutron masses are about three standard deviations smaller than in the pure QCD case, or roughly three percent. 
It is of interest to further investigate how large
the EM effect is on the nucleon masses themselves,
as well as on the mass difference.
Of course, in Nature there is no way to measure the nucleon mass due to QCD alone.}

\begin{figure}[ht]
  \includegraphics[scale=1.]{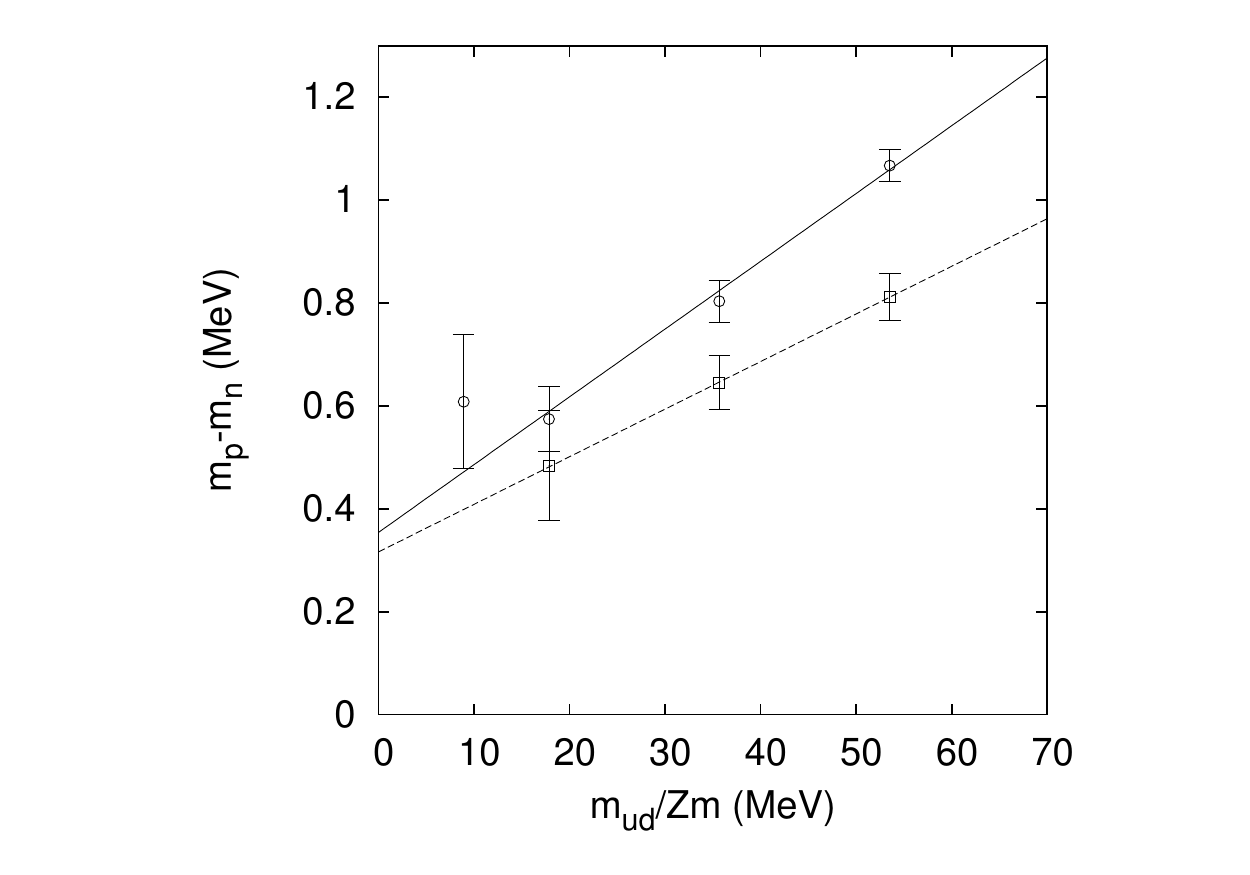}
  \caption{The proton-neutron mass difference due to the QED interaction computed for
    unitary points. $24^3$ (circle) and $16^3$ (square) lattice sizes.
    The solid (dashed) line corresponds to a linear fit to the $24^3$ ($16^3$) data points.}
  \label{fig:pn-em}
\end{figure}

Figure~\ref{fig:pn-em} shows the mass difference between the
proton and neutron due to the QED interaction for the unitary points. If
there is no QED interaction and $m_u=m_d$, then $m_n=m_p$, which is 
the result of isospin symmetry. When the QED interaction is included, the proton is
heavier than the neutron, and the mass difference decreases with quark mass
as observed in Fig.~\ref{fig:pn-em}. The $24^3$ result is larger than the 
$16^3$ result, once again signaling finite volume corrections. This 
simulation is on the unitary points, but $m_u\neq m_d$ in nature. 
When we extrapolate $(m_p-m_n)_\text{QED}$ to the physical point, we use the 
average light quark mass $m_{ud}$, as determined in the previous section.  
Finally we find that $(m_p-m_n)_\text{QED}$ is about 0.4 MeV (see Tab.~\ref{tab:pn-em}).
From Fig.~\ref{fig:pn-em} there is a visible flattening of the splitting at the lightest quark mass for the $24^3$ lattice size. Using only the lightest two quark masses in the extrapolation, we obtain $(m_p-m_n)_\text{QED}=0.63(23)$ MeV. The difference between the two results is used to estimate the systematic error in the chiral extrapolation.

Since the photon is not confined,
the EM proton-neutron mass difference
could suffer from a large finite volume effect.
In order to estimate this artifact,
we use the Cottingham formula~\cite{Cottingham:1963zz,Duncan:1996be},

\begin{eqnarray}
\delta m_{ele} &=& 2\pi \alpha m \frac{1}{L^3}
\sum_{q\neq 0} \frac{G_E(q)^2}{|q|}
\cdot\left[ \frac{2}{q^2 + 4m^2} + 
\frac{1}{2m^2}\left( \sqrt{1+\frac{4m^2}{q^2}} - 1 \right)\right] \\
\delta m_{mag} &=& - \frac{\pi \alpha}{2m^3} \frac{1}{L^3}
\sum_{q\neq 0} |q| G_M(q)^2
\cdot\left( 
\sqrt{1+\frac{4m^2}{q^2}}
-1 
-\frac{1}{2} \frac{1}{1+ q^2 / 4m^2} \right),
\end{eqnarray}
where $\delta m_{ele}$ ($\delta m_{mag}$) is 
the electric (magnetic) contribution to the nucleon mass $m$.
We evaluate the above formulae at the physical point, using the dipole form for
the nucleon electromagnetic form factors,
$G_E^p(Q^2) = G_M^p(Q^2) / \mu_p = G_M^n(Q^2) / \mu_n = G_D(Q^2)$,
where
$\mu_p (\mu_n)$, are proton(neutron) magnetic moment,
and $G_D(Q^2) = 1/(1+Q^2/\Lambda^2)^2$ with $\Lambda^2 = 0.71 {\rm GeV}^2$.
For $G_E^n(Q^2)$, we use the Galster parametrization of
$G_E^n(Q^2) = A Q^2 / (4m^2 + B Q^2) \cdot G_D(Q^2)$
with $A= 1.70, B=3.30$~\cite{Kelly:2004hm}.
We obtain
$(m_p-m_n)^{\text{(Cott.)}}_\text{QED} = 0.04$ MeV for $16^3$ volume, and
$(m_p-m_n)^{\text{(Cott.)}}_\text{QED} = 0.16$ MeV for $24^3$ volume.
Since the formula yields 
$(m_p-m_n)^{\text{(Cott.)}}_\text{QED} = 0.77$ MeV for the infinite volume limit,
the finite volume artifact corresponds to
an underestimate of 
0.73 MeV and 0.61 MeV for $16^3$ and $24^3$, respectively.
The tendency for the larger volume to correspond to
larger $(m_p - m_n)_{\rm QED}$ 
is qualitatively consistent with the lattice results presented here.

\begin{figure}[ht]
  \includegraphics[scale=1.]{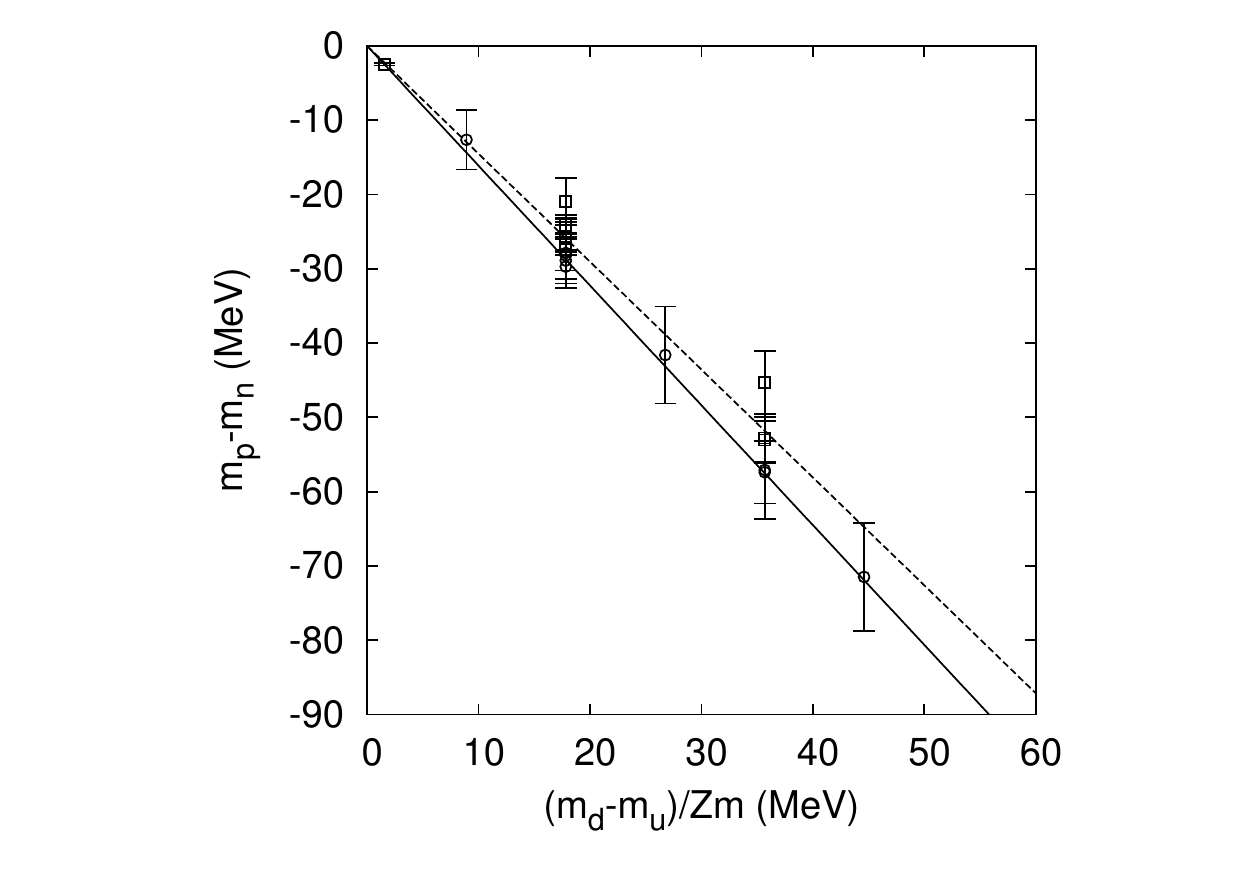}
  \caption{The proton-neutron mass difference for $m_u-m_d\neq 0$ and $e=0$. 
    The solid (dashed) line corresponds to a linear fit to the $24^3$ ($16^3$) data points, shown by circles (squares).}
  \label{fig:pn-isospin}
\end{figure}

Next we compute the mass splitting due to 
non-degenerate $u$ and $d$ quark masses, which is expected to switch the 
sign of the mass difference, in accord with Nature.
Figure~\ref{fig:pn-isospin} shows the fit of the proton and neutron 
mass difference due to non-degenerate $u,d$ quark masses computed on the QCD 
configurations. The LEC's and values of the splitting at the physical point are summarized in 
Tab.~\ref{tab:pn-isospin}. Figure~\ref{fig:pn-isospin} confirms that 
$(m_p-m_n)_{(m_d-m_u)}$ is proportional to $m_d-m_u$, which is predicted by 
baryon PQ$\chi$PT (Eq.~(\ref{eq:dif-np})). The slope is
extracted and the physical $(m_p-m_n)_{(m_d-m_u)}$ is estimated by setting 
$m_d-m_u$ to its physical value, again as determined in the previous section. Our result
is in good agreement with the one in Ref.~\cite{Beane:2006fk}.

The quark mass dependence of $m_p-m_n$ is simple in baryon
chiral perturbation theory~\cite{Beane:2006fk} to NLO in pure QCD, as seen in
Eq.~(\ref{eq:baryonchipt}). The leading quark mass dependence for
the EM splitting is unknown, so we assume that it is linear, and at this stage the measured values likely can not be used to discern a more complicated form anyway. In contrast, chiral perturbation theory for the nucleon mass itself predicts several non-analytic terms  at NLO, and the careful extrapolation to the physical point is an important topic of current calculations. Because we have few data points, and our quark masses are relatively heavy, we do not attempt such an extrapolation here. 

Combining the contributions from the EM interaction and non-degenerate $u,d$ 
quark masses, we give the physical $p$-$n$  mass splitting. 
We find $m_p-m_n=-1.93(12)$ and $-2.13(16)$ MeV, for $16^3$ and $24^3$ lattice sizes, respectively,
which is larger than the experimental result ($-1.293321~(4)$ MeV), but remarkable given that compared to the mass itself, the splitting is a 0.1\% effect in Nature. The errors above are statistical only, and their small size is due to the facts that the difference is calculated on exactly the same configurations and with the $\pm e$ averaging trick.  

To estimate the systematic error on the EM splitting from the chiral extrapolation we take the difference between the extrapolation using all of the data points (on the $24^3$ lattice) and the lightest two mass points, or roughly 0.3 MeV. The finite volume effects, while quite noticeable at the simulated quark masses, are smaller in the quark mass extrapolated result. To roughly estimate the finite volume effect, we consider the difference in the $16^3$ and $24^3$ results which is about 0.05 MeV when all of the data are used in the fits, and  roughly 0.3 MeV if only the lightest points on the $24^3$ lattice are used.
In light of the much larger artifact predicted by the Cottingham formula, we take the more conservative estimate of 0.3 MeV. The finite volume error on the pure QCD splitting appears to be under better control, and we simply take the difference of the two as an additional finite volume effect, or $\sim0.25$ MeV. The QCD splitting depends somewhat strongly on the value of $m_{u}-m_{d}$, and given the $\sim 20\%$ uncertainty in this quantity, we estimate the systematic error due to the extrapolation by varying $m_{u}-m_{d}$ over this range. This yields roughly a 0.5 MeV uncertainty. Adding all of these errors in quadrature, we find 
$m_p-m_n=-2.13(16)(70)$ MeV. 
The result and errors are summarized in Tab.~\ref{tab:mp-mn}.
Clearly further calculations are needed, at smaller quark masses to improve the extrapolation, with a different lattice spacing to take the continuum limit, and on a larger volume to improve the infinite volume extrapolation.

\begin{table}[h]
  \centering
  \begin{tabular}{ccccc}
    \hline  
    \hline  
    lattice size & $10^2A_0$ & $A_1$ & $\chi^2$/dof & $(m_p-m_n)_{\rm QED}$ (MeV)\\
    \hline
    $16^3$ & 2.42(95) & 1.26(38)  & 0.002(96) & 0.33(11)\\
    $24^3$ & 2.72(55) & 1.80(22)  & 0.7(1.2)  & 0.383(68) \\
    \hline
    \hline  
  \end{tabular}
  \caption{Proton and neutron mass difference due to the QED
    interaction. The LEC's are extracted from the nucleon data 
    on the unitary points. $(m_p-m_n)_{\rm QED}$ is given at the
    physical quark mass $m_{ud}$ determined in this work.}
  \label{tab:pn-em}
\end{table}

\begin{table}[h]
  \centering
  \begin{tabular}{cccc}
    \hline  
    \hline  
    lattice size & $-\frac{1}{3}(4\alpha-2\beta)$  & $\chi^2$/dof & $(m_p-m_n)_{(m_d-m_u)}$ (MeV)\\
    \hline
    $16^3$ & $-$1.452(45) & 1.1(1.2) & $-$2.265(70)\\
    $24^3$ & $-$1.612(92) & 0.06(24) & $-$2.51(14)\\
    \hline
    \hline  
  \end{tabular}
  \caption{Proton-neutron mass difference due to non-degenerate
    $u,d$ quark masses, computed on QCD configurations only. $(m_p-m_n)_{(m_d-m_u)}$ is calculated
    at the physical value of $(m_d-m_u)$ determined in this work.}
  \label{tab:pn-isospin}
\end{table}

\begin{table}
\centering
\begin{tabular}{cccc}
\hline
lattice size & $m_p-m_n$(MeV) & fit error (MeV) & finite vol. error (MeV)\\
\hline
$16^3$ & $-1.93(12)~$ &-&-\\
$24^3$ & $-2.13(16)~$ & 0.58 & 0.39 \\
\hline
\end{tabular}
\caption{Estimated result of the proton-neutron mass difference
in Nature (systematic errors as described in the text).}
\label{tab:mp-mn}
\end{table}

\clearpage

\section{Conclusion}\label{sec:conclusion}

In this paper we have investigated the EM mass splittings of the low-lying hadrons
from first principles in the framework of lattice QCD+QED.
Our simulations were based on 
the 2+1 flavor DWF QCD configurations generated by the RBC and UKQCD 
collaborations and quenched, non-compact, QED configurations generated by us.
 The mass splittings could be determined with very high statistical accuracy
since the QCD part of the fluctuations in the hadron masses largely
cancels in the splittings.
 The precision is further enhanced
by applying our $\pm e$ trick~\cite{Doi:2006xh,Blum:2007cy}
to cancel $O(e)$-noise on each configuration,
before averaging over the QCD ensemble.
 The statistical errors on the pseudo-scalar splittings
are at an impressive sub-one-percent level,
as are the errors on the masses themselves.
 
 The explicit chiral symmetry breaking induced
by the finite extra 5th-dimension of DWF was studied in detail
and shown to be under good control.
 This is important because the leading $O(\alpha_{\rm em} m_{\rm res})$
effect is comparable in size to the physical effects under investigation. 

 We fit the pseudo-scalar meson mass-squared splittings
to the theoretical predictions
of partially-quenched chiral perturbation theory, including photons,
to extract the EM low energy constants of the effective theory, up to NLO.
 We presented new analytic results for the kaon mass-squared
in Sec.~\ref{sec:chipt} and in the Appendix \ref{sec:PCHPT}.
 The fits were done to both ${\rm SU(3)_L \times SU(3)_R}$
and ${\rm SU(2)_L \times SU(2)_R}$-plus-kaon theories,
the latter being necessary to determine
the strange quark mass~\cite{Antonio:2007pb,Allton:2008pn,Kelly:2009fp,
Mawhinney:2009jy,RBC:2010,Aoki:2008sm,Kadoh:2008sq}.
 When using the finite volume PQ$\chi$PT formulas,
we found that the NLO corrections relative to LO are about 25\%
for the physical pion masses,
neglecting $O(\alpha_{\rm em}^{2})$-terms in the $\pi^{0}$ mass
that come from the axial anomaly (disconnected graphs)
and are expected to be small~\cite{Bijnens:2006mk}.
 Simple linear fits also work as well as the complicated NLO
chiral perturbation theory ones, as has been seen in the case of pure
QCD~\cite{Allton:2008pn,Kelly:2009fp,Mawhinney:2009jy,RBC:2010}.
 Indeed, our data do not show significant curvature,
so while they do not seem to require the presence of chiral logs
from a theoretical point of view, they are consistent with them.
 The EM splittings and LEC's are significantly affected
by the finite volume of the lattice,
as expected since the long range interactions of the photons are not confined.
 For our final values, we used the finite volume formulas
for the chiral logs computed in Ref.~\cite{Hayakawa:2008an}.
 { The lattice-extracted, ${\rm SU(3)_L\times SU(3)_R}$
LEC's were found to be somewhat inconsistent with the result of 
the phenomenological analysis in Ref.~\cite{Bijnens:2006mk}, although the latter were fit using an $ad
~hoc$ set of choices for the LEC's. This may also be due to a lack of convergence of SU(3) chiral perturbation theory in the range of quark masses used here, or finite volume effects, or both.}

The masses of the light quarks were also determined from our calculation. This is the first time EM interactions have been included directly in the quark masses determined from 2+1 flavor calculations.
We employed the physical masses of the $\pi^\pm$, $K^0$ 
and $K^\pm$ mesons as input to fix the quark masses in PQ$\chi$PT. 
The ${\rm SU(2)_L \times SU(2)_R}$-plus-kaon theory was used to quote 
our final values since the physical strange quark mass is outside the
range of convergence of ${\rm SU(3)_L\times SU(3)_R}$ chiral perturbation theory.
They are given in Eqs.~(\ref{eq:mu})-(\ref{eq:mud}), along with statistical and systematic errors.
The down-up mass difference and quark mass ratios are given 
in Eqs.~(\ref{eq:md-mu})-(\ref{eq:ms/mud}). These quark masses, up to EM
effects, are consistent with the pure QCD values given in Ref.~\cite{Allton:2008pn}, which is not surprising since the pure QCD LEC's were taken from an identical analysis of extended ensembles of configurations~\cite{RBC:2010} used there. Concerning the solution of the strong CP problem, it is of interest that our value for the up quark mass is different from zero by many ($\sim 6-7$) standard deviations.

{
The Dashen term, or LO EM contribution to the pion mass difference is 
$(m_{\pi^\pm}-m_{\pi^0})_{\rm QED}=3.38(23)$ MeV in our calculation, coming from the SU(2) chiral perturbation theory, finite-volume-corrected fit, which is our most reliable one. The error is statistical only. However, the value from the linear chiral fit agrees within errors. It is also consistent with the values of $m_{\pi^\pm}^2$ in the chiral limit recently reported in~\cite{Shintani:2008qe,Boyle:2009xi}, but somewhat smaller than the value from phenomenology and SU(3)
chiral perturbation theory~\cite{Bijnens:2006mk} and the value we
reported for two flavor QCD in Ref.~\cite{Blum:2007cy}.
This  suggests that NLO contributions at the physical quark masses may be as large as 25\% of the total pion mass difference, and approximating the $\pi^0$ mass from the LEC's computed here, we find the LO+NLO EM contribution at the physical point is $m_{\pi^{+}}-m_{\pi^{0}}=4.50(23)$ MeV.
Phenomenology predicts that a small part of the NLO correction is due to $m_{u}-m_{d}\neq0$, 0.17(3) MeV~\cite{Gasser:1984gg} and 0.32(20) MeV~\cite{Amoros:2001cp}. Similarly, we find for the kaons that the pure EM mass difference is $(m_{K^\pm}-m_{K^0})_{\rm QED}=1.87(10)$ MeV, while the contribution from $m_{u}-m_{d}\neq0$ is $-5.840(96)$ MeV. While these values are interesting, there is still systematic uncertainty in them which can only be removed by calculations with lighter quark masses and larger volumes.}

Finally, we also computed the proton-neutron mass difference, again for the first time in 2+1 flavor QCD+QED. Our result is somewhat bigger than the experimental one, but encouraging. { We found $m_p-m_n=0.383(68)$ MeV for the EM mass splitting, and $-2.51(14)$ MeV from $m_u\neq m_d$, both on the larger lattice (errors are statistical). Part of the systematic error, stemming mainly from finite volume and chiral extrapolations of the splittings, was estimated. The total splitting was found to be $m_{p}-m_{n}=-2.13(16)(70)$ MeV, where the first error is statistical and the second, part of the systematic error. The central value is from the $24^{3}$ lattice; we have not attempted either continuum limit or infinite volume extrapolations.
The sign and relative size of the EM effect compared to the $m_d-m_u$ mass difference effect is as expected.}

In this work, quenched QED configurations were used to 
account for the EM interactions of the valence quarks, {\it i.e.}, the sea quarks were neutral in our calculation.
The systematic error due to
this approximation can be removed by the re-weighing method~\cite{Duncan:2004ys,Jung:2010jt}.
We are now undertaking such a study.
In similar spirit to the most recent RBC/UKQCD pure QCD calculation~\cite{Kelly:2009fp,Mawhinney:2009jy,RBC:2010} on 
a finer lattice ensemble, $a\approx 0.086$ fm, the analysis
presented here is being replicated on those ensembles in order to take the continuum limit.
Similarly, calculations on a third set of ensembles being generated by the RBC and UKQCD collaborations are on-going, with a new modified Iwasaki gauge action~\cite{Renfrew:2009wu}, to better explore the chiral regime.

\section*{Acknowledgements}
 We thank Enno Scholz and the RBC and UKQCD collaborations 
 for providing us with the pure QCD LEC's used in
 this work. TB thanks Norman Christ for helpful discussions
on the EM induced part of the residual mass. TI thanks Claude Bernard,  
Mike Creutz, and Estia Eichten for illuminating discussions. TB and TI thank Gilberto Colangelo for helpful discussions on finite volume chiral perturbation theory.
TB and TI  thank the organizers of the CERN Theory Institute
``Future directions in Lattice Gauge Theory - LGT10", where a part of
this paper was finalized.
TB and TI  also appreciate discussions on  $f_K/f_\pi$ (Sec. \ref{sec:fK}) with G. Colangelo,  A. Juettner, Laurent Lellouch, C. Sachrajda,  and Y.Kuramashi held at
the workshop. TI also thanks W. Marciano concerning this section.
 We are grateful to USQCD and the RBRC for providing time on the DOE and RBRC 
QCDOC supercomputers at BNL for the computations reported here.
 TB and RZ were supported in part by the US DOE under contract DE-FG02-92ER40716,
 TD by
Grant-in-Aid for JSPS Fellows 21$\cdot$5985, MH by JSPS Grant-in-Aid
of Scientific Research (C) Grant No.~20540261 and (S) Grant No.~22224003,
TI and NY by Grant-in-Aid of the Japanese Ministry of Education (Nos. 20105001, 20105002, 22740183), and SU by the JSPS Grant-in-Aid No.~227180 and
Nagoya University Global COE program, Quest for Fundamental Principles in the Universe.
This manuscript has been authored by an employee (TI) of Brookhaven Science
Associates, LLC under Contract No. DE-AC02-98CH10886 with the U.S.
Department of Energy.

\clearpage

\appendix
 
\section{Tables of hadron masses and splittings}

\begin{longtable}{ccccccccc}
\caption{Summary of pseudo-scalar meson masses obtained from
 fits to the pseudo-scalar two-point correlation functions
 on the $16^3$ QCD+QED lattice configurations. Fit range
 is $9\le t \le N_{t}/2$ in each case. $m_{\text sea}$ is the mass 
 of the light quark in the sea sector. $m_1$ and $m_3$ are the 
 masses of the valence quarks. The mass of the strange sea quark 
 is fixed at 0.04. ``cov" and ``uncov" refer to  
 covariant and uncovariant fits, respectively.}\\

\hline\hline
$m_{sea}$ & $m_1$ & $m_3$ & $q_1(\frac{1}{3})$ & $q_3(\frac{1}{3})$ 
& $M$(cov) & $\chi^2/dof$(cov) & $M$(uncov) & $\chi^2/dof$(uncov) \\
\hline
\endfirsthead

\hline\hline
\multicolumn{9}{c}{ \tablename\ \thetable{} -- continued from previous page} \\
\hline
$m_{sea}$ & $m_1$ & $m_3$ & $q_1(\frac{1}{3})$ & $q_3(\frac{1}{3})$ 
& $M$(cov) & $\chi^2/dof$(cov) & $M$(uncov) & $\chi^2/dof$(uncov) \\
\hline
\endhead

\hline\hline
\endfoot
0.010 & 0.010 & 0.010 & 0 & 1 & 0.2430(14) & 1.21(89)  & 0.2420(16) & 0.02(4) \\ 
0.010 & 0.010 & 0.020 & 0 & 1 & 0.2837(13) & 1.39(96)  & 0.2828(15) & 0.03(5) \\ 
0.010 & 0.010 & 0.030 & 0 & 1 & 0.3196(12) & 1.50(99)  & 0.3188(14) & 0.04(6) \\ 
0.010 & 0.010 & 0.010 & 0 & 0 & 0.2427(14) & 1.21(89)  & 0.2417(16) & 0.02(4) \\ 
0.010 & 0.020 & 0.020 & 0 & 1 & 0.3193(12) & 1.55(1.01)  & 0.3185(13) & 0.04(6) \\ 
0.010 & 0.020 & 0.030 & 0 & 1 & 0.3518(11) & 1.62(1.03)  & 0.3510(12) & 0.04(6) \\ 
0.010 & 0.030 & 0.030 & 0 & 1 & 0.3817(11) & 1.65(1.05)  & 0.3809(11) & 0.04(6) \\ 
0.010 & 0.010 & 0.020 & 0 & 0 & 0.2834(13) & 1.39(96)  & 0.2825(15) & 0.03(5) \\
\end{longtable}

\clearpage
\begin{longtable}{ccccccccc}
\caption{Same as for Tab.~\ref{tab:mps16}, except for lattice size $24^3$.}\\

\hline\hline
$m_{sea}$ & $m_1$ & $m_3$ & $q_1(\frac{1}{3})$ & $q_3(\frac{1}{3})$ 
& $M$(cov) & $\chi^2/dof$(cov) & $M$(uncov) & $\chi^2/dof$(uncov) \\
\hline
\endfirsthead

\hline\hline
\multicolumn{9}{c}{ \tablename\ \thetable{} -- continued from previous page} \\
\hline
$m_{sea}$ & $m_1$ & $m_3$ & $q_1(\frac{1}{3})$ & $q_3(\frac{1}{3})$ 
& $M$(cov) & $\chi^2/dof$(cov) & $M$(uncov) & $\chi^2/dof$(uncov) \\
\hline
\endhead

\hline\hline
\endfoot
0.005 & 0.005 & 0.020 & -1 & 0 & 0.2605(4) & 0.92(40)  & 0.2605(5) & 0.07(7) \\ 
0.005 & 0.005 & 0.030 & -1 & 0 & 0.2984(4) & 0.99(42)  & 0.2985(5) & 0.05(5) \\ 
0.005 & 0.010 & 0.010 & -1 & 0 & 0.2393(4) & 0.97(41)  & 0.2393(5) & 0.08(8) \\ 
0.005 & 0.010 & 0.020 & -1 & 0 & 0.2798(4) & 0.94(41)  & 0.2799(5) & 0.06(5) \\ 
0.005 & 0.010 & 0.030 & -1 & 0 & 0.3154(4) & 0.98(42)  & 0.3156(5) & 0.04(4) \\ 
0.005 & 0.020 & 0.020 & -1 & 0 & 0.3153(3) & 0.89(40)  & 0.3154(5) & 0.04(3) \\ 
0.005 & 0.020 & 0.030 & -1 & 0 & 0.3475(3) & 0.91(40)  & 0.3477(4) & 0.04(3) \\ 
0.005 & 0.010 & 0.030 & 0 & 0 & 0.3152(4) & 0.98(42)  & 0.3154(5) & 0.04(4) \\
\end{longtable}

\clearpage

\begin{longtable}{ccccccc}
\caption{Summary of pseudo-scalar meson mass-squared splittings ($\times 10^3$)
  obtained from the masses in Tab.~\ref{tab:mps16}.
  Lattice size $16^3$.}\\

\hline\hline
$m_{sea}$ & $m_1$ & $m_3$ & $q_1(\frac{1}{3})$ & $q_3(\frac{1}{3})$ 
& $\Delta M^2 (\times 10^3) $(cov) & $\Delta M^2 (\times 10^3) $(uncov) \\
\hline
\endfirsthead

\hline\hline
\multicolumn{7}{c}{ \tablename\ \thetable{} -- continued from previous page} \\
\hline
$m_{sea}$ & $m_1$ & $m_3$ & $q_1(\frac{1}{3})$ & $q_3(\frac{1}{3})$ 
& $\Delta M^2 (\times 10^3) $(cov) & $\Delta M^2 (\times 10^3) $(uncov) \\
\hline
\endhead

\hline\hline
\endfoot
0.010 & 0.010 & 0.010 & 0 & 1 & 0.1210(14) &  0.1207(16) \\
0.010 & 0.010 & 0.020 & 0 & 1 & 0.1515(13) &  0.1510(14) \\
0.010 & 0.010 & 0.030 & 0 & 1 & 0.1820(14) &  0.1812(14) \\
0.010 & 0.020 & 0.020 & 0 & 1 & 0.1532(13) &  0.1529(13) \\
0.010 & 0.020 & 0.030 & 0 & 1 & 0.1844(13) &  0.1838(12) \\
0.010 & 0.030 & 0.030 & 0 & 1 & 0.1869(13) &  0.1865(12) \\
0.010 & 0.010 & 0.020 & 1 & 0 & 0.1222(15) &  0.1222(16) \\
0.010 & 0.010 & 0.030 & 1 & 0 & 0.1236(17) &  0.1237(17) \\
\end{longtable}

\clearpage
\begin{longtable}{ccccccc}
\caption{Summary of pseudo-scalar meson mass-squared splittings ($\times 10^3$)
  obtained from the masses in Tab.~\ref{tab:mps24}.
  Lattice size $24^3$.}\\
\hline\hline
$m_{sea}$ & $m_1$ & $m_3$ & $q_1(\frac{1}{3})$ & $q_3(\frac{1}{3})$ 
& $\Delta M^2 (\times 10^3) $(cov) & $\Delta M^2 (\times 10^3)$(uncov) \\
\hline
\endfirsthead

\hline\hline
\multicolumn{7}{c}{ \tablename\ \thetable{} -- continued from previous page} \\
\hline
$m_{sea}$ & $m_1$ & $m_3$ & $q_1(\frac{1}{3})$ & $q_3(\frac{1}{3})$ 
& $\Delta M^2 (\times 10^3)$(cov) & $\Delta M^2 (\times 10^3) $(uncov) \\
\hline
\endhead

\hline\hline
\endfoot

0.005 & 0.005 & 0.020 & -1 & 0 & 0.1148(11) &  0.1147(11) \\
0.005 & 0.005 & 0.030 & -1 & 0 & 0.1161(15) &  0.1166(15) \\
0.005 & 0.010 & 0.010 & -1 & 0 & 0.12945(77) &  0.12939(78) \\
0.005 & 0.010 & 0.020 & -1 & 0 & 0.13090(92) &  0.13096(87) \\
0.005 & 0.010 & 0.030 & -1 & 0 & 0.1327(10) &  0.1331(10) \\
0.005 & 0.020 & 0.020 & -1 & 0 & 0.16304(85) &  0.16335(76) \\
0.005 & 0.020 & 0.030 & -1 & 0 & 0.16615(87) &  0.16638(80) \\
0.005 & 0.030 & 0.030 & -1 & 0 & 0.19980(86) &  0.19990(80) \\
\end{longtable}

\begin{table}[h]
  \centering
  \begin{tabular}{cccccc}
    \hline  
    \hline  
    lattice size & $m_f$ & $m_p$ & $\chi^2$/dof & $m_n$ & $\chi^2$/dof \\
    \hline
    $16^3$ &  0.010 &   0.7125(57) &   0.70(85) &   0.7122(57) &   0.70(85)   \\
    $16^3$ &  0.020 &   0.7986(40) &   1.8(1.3) &   0.7982(40) &   1.8(1.3)   \\
    $16^3$ &  0.030 &   0.8747(36) &   2.2(1.5) &   0.8742(36) &   2.2(1.5)   \\
    \hline
    $24^3$ &  0.005 &   0.6477(53) &   0.85(94) &   0.6474(53) &   0.85(94)   \\
    $24^3$ &  0.010 &   0.7121(31) &   0.20(46) &   0.7118(32) &   0.21(46)   \\
    $24^3$ &  0.020 &   0.8065(25) &   0.82(92) &   0.8060(25) &   0.84(93)   \\
    $24^3$ &  0.030 &   0.8871(23) &   0.53(72) &   0.8865(23) &   0.53(72)   \\
    \hline
    \hline  
  \end{tabular}
  \caption{Proton and neutron masses on unitary points for QCD+QED
    configurations. The masses are from size $16^3$ box source, point sink
    correlation functions for both $16^3$ and $24^3$ lattices. The fit range 
    for $16^3$ lattices is 5-10. The fit range for $24^3$ lattices 
    is 6-11. The $\chi^2$/dof is from a covariant fit.}
  \label{tab:mass-em}
\end{table}

\begin{table}[h]
  \centering
  \begin{tabular}{cccccccc}
    \hline  
    \hline  
    lattice size & $m_{sea}$ & $m_u$ & $m_d$ & $m_p$ & $\chi^2$/dof & $m_n$ & $\chi^2$/dof \\
    \hline
    $16^3$& 0.010 & 0.010 & 0.020 &   0.7416(49) &   0.59(78) &   0.7562(43) &0.83(92)    \\
    $16^3$& 0.010 & 0.010 & 0.030 &   0.7684(45) &   0.77(88) &   0.7981(36) &1.3(1.1)    \\
    $16^3$& 0.010 & 0.020 & 0.030 &   0.8086(36) &   1.7(1.2) &   0.8238(33) &2.0(1.4)    \\
    $16^3$& 0.020 & 0.010 & 0.020 &   0.7553(51) &   0.88(95) &   0.7698(46) &1.5(1.2)    \\
    $16^3$& 0.020 & 0.010 & 0.030 &   0.7825(46) &   1.0(1.0) &   0.8120(40) &2.4(1.5)    \\
    $16^3$& 0.020 & 0.020 & 0.030 &   0.8230(38) &   2.2(1.4) &   0.8380(36) &2.9(1.7)    \\
    $16^3$& 0.030 & 0.010 & 0.020 &   0.7721(60) &   1.6(1.2) &   0.7839(51) &1.3(1.1)    \\
    $16^3$& 0.030 & 0.010 & 0.030 &   0.7988(55) &   1.6(1.3) &   0.8241(42) &1.2(1.1)    \\
    $16^3$& 0.030 & 0.020 & 0.030 &   0.8361(41) &   1.7(1.3) &   0.8496(38) &1.7(1.3)    \\
    \hline
    $24^3$& 0.005 & 0.005 & 0.010 &   0.6676(85) &   1.3(1.1) &   0.6747(73) &   1.0(1.0)    \\
    $24^3$& 0.005 & 0.005 & 0.020 &   0.6992(68) &   1.5(1.2) &   0.7225(51) &   1.0(1.0)    \\
    $24^3$& 0.005 & 0.005 & 0.030 &   0.7279(59) &   1.5(1.2) &   0.7680(41) &   1.2(1.1)    \\
    $24^3$& 0.005 & 0.010 & 0.020 &   0.7225(51) &   1.6(1.2) &   0.7383(44) &   1.4(1.2)    \\
    $24^3$& 0.005 & 0.010 & 0.030 &   0.7502(44) &   1.7(1.3) &   0.7824(37) &   1.7(1.3)    \\
    $24^3$& 0.005 & 0.020 & 0.030 &   0.7928(33) &   2.4(1.5) &   0.8090(32) &   2.5(1.6)    \\
    $24^3$& 0.010 & 0.010 & 0.020 &   0.7304(77) &   0.70(96) &   0.7461(61) &   0.9(1.1)    \\
    $24^3$& 0.010 & 0.010 & 0.030 &   0.7575(71) &   0.63(94) &   0.7895(48) &   1.3(1.2)    \\
    $24^3$& 0.010 & 0.020 & 0.030 &   0.7980(46) &   1.0(1.2) &   0.8146(40) &   1.8(1.5)    \\
    \hline    \hline  
  \end{tabular}
 \caption{Proton and neutron masses on QCD configurations with
    non-degenerate $u,d$ quark masses. The $p$-$n$ masses are from wall source, 
    point sink correlation functions for $16^3$ and $24^3$ lattices. 
    The fit range for $16^3$ lattices is 5-10. The fit range for 
    $24^3$ lattices is 7-12. The $\chi^2$/dof is from a covariant fit.}
  \label{tab:mass-isospin}
\end{table}

\clearpage

\section{Partially quenched chiral perturbation theory framework}
\label{sec:PCHPT}

 The aim of this appendix is twofold:
(1) to obtain all possible terms in the chiral Lagrangian
relevant to the kaon mass-squared at order $O(e^2)$ and $O(e^{2}p^{2})$
for the partially quenched ${\rm SU(2)}$ + kaon system, and 
(2) to derive the expression for the EM correction to the kaon mass-squared
to order $O(e^2 p^2)$.
 The appendix is compact,
mainly summarizing results and defining notation.
 Much of the machinery, of course, has been worked out before,
and we refer the interested reader to the literature.
 Here we follow closely the works
in Refs.~\cite{Roessl:1999iu,Ouellette:2001ib,Bijnens:2006mk,Hayakawa:2008an}.
 The new contributions in this work are
$O(e^2)$-terms and electromagnetic one-loop chiral logarithmic 
correction to the kaon mass-squared.
 We also list the $O(e^2 p^2)$-operators
relevant to the kaon-mass-squared, 
which serves as a check of the possible dependence of $O(e^2 p^2)$ corrections
on charges and masses. 

 We begin by reminding the reader of the important details and notation,
then construct the Lagrangian density,
and finally compute the corrections to the kaon mass-squared
to the order of our interest.

\subsection{SU(2) pion sector}
\label{subsec:su2_part}

 In the partially quenched system composed of $N_V$ valence quarks,
$N_S$ sea quarks and $N_V$ ghost quarks, the field $\Pi(x)$ 
representing the Nambu-Goldstone multiplet is the local coordinate of
the coset space $G /H$ 
($G \equiv {\rm SU}(N_S + N_V | N_V)_L \times {\rm SU}(N_S + N_V | N_V)_R$,
$H \equiv {\rm SU}(N_S + N_V | N_V)_V$) at each $x$
\begin{eqnarray}
 u[\Pi(x)] \equiv 
 \exp\left(i\,\frac{\Pi(x)}{\sqrt{2}\,F}\right)\,.
\end{eqnarray}
 With the normalization of $F$ such that $F \simeq 92$ MeV,
the leading-order (LO) chiral Lagrangian reads
\begin{eqnarray}
 \mathcal{L}_{{\rm QCD},\,2} &=&
 \frac{F^2}{4}\,
 \left< u_\mu u^\mu + \chi_+ \right>\,,\label{eq:chiLag_QCD2}
\end{eqnarray}
where $\left<,\right>$ denotes the supertrace
in the partially quenched light quark sector,
whose flavors can be indexed as
\begin{eqnarray}
 &&
 I = 1,\,\cdots,\,N_V\,:\,\mbox{light valence quark flavors}\,,\nonumber\\
 &&
 I = N_V + 1,\,\cdots,\,N_V + N_S\,:\,
  \mbox{light sea quark flavors}\,,\nonumber\\
 &&
 I = N_V + N_S + 1,\,\cdots,\,2 N_V + N_S\,:\,
  \mbox{light ghost quark flavors}\,.
 \label{eq:PQ_flavorBasis}
\end{eqnarray}
 The variables appearing in Eq.~(\ref{eq:chiLag_QCD2}),
\begin{eqnarray}
 &
 \displaystyle{ 
  u_\mu \equiv 
  i \left\{
   u^\dagger \left(\del_\mu u - i R_\mu u\right)
   - u \left(\del_\mu u^\dagger - i L_\mu u^\dagger\right)
  \right\}\,,
 }&\nonumber\\
 &\displaystyle{
  \chi_\pm \equiv 
  u^\dagger\chi u^\dagger \pm u\chi^\dagger u\,,
  \quad \chi \equiv 2 B_0 \mathcal{M}\,,
 }& \label{eq:def_uMu_chiPM}
\end{eqnarray}
are given in terms of the spurion field $\mathcal{M}$
in place of the ordinary quark mass matrix,
and the external fields $R_\mu$, $L_\mu$, 
which transform
under the local chiral rotation $(g_L(x),\,g_R(x)) \in G$ as
\begin{eqnarray}
 &
 \displaystyle{
  R_\mu \mapsto
  R^\prime_\mu = g_R R_\mu g_R^\dagger + i g_R\,\del_\mu g_R\,,
 }&\nonumber\\
 &
 \displaystyle{
  L_\mu \mapsto
  L^\prime_\mu = g_L L_\mu g_L^\dagger + i g_L\,\del_\mu g_L\,,
 }&\nonumber\\
 &
 \displaystyle{
  \mathcal{M} \mapsto
  \mathcal{M}^\prime = g_R \mathcal{M} g_L^\dagger\,.
 }&\
\end{eqnarray}
 For  
\begin{eqnarray}
 u[\Pi] \mapsto u[\Pi^\prime] = 
 g_R\,u[\Pi]\,h((g_L,\,g_R);\,\Pi)^\dagger 
 = h((g_L,\,g_R);\,\Pi)\,u[\Pi]\,g_L^\dagger\,,
\end{eqnarray}
with $h((g_L,\,g_R);\Pi) \in H$, it turns out
that $u_\mu$ and $\chi_\pm$ transform covariantly with respect to $h$
\begin{eqnarray}
 A \mapsto A^\prime = h A h^\dagger\,,
  \label{eq:h-cov-trans}
\end{eqnarray}
and that $\mathcal{L}_{{\rm QCD},\,2}$ is invariant under
the local chiral transformation.

 The high frequency modes of photons coupled to quarks also generate 
local interactions in the low-energy effective Lagrangian of QCD.
The coupling of quarks to photons preserves chiralities;
\begin{eqnarray}
 A_\mu 
 \left(
  \overline{q}_L\gamma^\mu\,\mathcal{Q} q_L
  +
  \overline{q}_R\gamma^\mu\,\mathcal{Q} q_R
 \right)\,,
\end{eqnarray}
where $\mathcal{Q}$ represents the charge matrix and takes the form
\begin{eqnarray}
 \mathcal{Q} &=&
 e\,{\rm diag}
    \left(q_{uV},\,q_{dV},\,q_{uS},\,q_{dS},\,q_{uV},\,q_{dV}\right)\,,
 \label{eq:CM_light}
\end{eqnarray}
for two-light flavors.
 The systematic dependence on these quark charges 
can hence be traced back once $\mathcal{Q}$ is promoted
to a set of hermitian spurion fields, 
$\mathcal{Q}_{R,\,L}$, that transform under chiral rotations as 
\begin{eqnarray}
 \mathcal{Q}_L \mapsto 
 \mathcal{Q}_L^\prime = g_L\,\mathcal{Q}_L\,g_L^\dagger\,,\quad
 \mathcal{Q}_R \mapsto 
 \mathcal{Q}_R^\prime = g_R\,\mathcal{Q}_R\,g_R^\dagger\,.
\end{eqnarray}
 On the other hand, to construct the effective Lagrangian,
it is convenient to define quantities that transform covariantly by
$h((g_L,\,g_R);\Pi)$, i.e., as in Eq.~(\ref{eq:h-cov-trans}) 
\begin{eqnarray}
 \widetilde{\mathcal{Q}}_L \equiv u \mathcal{Q}_L u^\dagger\,,\quad
 \widetilde{\mathcal{Q}}_R \equiv u^\dagger \mathcal{Q}_R u\,.\quad
\end{eqnarray}
 Since we will set $\mathcal{Q}_L,\,\mathcal{Q}_R$ to 
the diagonal EM charge matrix $\mathcal{Q}$ 
after constructing the effective Lagrangian,
we impose the chiral-invariant condition
\begin{eqnarray}
 \left<\mathcal{Q}_R\right> = \left<\mathcal{Q}_L\right>
 \equiv \left<\mathcal{Q}\right>\,,
\end{eqnarray}
which reduces just to the charge matrix in the end.
 The leading order ($O(p^2) \sim O(e^2)$) Lagrangian
involving Nambu-Goldstone bosons is thus given by
\begin{eqnarray}
 L_{\pi,\,2} &=&
 \frac{F^2}{4}\,\left<u_\mu\,u^\mu + \chi_+\right>
 +
 C\,\left<\widetilde{\mathcal{Q}}_R\,\widetilde{\mathcal{Q}}_L\right>\,,
\end{eqnarray}
 The QED corrections from the low frequency photons can also be included 
by coupling the Nambu-Goldstone bosons to 
the $U(1)$-gauge potential $A_\mu(x)$
and by setting the external fields $L_\mu,\,R_\mu$
along the direction of $\mathcal{Q}$ in the end;
\begin{eqnarray}
 L_\mu = \mathcal{Q}\,A_\mu = R_\mu\,.
\end{eqnarray}

\subsection{The kaon sector}
\label{subsec:kaon_su2}

 In SU(2) chiral perturbation theory, the strange quark
is treated as being heavy, and hence the kaons are no longer treated as Nambu-Goldstone bosons.
 Since the EM charge of the sea strange quark, $s_S$, 
differs from that of the valence strange quark, $s_V$, in our simulation,
these together with the ghost strange quark, $\widetilde{s}$,
are regarded as constituting the partially-quenched strange sector.
 Nevertheless, for the purpose of the analysis of our lattice data,
it suffices to write down the effective Lagrangian
with respect to the kaon multiplet
including the valence anti-strange quark $\overline{s}_V$,
keeping track explicitly of the dependence on the electric charges $Q_{s,\,V}$,
$Q_{s,\,S}$ (including $e$) of $s_V$ and $s_S$, respectively,
with the low energy constants having implicit dependence on
the sea strange quark mass.
 The relativistic form of the kinetic and mass terms of
the kaon multiplet $K$ ($U$, $D$ denote {\it constituent quarks})
\begin{eqnarray}
 &&
 K \sim
 \left(
  \begin{array}{c}
   \left[U_V \overline{s}_V\right] \\ \left[D_V \overline{s}_V\right] \\
   \left[U_S \overline{s}_V\right] \\ \left[D_S \overline{s}_V\right] \\
   \left[U_G \overline{s}_V\right] \\ \left[D_G \overline{s}_V\right] \\
  \end{array}
  \right)\,,
\end{eqnarray}
which is subject to the chiral rotation
\begin{eqnarray}
 K \mapsto
 h[\Pi(x),\,\left(g_L,\,g_R\right)]\,K\,,
\end{eqnarray}
is given by
\begin{eqnarray}
 \mathcal{L}_{K,\,{\rm kin}} &=&
 \nabla_\mu K^\dagger\,\nabla^\mu K
  - M^2 K^\dagger K\,, \label{eq:relativistic_form}
\end{eqnarray}
where $M$ is the LO mass of the kaon
and the covariant derivative $\nabla_\mu K$ is
with respect to the Maurer-Cartan form $\Gamma_\mu$ 
\begin{eqnarray}
 \nabla_\mu K &\equiv& \del_\mu K - i \Gamma_\mu K\,,\nonumber\\
 \Gamma_\mu &\equiv&
 - \frac{1}{2 i} 
   \left\{
    u^\dagger \left(\del_\mu u - i R_\mu u\right)
    +
    u \left(\del_\mu u^\dagger - i L_\mu u^\dagger\right)
   \right\}\,.
 \label{eq:def_DK_GammaMu}
\end{eqnarray}
 As is well-known \cite{Jenkins:1990jv},
$K$ is not suitable for the chiral order counting 
since that variable also carries the high frequency modes.
 The fluctuation is decomposed into the high frequency modes
originating from $M$
and the low frequency modes represented by $k \equiv k_v$ 
\begin{eqnarray}
 \mathcal{K}(x) &=& e^{i M\,v \cdot x}\,k(x)\,,\label{eq:rel_nonrel}
\end{eqnarray}
where $v$ is a light-like four-vector.
 In terms of $k$, Eq.~(\ref{eq:relativistic_form}) becomes
\begin{eqnarray}
 \mathcal{L}_{K,\,{\rm kin}} &=&
 - i M\,v^\mu
 \left(
  k^\dagger \nabla_\mu k 
  - 
  \left(\nabla_\mu k^\dagger\right) k
 \right) 
 +
 \nabla_\mu k^\dagger\,\nabla^\mu k\,.\label{eq:kaonKineticTerm_k}
\end{eqnarray}
 The field $k$ carries
the momentum of the order $p \lesssim 4 \pi F_\pi,\,M$,
and the above Lagrangian is $O(p)$.
 In the succeeding sections
the effective Lagrangian is constructed in terms of $k$ and
is converted to the relativistic form described by $K$.

\subsection{$O(e^2)$ and $O(e^2 p^2)$ Lagrangian for the kaon sector}
\label{subsec:kaonTreeLevel}

 Having established the partially quenched framework and notation,
we construct the electromagnetic part of the chiral Lagrangian
by writing down all possible $O(e^2)$-terms 
possessing the symmetries of massless (QCD + QED)
in the non-relativistic theory and their relativistic counterparts, 
and $O(e^2 p^2)$-terms that can induce the tree-level contribution
to the kaon mass-squared.

\begin{table}[htb]
\begin{center}
\begin{tabular}{ccccc}
\hline\hline
building block & definition & order & $P$ & $C$ \\
\hline 
%
$\chi_\pm$ & Eq.~(\ref{eq:def_uMu_chiPM}) & $O(p^2)$ &
 $\pm \chi_\pm(\widetilde{x})$ & $\left(\chi_\pm\right)^T$ \\
%
%
%
$\left<\mathcal{Q}\right>\,\widetilde{\mathcal{Q}}_\pm$ & 
 $\left<\mathcal{Q}\right> 
 \left(
  \widetilde{\mathcal{Q}}_R \pm \widetilde{\mathcal{Q}}_L 
 \right)$ & $O(e^2)$ & 
 $\pm \left<\mathcal{Q}\right> 
  \widetilde{\mathcal{Q}}_\pm(\widetilde{x})$ &
 $\pm \left<\mathcal{Q}\right> 
  \left(\widetilde{\mathcal{Q}}_\pm\right)^T$ \\
$\widetilde{\mathcal{Q}}^2_{\left(\pm\right)}$ & 
 $\left(\widetilde{\mathcal{Q}}_R\right)^2 
  \pm \left(\widetilde{\mathcal{Q}}_L\right)^2$ & $O(e^2)$ & 
 $\pm \widetilde{\mathcal{Q}}_{\left(\pm\right)}^2(\widetilde{x})$ &
 $\pm \left(\widetilde{\mathcal{Q}}^2_{\left(\pm\right)}\right)^T$ \\
%
$\widetilde{\mathcal{Q}}_{RL,\,\pm}$ &
 $\widetilde{\mathcal{Q}}_R \widetilde{\mathcal{Q}}_L
  \pm \widetilde{\mathcal{Q}}_L \widetilde{\mathcal{Q}}_R$ & $O(e^2)$ &
 $\pm \widetilde{\mathcal{Q}}_{RL,\,\pm}(\widetilde{x})$ &
 $\left(\widetilde{\mathcal{Q}}_{RL,\,\pm}\right)^T$ \\
\hline\hline
\end{tabular}
\caption{Parity ($P$) and charge conjugation ($C$) transformation properties 
for operators at chiral order
$O(p^2),\,O(e^2)$ that do not contain kaon fields
and transform as $A \mapsto h\,A\,h^\dagger$.
 Under $P$,
$x = (x^0,\,{\bf x})$ transforms to $\widetilde{x} = (x^0,\,-{\bf x})$.
}
\label{tab:QCD_buildingBlocks}
\end{center}
\end{table}

\begin{table}[htb]
\begin{center}
\begin{tabular}{ccccc}
\hline\hline
& definition & order & $P$ & $C$ \\
\hline
 $k k^\dagger$ & & $O(1)$ &
 $k k^\dagger(\widetilde{x})$ &
 $\left(k k^\dagger\right)^T$ \\
$k_{\pm,\,\mu}$ & Eq.~(\ref{eq:kdk}) &
 $O(p)$ & $k^\mu_\pm(\widetilde{x})$ &
 $\pm \left(k_{\pm,\,\mu}\right)^T$ \\
$k_{\left(\mu\nu\right]}$ & Eq.~(\ref{eq:k(MuNu]k}) & $O(p^2)$ &
$k^{\left(\mu\nu\right]}(\widetilde{x})$ &
$\pm \left(k_{\left(\mu\nu\right]}\right)^T$ \\ 
$k_{\pm,\,\mu\nu}$ & Eq.~(\ref{eq:kMuNu2}) & $O(p^2)$ &
$k^{\mu\nu}_\pm(\widetilde{x})$ &
$\pm \left(k_{\pm,\,\mu\nu}\right)^T$ \\ 
\hline
 $k^{W,\,\widetilde{\mathcal{Q}}_+}_\pm$ & Eq.~(\ref{eq:kWQ})
 & $O(e^2)$ &
 $k^{W,\,\widetilde{\mathcal{Q}}_+}_\pm(\widetilde{x})$ &
 $\pm
  \left(k^{W,\,\widetilde{\mathcal{Q}}_+}_\pm\right)^T$ \\
 $k^{W,\,\widetilde{\mathcal{Q}}_-}_\pm$ & Eq.~(\ref{eq:kWQ})
 & $O(e^2)$ &
 $- k^{W,\,\widetilde{\mathcal{Q}}_-}_\pm(\widetilde{x})$ &
 $\mp
  \left(k^{W,\,\widetilde{\mathcal{Q}}_-}_\pm\right)^T$ \\
\hline\hline
\end{tabular}
\caption{Parity ($P$) and charge conjugation ($C$) transformation properties of operators
at chiral order $O(p^2), \,O(e^2)$ that are bilinear in kaon fields
and transform as
$A \mapsto h\,A\,h^\dagger$.
 $W$ is either one of $Q_{s,\,V}$ or $Q_{s,\,S}$.
 Under $P$,
$x = (x^0,\,{\bf x})$ transforms to $\widetilde{x} = (x^0,\,-{\bf x})$.
}
\label{tab:kaonTable}
\end{center}
\end{table} 

 To this end, Table \ref{tab:QCD_buildingBlocks} and \ref{tab:kaonTable} list
the building blocks of chiral order 
$O(p^2)$ and $O(e^2)$
that transform as Eq.~(\ref{eq:h-cov-trans}).
 The definition of various variables appearing in Table \ref{tab:kaonTable}
are as follows;
\begin{eqnarray}
 k_{\pm,\,\mu}
 &\equiv&
 i 
 \left(
  (\nabla_\mu k) k^\dagger
  \pm
  k \left(\nabla_\mu k\right)^\dagger
 \right)\,,\label{eq:kdk}\\
 k_{\left(\mu\nu\right)}
 &\equiv&
 \nabla_{\left(\mu\right.} k \nabla_{\left.\nu\right)} k^\dagger
 \nonumber\\
 &=&
 \frac{1}{2}
 \left(
  \nabla_\mu k \nabla_\nu k^\dagger 
  +
  \nabla_\nu k \nabla_\mu k^\dagger
 \right)\nonumber\\
 k_{\left[\mu\nu\right]}
 &\equiv&
 \nabla_{\left[\mu\right.} k \nabla_{\left.\nu\right]} k^\dagger\nonumber\\
 &=&
 \frac{1}{2}
 \left(
  \nabla_\mu k \nabla_\nu k^\dagger
  -
  \nabla_\nu k \nabla_\mu k^\dagger
 \right)\,,\label{eq:k(MuNu]k} \\
 k_{\pm,\,\mu\nu}
 &\equiv&
 \left(\nabla_{\mu\nu} k\right) k^\dagger
 \pm
 k \left(\nabla_{\mu\nu} k\right)^\dagger\,,
 \quad
 \nabla_{\mu\nu} \equiv \nabla_\mu \nabla_\nu + \nabla_\nu \nabla_\mu\,,
  \label{eq:kMuNu2}\\
 k^{W,\,\widetilde{\mathcal{Q}}_\pm}_\pm
 &\equiv&
 W
 \left(
  k k^\dagger \widetilde{\mathcal{Q}}_\pm \pm
  \widetilde{\mathcal{Q}}_\pm k k^\dagger 
 \right)\,.\label{eq:kWQ}
\end{eqnarray}
 In Table \ref{tab:QCD_buildingBlocks}, $u_\mu$, for instance,
is omitted, as it will be not be used hereafter.
 Table \ref{tab:QCD_buildingBlocks} and \ref{tab:kaonTable}
include $O(e^2)$-terms but not $O(e)$ and $O(e p)$, 
because EM charges are left in pairs in the low-energy effective theory 
after the high frequency photon modes are integrated out.

 From Table \ref{tab:QCD_buildingBlocks} and \ref{tab:kaonTable},
we find that there are 13 $O(e^2)$-operators bilinear in kaon fields
that are invariant under chiral, $P$ and $C$ transformations
\begin{eqnarray}
 \left< k k^\dagger \mathcal{A} \right>\,,\quad
 W \left< k k^\dagger \mathcal{B} \right>\,,\quad
 W_1 W_2 \left< k k^\dagger \right>\,,\label{eq:kaon_e2}
\end{eqnarray}
where $W$ is $Q_{s,\,V}$ or $Q_{s,\,S}$, and
\begin{eqnarray}
 &
 \displaystyle{
  \mathcal{A}
  \in
  \left\{
   \widetilde{\mathcal{Q}}^{\,2}_{(+)}\,,\,
   \widetilde{\mathcal{Q}}_{RL,\,+}\,,\,
   \left<\mathcal{Q}\right> \widetilde{\mathcal{Q}}_+\,,\,
   \left<\mathcal{Q}^2\right>\,,\,
   \left<\widetilde{\mathcal{Q}}_{RL,\,+}\right>\,,\,
   \left(\left<\mathcal{Q}\right>\right)^2
  \right\}\,,
 }&\nonumber\\
 &
 \displaystyle{
  \mathcal{B}
  \in
  \left\{
   \widetilde{\mathcal{Q}}_+,\,\left<\mathcal{Q}\right>
  \right\}\,,
 }&\nonumber\\
 &
 \displaystyle{
  \left(W_1,\,W_2\right)
  \in
  \left\{
   \left(Q_{s,\,V},\,Q_{s,\,V}\right),\,
   \left(Q_{s,\,V},\,Q_{s,\,S}\right),\,
   \left(Q_{s,\,S},\,Q_{s,\,S}\right)
  \right\}\,.
 }\label{eq:defOfcalA_calB}
\end{eqnarray}
 The relativistic forms of the individual operators are read
from the relation (\ref{eq:rel_nonrel}), and $O(e^2)$-Lagrangian density
in the kaon sector is hence given by
\begin{eqnarray}
 \mathcal{L}_{K,\,e^2} &=&
 - A_K^{(1,\,1)}\,
  K^\dagger 
  \left(
   \left(\widetilde{\mathcal{Q}}_R\right)^2 
    + \left(\widetilde{\mathcal{Q}}_L\right)^2
  \right) K \nonumber\\
 &&
 - A_K^{(1,\,2)}\,
   \left<
       \left(\widetilde{\mathcal{Q}}_R\right)^2 
     + \left(\widetilde{\mathcal{Q}}_L\right)^2
   \right> K^\dagger K \nonumber\\
 && 
 - A_K^{(2,\,1)}\,
  K^\dagger
  \left( 
     \widetilde{\mathcal{Q}}_R \widetilde{\mathcal{Q}}_L 
   + \widetilde{\mathcal{Q}}_L \widetilde{\mathcal{Q}}_R 
  \right) K
 -
 A_K^{(2,\,2)}\,
 \left<
    \widetilde{\mathcal{Q}}_R \widetilde{\mathcal{Q}}_L 
  + \widetilde{\mathcal{Q}}_L \widetilde{\mathcal{Q}}_R 
 \right> K^\dagger K \nonumber\\
 &&
 - 
 A_K^{(3)}\,\left<\mathcal{Q}\right>
 K^\dagger
 \left(\widetilde{\mathcal{Q}}_R + \widetilde{\mathcal{Q}}_L\right) K 
 -
 A_K^{(4)}\,\left<\mathcal{Q}\right>^2 K^\dagger K \nonumber\\
 &&
 - A_K^{(s,\,1)}\,Q_{s,\,V}^2\,K^\dagger K
 - A_K^{(s,\,1,\,2)}\,Q_{s,\,S}^2\,K^\dagger K
 - A_K^{(s,\,1,\,3)}\,Q_{s,\,V}\,Q_{s,\,S}\,K^\dagger K \nonumber\\
 &&
 -
 A_K^{(s,\,2)}\,Q_{s,\,V}
  K^\dagger
  \left(\widetilde{\mathcal{Q}}_R + \widetilde{\mathcal{Q}}_L\right) K
 \nonumber\\
 &&
 -
 A_K^{(s,\,3)}\,\left<\mathcal{Q}\right> 
 Q_{s,\,V}\,K^\dagger K
 -
 A_K^{(s,\,3,\,2)}\,
  \left<\mathcal{Q}\right>\,Q_{s,\,S}\,K^\dagger K \nonumber \\
 &&
 -A_{K}^{(s,\,3,\,3)}\,
  Q_{s,\,S}\,K^\dagger \left(\widetilde{Q}_R + \widetilde{Q}_L\right) K\,.
 \label{eq:kaon lag e2}
\end{eqnarray}
 There are $O(e^2 p)$-terms bilinear in kaon fields 
that are allowed from the symmetries.
 All possible terms are obtained from Eq.~(\ref{eq:defOfcalA_calB})
by the replacement $k k^\dagger \rightarrow v_\mu k_-^\mu$.
 These operators generate no $O(e^2)$-contribution
and chiral-logarithmic corrections to the kaon mass-squared.
 They, however, induce $O(e^2 p^2)$-contribution
to the kaon mass-squared after the renormalization of the kaon field.
 We shall come back to this point in Section \ref{subsec:EM_kaonMass}.

\begin{table}[htb]
\begin{center}
\begin{tabular}{ccccc}
\hline\hline
& definition & $P$ & $C$ \\
\hline
$k^{W,\,
    \widetilde{\mathcal{Q}}_\pm}_{\left(\mu\nu\right]_{(1)},\,\pm_{(2)}}$ &
Eq.~(\ref{eq:(MuNu]QW}) &
$k^{W,\,
    \widetilde{\mathcal{Q}}_\pm,\,\left(\mu\nu\right]_{(1)}}_{\pm_{(2)}}
 \left(\widetilde{x}\right)$ &
$\left(\pm_{(1)} 1\right)\left(\pm_{(2)} 1\right)\left(\pm 1\right)
\left(
  k^{W,\,
  \widetilde{\mathcal{Q}}_\pm}_{\left(\mu\nu\right]_{(1)},\,\pm_{(2)}}
\right)^T$
\\
$k^{W,\,\widetilde{Q}_+}_{\pm_{(1)},\,\pm_{(2)},\,\mu\nu}$ &
Eq.~(\ref{eq:(MuNuk)kQlight}) &
$k^{W,\,\widetilde{Q}_+,\,\mu\nu}_{\pm_{(1)},\,\pm_{(2)}}
 (\widetilde{x})$ &
$\left(\pm_{(1)} 1\right) \left(\pm_{(2)} 1\right)
\left(
 k^{W,\,\widetilde{Q}_+}_{\pm_{(1)},\,\pm_{(2)},\,\mu\nu}
\right)^T$
\\
$k^{W,\,\widetilde{Q}_-}_{\pm_{(1)},\,\pm_{(2)},\,\mu\nu}$ &
Eq.~(\ref{eq:(MuNuk)kQlight}) &
$k^{W,\,\widetilde{Q}_-,\,\mu\nu}_{\pm_{(1)},\,\pm_{(2)}}
 (\widetilde{x})$ &
$-\left(\pm_{(1)} 1\right) \left(\pm_{(2)} 1\right)
\left(
 k^{W,\,\widetilde{Q}_-}_{\pm_{(1)},\,\pm_{(2)},\,\mu\nu}
\right)^T$
\\
$k^{W,\,
    \nabla_\nu \widetilde{Q}_+}_{\pm_{(1)},\,\pm_{(2)},\,\mu}$ &
Eq.~(\ref{eq:kQsdkdQlight}) &
$k^{W,\,\nabla^\nu \widetilde{Q}_+,\,
    \mu}_{\pm_{(1)},\,\pm_{(2)}}(\widetilde{x})$ &
$ \left(\pm_{(1)} 1\right) \left(\pm_{(2)} 1\right)
\left(
 k^{W,\,
    \nabla_\nu \widetilde{Q}_+}_{\pm_{(1)},\,\pm_{(2)},\,\mu}
\right)^T$
\\
$k^{W,\,
    \nabla_\nu \widetilde{Q}_-}_{\pm_{(1)},\,\pm_{(2)},\,\mu}$ &
Eq.~(\ref{eq:kQsdkdQlight}) &
$k^{W,\,\nabla^\nu \widetilde{Q}_-,\,
    \mu}_{\pm_{(1)},\,\pm_{(2)}}(\widetilde{x})$ &
$- \left(\pm_{(1)} 1\right) \left(\pm_{(2)} 1\right)
\left(
 k^{W,\,
    \nabla_\nu \widetilde{Q}_-}_{\pm_{(1)},\,\pm_{(2)},\,\mu}
\right)^T$
\\
$k^{W,\,\nabla_{\mu\nu} \widetilde{Q}_\pm}$ &
Eq.~(\ref{eq:kQskddQlight}) &
$k^{W,\,\nabla^{\mu\nu} \widetilde{Q}_\pm}(\widetilde{x})$ &
$\pm
\left(
 k^{W,\,\nabla_{\mu\nu} \widetilde{Q}_\pm}
\right)^T$
\\
\hline\hline
\end{tabular}
\caption{Parity ($P$) and charge conjugation ($C$) transformation properties of
operators at chiral order $O(e^2 p^2)$
that are bilinear in kaon fields
and transform as $A \mapsto h\,A\,h^\dagger$.}
\label{tab:e2p2_kaon}
\end{center}
\end{table}

 Next we turn to listing up $O(e^2 p^2)$-terms
that induce the corrections to the kaon mass-squared at the tree level.
 The building blocks are those in Table
\ref{tab:QCD_buildingBlocks}, \ref{tab:kaonTable}
and \ref{tab:e2p2_kaon}.
 The definition of the quantities in Table \ref{tab:e2p2_kaon}
is as follows;
\begin{eqnarray}
 k^{W,\,
    \widetilde{\mathcal{Q}}_\pm}_{\left(\mu\nu\right],\,\pm}
 &=&
 W
 \left(
  k_{\left(\mu\nu\right]} \widetilde{\mathcal{Q}}_\pm
  \pm
  \widetilde{\mathcal{Q}}_\pm k_{\left(\mu\nu\right]}
 \right)\,,\label{eq:(MuNu]QW}\\
 k^{W,\,\widetilde{\mathcal{Q}}_\pm}_{\pm_{(1)},\,\pm_{(2)},\,\mu\nu}
 &=&
 W
 \left(
  k_{\pm_{(1)},\,\mu\nu} \widetilde{\mathcal{Q}}_\pm
  \pm_{(2)}
  \widetilde{\mathcal{Q}}_\pm k_{\pm_{(1)},\,\mu\nu}
 \right)\,,\label{eq:(MuNuk)kQlight}\\
 k^{W,\,
    \nabla_\nu \widetilde{\mathcal{Q}}_\pm}_{\pm_{(1)},\,\pm_{(2)},\,\mu}
 &\equiv&
 W
 \left(
  k_{\pm_{(1)},\,\mu} \nabla_\nu \widetilde{\mathcal{Q}}_\pm
  \pm_{(2)}
  \left(\nabla_\nu \widetilde{\mathcal{Q}}_\pm\right) k_{\pm_{(1)},\,\mu}
 \right)\,,\label{eq:kQsdkdQlight}\\
 k^{W,\,\nabla_{\mu\nu} \widetilde{\mathcal{Q}}_\pm}_\pm
 &\equiv&
 W
 \left(
  k k^\dagger\,\nabla_{\mu\nu} \widetilde{\mathcal{Q}}_\pm
  \pm
  \left(\nabla_{\mu\nu} \widetilde{\mathcal{Q}}_\pm\right) k k^\dagger
 \right)\,.\label{eq:kQskddQlight}
\end{eqnarray}
 The $O(e^2 p^2)$-terms with no derivatives are
\begin{eqnarray}
 &
 \displaystyle{
  \left<
   k k^\dagger \left\{ \chi_+,\,\mathcal{C}_+ \right\}
  \right>\,,\quad
  \left<
   k k^\dagger \widetilde{\mathcal{Q}}_+ \chi_+ \widetilde{\mathcal{Q}}_+
  \right>\,,\quad
  \left<
   k k^\dagger \widetilde{\mathcal{Q}}_- \chi_+ \widetilde{\mathcal{Q}}_-
  \right>\,,
 }& \nonumber\\
 &
 \displaystyle{
  \left<
   k k^\dagger \left[ \chi_-,\,\mathcal{C}_-\right]
  \right>\,,\quad
  \left<
   k k^\dagger \left\{ \chi_-\,,\,\widetilde{\mathcal{Q}}_{RL,\,-}\right\}
  \right>\,,
 }& \nonumber\\
 &
 \displaystyle{
  \left<
   k k^\dagger
   \left(
    \widetilde{\mathcal{Q}}_+ \chi_- \widetilde{\mathcal{Q}}_-
    -
    \widetilde{\mathcal{Q}}_- \chi_- \widetilde{\mathcal{Q}}_+
   \right)
  \right>\,,
 }&\nonumber\\
 &
 \displaystyle{
  k^\dagger k\,\left< \chi_+ \mathcal{C}_+ \right>\,,
  \quad
  \left< k k^\dagger \mathcal{C}_+ \right> \left<\chi_+\right>\,,
  \quad
  \left< k k^\dagger \chi_+ \right> \left<\mathcal{C}_+\right>\,,
  \quad
  k^\dagger k\,\left<\chi_+\right>
  \left< \mathcal{C}_+ \right>\,,
 }&\nonumber\\
 &
 \displaystyle{
  \left<
   k k^\dagger \widetilde{\mathcal{Q}}_+
  \right> \left<\chi_+ \widetilde{\mathcal{Q}}_+\right>\,,\quad
  \left<
   k k^\dagger \widetilde{\mathcal{Q}}_-
  \right> \left<\chi_+ \widetilde{\mathcal{Q}}_-\right>\,,\quad
 }&\nonumber\\
 &
 \displaystyle{
  \left< k k^\dagger \widetilde{\mathcal{Q}}_{RL,\,-} \right>
  \left< \chi_- \right>\,,\quad
  k^\dagger k\,
  \left< \chi_- \widetilde{\mathcal{Q}}_{RL,\,-} \right>\,,
 }&\nonumber\\
 &
 \displaystyle{
  W
  \left<
   k k^\dagger \left\{\chi_+,\,\widetilde{\mathcal{Q}}_+\right\}
  \right>\,,
  \quad
  W
  \left<
   k k^\dagger \left[\chi_-,\,\widetilde{\mathcal{Q}}_-\right]
  \right>\,,
 }&\nonumber\\
 &
 \displaystyle{
  W k^\dagger k\,
  \left< \chi_+ \widetilde{\mathcal{Q}}_+ \right>\,,\quad
  W \left< k k^\dagger \chi_+ \right> \left<\mathcal{Q}\right>\,,\quad
  W \left< k k^\dagger \widetilde{\mathcal{Q}}_+ \right>
   \left<\chi_+\right>\,,
 }&\nonumber\\
 &
 \displaystyle{
  W k^\dagger k \left<\chi_+\right>
  \left< \mathcal{Q} \right>\,,
 }&\nonumber\\
 &
 \displaystyle{
  W_1 W_2 \left< k k^\dagger \chi_+ \right>\,,\quad
  W_1 W_2\,k^\dagger k\,\left< \chi_+ \right>\,,
 }& \label{eq:e^2p^2_nodrv}
\end{eqnarray}
where
\begin{eqnarray}
 \mathcal{C}_+
 \in
 \left\{
  \left<\mathcal{Q}\right> \widetilde{\mathcal{Q}}_+\,,\
  \widetilde{\mathcal{Q}}^{\,2}_{(+)}\,,\ 
  \widetilde{\mathcal{Q}}_{RL,\,+}
 \right\}\,,\quad
 \mathcal{C}_-
 \in
 \left\{
  \left<\mathcal{Q}\right> \widetilde{\mathcal{Q}}_-\,,\ 
  \widetilde{\mathcal{Q}}^{\,2}_{(-)}
 \right\}\,.
\end{eqnarray}
 The $O(e^2 p^2)$-terms with two derivatives
that contribute to the kaon mass-squared at the tree level are
($\eta^{\mu\nu}$ is the metric in Minkowski space)
\begin{eqnarray}
 &
 \displaystyle{
  \left<
   \eta^{\mu\nu} k_{\left(\mu\nu\right)} \mathcal{A}
  \right>\,,\quad
  \left< \eta^{\mu\nu} k_{+,\,\mu\nu} \mathcal{A} \right>\,,
 }&\nonumber\\
 &
 \displaystyle{
  W
  \left<
   \eta^{\mu\nu}\,k_{\left(\mu\nu\right)}\,\mathcal{B}
  \right>\,,\quad
  W
  \left<
   \eta^{\mu\nu}\,k_{-,\,\mu\nu}\,\widetilde{\mathcal{Q}}_-
  \right>\,,\quad
  W
  \left<
   \eta^{\mu\nu}\,k_{+,\,\mu\nu}\,\mathcal{B}
  \right>\,,
 }&\nonumber\\
 &
 \displaystyle{
  W_1 W_2 \left< \eta^{\mu\nu} k_{\left(\mu\nu\right)}\right>
  \,,\quad
  W_1 W_2 \left< \eta^{\mu\nu} k_{+,\,\mu\nu} \right>\,,
 }& \label{eq:e^2p^2_2drv}
\end{eqnarray}
where $\mathcal{A}$, $\mathcal{B}$ are the same as in
Eq.~(\ref{eq:defOfcalA_calB}).

\subsection{EM correction to kaon mass-squared}
\label{subsec:EM_kaonMass}
 In this subsection, the explicit expression 
for the $O(e^2)$ and the $O(e^2 p^2)$ chiral-logarithmic correction
to the kaon mass-squared is obtained by
setting the charge matrices $\mathcal{Q}$ in light flavor 
partially quenched system as in Eq.~(\ref{eq:CM_light}), 
$Q_{s,\,V} = e\,q_{sV}$ and $Q_{s,\,S} = e\,q_{sS}$.
 The EM contribution to the $K^{+}$ mass-squared at order $e^{2}$ is
\begin{eqnarray}
 (M^{e^2}_{K^{+}})^{2} &=&
 2e^{2}((A_K^{(1,1)}+A_K^{(2,1)}) q_{uV}^{2}
 + e^2(2 A_K^{(1,2)}+2 A_K^{(2,2)} + e^2A_K^{(4)})
 (q_{uS}^{2}+q_{dS}^{2})\non\\
 &&+ e^2A_K^{(3)}(q_{uS}+q_{dS})q_{uV}\nonumber\\
 &&
  + e^2A_K^{(s,1)}q_{sV}^{2}
  + e^2A_K^{(s,1,2)}q_{sS}^{2}
  + e^2A_K^{(s,1,3)}q_{sV}q_{sS}\non\\
 &&
  + 2e^2 A_K^{(s,2)} q_{uV}q_{sV} \nonumber\\
 &&
  + e^2A_K^{(s,3)} (q_{uS}+q_{dS})q_{sV}
  + e^2A_K^{(s,3,2)} (q_{uS}+q_{dS})q_{sS}\nonumber\\
 &&
  + 2e^2A_K^{(s,3,3)}q_{uV}\,q_{sS}\,.
\label{eq:kaon mass e2}
\end{eqnarray}
 In quenched QED this becomes
\begin{eqnarray}
(M^{e^2}_{K^{+}})^{2} &=& 2e^{2}(A_K^{(1,1)}+A_K^{(2,1)}) q_{uV}^{2}
 + e^{2}A_K^{(s,1)}q_{sV}^{2} + 2 e^{2}A_K^{(s,2)} q_{uV}q_{sV}\,.
\label{eq:esq kaon mass sq}
\end{eqnarray}
 The $O(e^2)$-correction to the neutral kaon mass-squared,
$(M_{K^0}^{e^2})^2$,
is given by substituting $q_{dV}$ for $q_{uV}$ in Eq.~(\ref{eq:kaon mass e2}).

 We next consider the one-loop contribution to kaon mass squared.
 The scalar QED Lagrangian density (\ref{eq:relativistic_form}) 
gives the correction from the diagrams, in each of which a photon
propagates explicitly,
but these contributions are absorbed by the redefinition of
the coefficients in Eq.~(\ref{eq:kaon lag e2}) and
those of $O(e^2 p)$-operators in the infinite volume.
 The leading EM chiral-logarithmic correction comes only
from the tadpole diagrams induced by Eq.~(\ref{eq:kaon lag e2})
\begin{eqnarray}
 (M_{K,\,i}^{\rm log})^2 &=&
 - \frac{e^2}{16\pi^2}\,\frac{A_K^{(1,\,1)}}{F_0^2}
 \sum_{n\,:\,{\rm sea}} \left(q_{iV}^2 - q_{nS}^2\right)
 \chi_{in}\,
  {\rm ln}\left(\frac{\chi_{in}}{\mu^2}\right)\nonumber\\
 &&
 - 
 \frac{e^2}{16\pi^2}\,\frac{A_K^{(2,\,1)}}{F_0^2}
 \sum_{n\,:\,{\rm sea}} 
 \left\{
  2 q_{iV} \left(q_{iV} - q_{nS}\right) 
  + \left(q_{iV} - q_{nS}\right)^2
 \right\}\,
 \chi_{in}\,
  {\rm ln}\left(\frac{\chi_{in}}{\mu^2}\right)\nonumber\\
 &&
 - 
 \frac{e^2}{16\pi^2}\,\frac{2\,A_K^{(2,\,2)}}{F_0^2}
 \sum_{n,\,m\,:\,{\rm sea},\,n \ne m} 
 \left(q_{nS} - q_{mS}\right)^2\,
 \chi_{mn}\,
  {\rm ln}\left(\frac{\chi_{mn}}{\mu^2}\right)\nonumber\\
 &&
 -
 \frac{e^2}{16\pi^2}\,
 \frac{A_K^{(3)}\,N_S\,\overline{Q} + A_K^{(s,\,2)}\,q_{sV}
       + A_K^{(s,\,3,\,3)}\,q_{sS}}{F_0^2}\nonumber\\
 &&\qquad
 \times
 \sum_{n\,:\,{\rm sea}} 
 \left(q_{iV} - q_{nS}\right)
 \chi_{in}\,
  {\rm ln}\left(\frac{\chi_{in}}{\mu^2}\right)\,,
  \label{eq:chiral_logs_K_unquencehdQED}
\end{eqnarray}
where $i = u$ or $d$, $\mu$ is the renormalization scale, and  
\begin{eqnarray}
 &
 \displaystyle{
  \chi_{mn} = \frac{\chi_m + \chi_n}{2}\,,\quad
  \chi_n = 2 B_0 m_n\,,
 }&\nonumber\\
 &
 \displaystyle{
  \overline{Q} = \frac{1}{N_S} \sum_{n\,:\,{\rm sea}} q_{nS}\,.
 }&
\end{eqnarray}
 In our simulation, all sea quarks are neutral
and the two light sea quarks are degenerate in mass $m_{(S)}$.
 Hence $(M_{K,\,i}^{\rm log})^2$ reduces to
\begin{eqnarray}
 (M_{K,\,i}^{\rm log})^2 &=&
 - 
 2\,\frac{e^2}{16\pi^2}\,\frac{1}{F_0^2} \nonumber\\
 &&\quad
 \times
 \left\{
  q_{iV}^2
  \left(A_K^{(1,\,1)} + 3\,A_K^{(2,\,1)}\right)
  +
  q_{iV} q_{sV}\,A_K^{(s,\,2)}
 \right\} 
 \chi_{i{(S)}}\,
  {\rm ln}\left(\frac{\chi_{i{(S)}}}{\mu^2}\right)
   \,,\label{eq:chiral_logs_L_quenchedQED}
\end{eqnarray}
where $\chi_{i{(S)}} \equiv B_0 \left(m_i + m_{(S)}\right)$.

 There are two types of finite volume corrections
induced at the one-loop level.
 The first type is given by the scalar QED diagrams
\begin{eqnarray}
 \left.\Delta \left(M_{K^+}\right)^2\right|_{\rm EM,\,photonic}(L)
 &=&
 \left(q_K\right)^2 e^2
 \left\{
  - 3\,\frac{\kappa}{4\pi}\,\frac{1}{L^2}
  + \frac{1}{\left(4\pi\right)^2}\,\frac{\mathcal{K}(m_K L)}{L^2}
 \right.\nonumber\\
 &&\qquad\qquad
 \left.
  - 4\,\frac{1}{\left(4\pi\right)^2}\,\frac{m_K}{L}\,\mathcal{H}(m_K L)
 \right\}\,,
  \label{eq:kaon fv 1}
\end{eqnarray}
where $\kappa$ and various functions are defined  
in Eqs.~(\ref{eq:def_kappa}), (\ref{eq:fun_H}) and (\ref{eq:fun_K}).
 Another type is the finite volume correction to the terms
(\ref{eq:chiral_logs_K_unquencehdQED}), 
$\Delta \left(M_{K,\,i}^{\rm log}\right)^2(L)$,
whose expression is obtained by making the following substitution 
to each logarithm in Eq.~(\ref{eq:chiral_logs_K_unquencehdQED})
\begin{eqnarray}
 m^2\,{\rm ln}\left(\frac{m^2}{\mu^2}\right)
 \,\Rightarrow\,
 \frac{\mathcal{M}(mL)}{L^2}\,,
 \label{eq:kaon fv 2}
\end{eqnarray}
with $\mathcal{M}(x)$ in Eq.~(\ref{eq:def_fun_M}).
 The finite size scaling effect
on the $O(e^2)$ wave function renormalization
could induce $O(e^2 p^2)$-correction to kaon mass squared
after the renormalization of the kaon field.
 The explicit calculation, however, shows that such effects do not exist.

 There are as many LEC's as $O(e^2 p^2)$-operators
in Eqs.~(\ref{eq:e^2p^2_nodrv}) and (\ref{eq:e^2p^2_2drv})
participating in the $O(e^2 p^2)$-contribution to kaon mass-squared,
while our lattice study here can determine at best
the linear combinations of LEC's of terms with the same charge and light
quark mass dependence of order $e^2 m$,
from the response of the data
to the variation of these parameters in the (QCD + QED) action.
 The dependence on those parameters can be read off from
Eqs.~(\ref{eq:e^2p^2_nodrv}) and (\ref{eq:e^2p^2_2drv}).
 In effect, Eq.~(\ref{eq:e^2p^2_nodrv}) alone leads to 
the following form of the charge and mass dependence 
of the $O(e^2 m)$-correction ($i = u,\,d$) in quenched QED, as anticipated,
\begin{eqnarray}
 \left(M_{K,\,i}^{e^2 p^2}\right) &=&
 e^2\,m_{iV}
 \left(
  x^{(K)}_3 \left(q_{iV} + q_{sV}\right)^2
  +
  x^{(K)}_4 \left(q_{iV} - q_{sV}\right)^2
  +
  x^{(K)}_5 \left(q_{iV}^2 - q_{sV}^2\right)
 \right)\nonumber\\
 &&
 +
 e^2\,m_{(S)}
 \left(
  x^{(K)}_6 \left(q_{iV} + q_{sV}\right)^2
  +
  x^{(K)}_7 \left(q_{iV} - q_{sV}\right)^2
  +
  x^{(K)}_8 \left(q_{iV}^2 - q_{sV}^2\right)
 \right)\,.\nonumber\\
 \label{eq:esq-psq kaon mass sq}
\end{eqnarray}
 We note that $m_{iV}$ and $m_{(S)}$
are denoted by $m_1$ and $m_4 = m_5$, respectively, 
in Eq.~(\ref{eq:kaon mass sq}).

\clearpage

\bibliography{paper}

\begin{thebibliography}{10}%
\makeatletter
\providecommand \@ifxundefined [1]{%
 \ifx #1\undefined \expandafter \@firstoftwo
 \else \expandafter \@secondoftwo
\fi
}%
\providecommand \@ifnum [1]{%
 \ifnum #1\expandafter \@firstoftwo
 \else \expandafter \@secondoftwo
\fi
}%
\providecommand \enquote [1]{``#1''}%
\providecommand \bibnamefont  [1]{#1}%
\providecommand \bibfnamefont [1]{#1}%
\providecommand \citenamefont [1]{#1}%
\providecommand\href[0]{\@sanitize\@href}%
\providecommand\@href[1]{\endgroup\@@startlink{#1}\endgroup\@@href}%
\providecommand\@@href[1]{#1\@@endlink}%
\providecommand \@sanitize [0]{\begingroup\catcode`\&12\catcode`\#12\relax}%
\@ifxundefined \pdfoutput {\@firstoftwo}{%
 \@ifnum{\z@=\pdfoutput}{\@firstoftwo}{\@secondoftwo}%
}{%
 \providecommand\@@startlink[1]{\leavevmode\special{html:<a href="#1">}}%
 \providecommand\@@endlink[0]{\special{html:</a>}}%
}{%
 \providecommand\@@startlink[1]{%
  \leavevmode
  \pdfstartlink
   attr{/Border[0 0 1 ]/H/I/C[0 1 1]}%
   user{/Subtype/Link/A<</Type/Action/S/URI/URI(#1)>>}%
  \relax
 }%
 \providecommand\@@endlink[0]{\pdfendlink}%
}%
\providecommand \url  [0]{\begingroup\@sanitize \@url }%
\providecommand \@url [1]{\endgroup\@href {#1}{\urlprefix}}%
\providecommand \urlprefix [0]{URL }%
\providecommand \Eprint[0]{\href }%
\@ifxundefined \urlstyle {%
  \providecommand \doi [1]{doi:\discretionary{}{}{}#1}%
}{%
  \providecommand \doi [0]{doi:\discretionary{}{}{}\begingroup
  \urlstyle{rm}\Url }%
}%
\providecommand \doibase [0]{http://dx.doi.org/}%
\providecommand \Doi[1]{\href{\doibase#1}}%
\providecommand \bibAnnote [3]{%
  \BibitemShut{#1}%
  \begin{quotation}\noindent
    \textsc{Key:}\ #2\\\textsc{Annotation:}\ #3%
  \end{quotation}%
}%
\providecommand \bibAnnoteFile [2]{%
  \IfFileExists{#2}{\bibAnnote {#1} {#2} {\input{#2}}}{}%
}%
\providecommand \typeout [0]{\immediate \write \m@ne }%
\providecommand \selectlanguage [0]{\@gobble}%
\providecommand \bibinfo [0]{\@secondoftwo}%
\providecommand \bibfield [0]{\@secondoftwo}%
\providecommand \translation [1]{[#1]}%
\providecommand \BibitemOpen[0]{}%
\providecommand \bibitemStop [0]{}%
\providecommand \bibitemNoStop [0]{.\EOS\space}%
\providecommand \EOS [0]{\spacefactor3000\relax}%
\providecommand \BibitemShut [1]{\csname bibitem#1\endcsname}%
\bibitem{Scholz:2009yz}%
  \BibitemOpen
  \bibfield{author}{%
  \bibinfo {author} {\bibfnamefont{E.~E.}\ \bibnamefont{Scholz}}}%
   (\bibinfo {year} {2009}),\
  \Eprint{http://arxiv.org/abs/0911.2191}{arXiv:0911.2191 [hep-lat]}%
  \bibAnnoteFile{NoStop}{Scholz:2009yz}%
\bibitem{Blum:2007cy}%
  \BibitemOpen
  \bibfield{author}{%
  \bibinfo {author} {\bibfnamefont{T.}~\bibnamefont{Blum}}, \bibinfo {author}
  {\bibfnamefont{T.}~\bibnamefont{Doi}}, \bibinfo {author}
  {\bibfnamefont{M.}~\bibnamefont{Hayakawa}}, \bibinfo {author}
  {\bibfnamefont{T.}~\bibnamefont{Izubuchi}},\ and\ \bibinfo {author}
  {\bibfnamefont{N.}~\bibnamefont{Yamada}},\ }%
  \bibfield{journal}{%
  \bibinfo {journal} {Phys. Rev.}\ }%
  \textbf{\bibinfo {volume} {D76}},\ \bibinfo {pages} {114508} (\bibinfo {year}
  {2007}),\ \Eprint{http://arxiv.org/abs/arXiv:0708.0484
  [hep-lat]}{arXiv:0708.0484 [hep-lat]}%
  \bibAnnoteFile{NoStop}{Blum:2007cy}%
\bibitem{Dashen:1969eg}%
  \BibitemOpen
  \bibfield{author}{%
  \bibinfo {author} {\bibfnamefont{R.~F.}\ \bibnamefont{Dashen}},\ }%
  \bibfield{journal}{%
  \Doi{10.1103/PhysRev.183.1245}{\bibinfo {journal} {Phys. Rev.}}\ }%
  \textbf{\bibinfo {volume} {183}},\ \bibinfo {pages} {1245} (\bibinfo {year}
  {1969})%
  \bibAnnoteFile{NoStop}{Dashen:1969eg}%
\bibitem{Amsler:2008zzb}%
  \BibitemOpen
  \bibfield{author}{%
  \bibinfo {author} {\bibfnamefont{C.}~\bibnamefont{Amsler}} \emph{et~al.}
  (\bibinfo {collaboration} {Particle Data Group}),\ }%
  \bibfield{journal}{%
  \Doi{10.1016/j.physletb.2008.07.018}{\bibinfo {journal} {Phys. Lett.}}\ }%
  \textbf{\bibinfo {volume} {B667}},\ \bibinfo {pages} {1} (\bibinfo {year}
  {2008})%
  \bibAnnoteFile{NoStop}{Amsler:2008zzb}%
\bibitem{Kaplan:1992bt}%
  \BibitemOpen
  \bibfield{author}{%
  \bibinfo {author} {\bibfnamefont{D.~B.}\ \bibnamefont{Kaplan}},\ }%
  \bibfield{journal}{%
  \bibinfo {journal} {Phys. Lett.}\ }%
  \textbf{\bibinfo {volume} {B288}},\ \bibinfo {pages} {342} (\bibinfo {year}
  {1992}),\ \Eprint{http://arxiv.org/abs/hep-lat/9206013}{hep-lat/9206013}%
  \bibAnnoteFile{NoStop}{Kaplan:1992bt}%
\bibitem{Shamir:1993zy}%
  \BibitemOpen
  \bibfield{author}{%
  \bibinfo {author} {\bibfnamefont{Y.}~\bibnamefont{Shamir}},\ }%
  \bibfield{journal}{%
  \bibinfo {journal} {Nucl. Phys.}\ }%
  \textbf{\bibinfo {volume} {B406}},\ \bibinfo {pages} {90} (\bibinfo {year}
  {1993}),\ \Eprint{http://arxiv.org/abs/hep-lat/9303005}{hep-lat/9303005}%
  \bibAnnoteFile{NoStop}{Shamir:1993zy}%
\bibitem{Allton:2008pn}%
  \BibitemOpen
  \bibfield{author}{%
  \bibinfo {author} {\bibfnamefont{C.}~\bibnamefont{Allton}} \emph{et~al.}
  (\bibinfo {collaboration} {RBC-UKQCD}),\ }%
  \bibfield{journal}{%
  \Doi{10.1103/PhysRevD.78.114509}{\bibinfo {journal} {Phys. Rev.}}\ }%
  \textbf{\bibinfo {volume} {D78}},\ \bibinfo {pages} {114509} (\bibinfo {year}
  {2008}),\ \Eprint{http://arxiv.org/abs/0804.0473}{arXiv:0804.0473 [hep-lat]}%
  \bibAnnoteFile{NoStop}{Allton:2008pn}%
\bibitem{Duncan:1996xy}%
  \BibitemOpen
  \bibfield{author}{%
  \bibinfo {author} {\bibfnamefont{A.}~\bibnamefont{Duncan}}, \bibinfo {author}
  {\bibfnamefont{E.}~\bibnamefont{Eichten}},\ and\ \bibinfo {author}
  {\bibfnamefont{H.}~\bibnamefont{Thacker}},\ }%
  \bibfield{journal}{%
  \Doi{10.1103/PhysRevLett.76.3894}{\bibinfo {journal} {Phys. Rev. Lett.}}\ }%
  \textbf{\bibinfo {volume} {76}},\ \bibinfo {pages} {3894} (\bibinfo {year}
  {1996}),\
  \Eprint{http://arxiv.org/abs/hep-lat/9602005}{arXiv:hep-lat/9602005}%
  \bibAnnoteFile{NoStop}{Duncan:1996xy}%
\bibitem{Duncan:1996be}%
  \BibitemOpen
  \bibfield{author}{%
  \bibinfo {author} {\bibfnamefont{A.}~\bibnamefont{Duncan}}, \bibinfo {author}
  {\bibfnamefont{E.}~\bibnamefont{Eichten}},\ and\ \bibinfo {author}
  {\bibfnamefont{H.}~\bibnamefont{Thacker}},\ }%
  \bibfield{journal}{%
  \Doi{10.1016/S0370-2693(97)00850-2}{\bibinfo {journal} {Phys. Lett.}}\ }%
  \textbf{\bibinfo {volume} {B409}},\ \bibinfo {pages} {387} (\bibinfo {year}
  {1997}),\
  \Eprint{http://arxiv.org/abs/hep-lat/9607032}{arXiv:hep-lat/9607032}%
  \bibAnnoteFile{NoStop}{Duncan:1996be}%
\bibitem{Hayakawa:2008an}%
  \BibitemOpen
  \bibfield{author}{%
  \bibinfo {author} {\bibfnamefont{M.}~\bibnamefont{Hayakawa}}\ and\ \bibinfo
  {author} {\bibfnamefont{S.}~\bibnamefont{Uno}},\ }%
  \bibfield{journal}{%
  \Doi{10.1143/PTP.120.413}{\bibinfo {journal} {Prog. Theor. Phys.}}\ }%
  \textbf{\bibinfo {volume} {120}},\ \bibinfo {pages} {413} (\bibinfo {year}
  {2008}),\ \Eprint{http://arxiv.org/abs/0804.2044}{arXiv:0804.2044 [hep-ph]}%
  \bibAnnoteFile{NoStop}{Hayakawa:2008an}%
\bibitem{Bijnens:2006mk}%
  \BibitemOpen
  \bibfield{author}{%
  \bibinfo {author} {\bibfnamefont{J.}~\bibnamefont{Bijnens}}\ and\ \bibinfo
  {author} {\bibfnamefont{N.}~\bibnamefont{Danielsson}},\ }%
  \bibfield{journal}{%
  \Doi{10.1103/PhysRevD.75.014505}{\bibinfo {journal} {Phys. Rev.}}\ }%
  \textbf{\bibinfo {volume} {D75}},\ \bibinfo {pages} {014505} (\bibinfo {year}
  {2007}),\
  \Eprint{http://arxiv.org/abs/hep-lat/0610127}{arXiv:hep-lat/0610127}%
  \bibAnnoteFile{NoStop}{Bijnens:2006mk}%
\bibitem{Roessl:1999iu}%
  \BibitemOpen
  \bibfield{author}{%
  \bibinfo {author} {\bibfnamefont{A.}~\bibnamefont{Roessl}},\ }%
  \bibfield{journal}{%
  \bibinfo {journal} {Nucl. Phys.}\ }%
  \textbf{\bibinfo {volume} {B555}},\ \bibinfo {pages} {507} (\bibinfo {year}
  {1999}),\ \Eprint{http://arxiv.org/abs/hep-ph/9904230}{hep-ph/9904230}%
  \bibAnnoteFile{NoStop}{Roessl:1999iu}%
\bibitem{Ouellette:2001ib}%
  \BibitemOpen
  \bibfield{author}{%
  \bibinfo {author} {\bibfnamefont{S.~M.}\ \bibnamefont{Ouellette}}}%
   (\bibinfo {year} {2001}),\
  \Eprint{http://arxiv.org/abs/hep-ph/0101055}{arXiv:hep-ph/0101055}%
  \bibAnnoteFile{NoStop}{Ouellette:2001ib}%
\bibitem{Antonio:2007pb}%
  \BibitemOpen
  \bibfield{author}{%
  \bibinfo {author} {\bibfnamefont{D.~J.}\ \bibnamefont{Antonio}} \emph{et~al.}
  (\bibinfo {collaboration} {RBC}),\ }%
  \bibfield{journal}{%
  \bibinfo {journal} {Phys. Rev. Lett.}\ }%
  \textbf{\bibinfo {volume} {100}},\ \bibinfo {pages} {032001} (\bibinfo {year}
  {2008}),\ \Eprint{http://arxiv.org/abs/hep-ph/0702042}{hep-ph/0702042}%
  \bibAnnoteFile{NoStop}{Antonio:2007pb}%
\bibitem{Kelly:2009fp}%
  \BibitemOpen
  \bibfield{author}{%
  \bibinfo {author} {\bibfnamefont{C.}~\bibnamefont{Kelly}}, \bibinfo {author}
  {\bibfnamefont{P.~A.}\ \bibnamefont{Boyle}},\ and\ \bibinfo {author}
  {\bibfnamefont{C.~T.}\ \bibnamefont{Sachrajda}},\ }%
  \bibfield{journal}{%
  \bibinfo {journal} {PoS}\ }%
  \textbf{\bibinfo {volume} {LAT2009}},\ \bibinfo {pages} {087} (\bibinfo
  {year} {2009}),\ \Eprint{http://arxiv.org/abs/0911.1309}{arXiv:0911.1309
  [hep-lat]}%
  \bibAnnoteFile{NoStop}{Kelly:2009fp}%
\bibitem{Mawhinney:2009jy}%
  \BibitemOpen
  \bibfield{author}{%
  \bibinfo {author} {\bibfnamefont{R.}~\bibnamefont{Mawhinney}} (\bibinfo
  {collaboration} {RBC}),\ }%
  \bibfield{journal}{%
  \bibinfo {journal} {PoS}\ }%
  \textbf{\bibinfo {volume} {LAT2009}},\ \bibinfo {pages} {081} (\bibinfo
  {year} {2009}),\ \Eprint{http://arxiv.org/abs/0910.3194}{arXiv:0910.3194
  [hep-lat]}%
  \bibAnnoteFile{NoStop}{Mawhinney:2009jy}%
\bibitem{RBC:2010}%
  \BibitemOpen
  \bibfield{journal}{%
  \bibinfo {journal} {RBC and UKQCD collaborations, In preparation}}%
   (\bibinfo {year} {2010})%
  \bibAnnoteFile{NoStop}{RBC:2010}%
\bibitem{Aoki:2008sm}%
  \BibitemOpen
  \bibfield{author}{%
  \bibinfo {author} {\bibfnamefont{S.}~\bibnamefont{Aoki}} \emph{et~al.}
  (\bibinfo {collaboration} {PACS-CS}),\ }%
  \bibfield{journal}{%
  \Doi{10.1103/PhysRevD.79.034503}{\bibinfo {journal} {Phys. Rev.}}\ }%
  \textbf{\bibinfo {volume} {D79}},\ \bibinfo {pages} {034503} (\bibinfo {year}
  {2009}),\ \Eprint{http://arxiv.org/abs/0807.1661}{arXiv:0807.1661 [hep-lat]}%
  \bibAnnoteFile{NoStop}{Aoki:2008sm}%
\bibitem{Kadoh:2008sq}%
  \BibitemOpen
  \bibfield{author}{%
  \bibinfo {author} {\bibfnamefont{D.}~\bibnamefont{Kadoh}} \emph{et~al.}
  (\bibinfo {collaboration} {PACS-CS}),\ }%
  \bibfield{journal}{%
  \bibinfo {journal} {PoS}\ }%
  \textbf{\bibinfo {volume} {LATTICE2008}},\ \bibinfo {pages} {092} (\bibinfo
  {year} {2008}),\ \Eprint{http://arxiv.org/abs/0810.0351}{arXiv:0810.0351
  [hep-lat]}%
  \bibAnnoteFile{NoStop}{Kadoh:2008sq}%
\bibitem{Zhou:2008gb}%
  \BibitemOpen
  \bibfield{author}{%
  \bibinfo {author} {\bibfnamefont{R.}~\bibnamefont{Zhou}} \emph{et~al.},\ }%
  \bibfield{journal}{%
  \bibinfo {journal} {PoS}\ }%
  \textbf{\bibinfo {volume} {LATTICE2008}},\ \bibinfo {pages} {131} (\bibinfo
  {year} {2008}),\ \Eprint{http://arxiv.org/abs/0810.1302}{arXiv:0810.1302
  [hep-lat]}%
  \bibAnnoteFile{NoStop}{Zhou:2008gb}%
\bibitem{Zhou:2009ku}%
  \BibitemOpen
  \bibfield{author}{%
  \bibinfo {author} {\bibfnamefont{R.}~\bibnamefont{Zhou}}\ and\ \bibinfo
  {author} {\bibfnamefont{S.}~\bibnamefont{Uno}},\ }%
  \bibfield{journal}{%
  \bibinfo {journal} {PoS}\ }%
  \textbf{\bibinfo {volume} {LAT2009}},\ \bibinfo {pages} {182} (\bibinfo
  {year} {2009}),\ \Eprint{http://arxiv.org/abs/0911.1541}{arXiv:0911.1541
  [hep-lat]}%
  \bibAnnoteFile{NoStop}{Zhou:2009ku}%
\bibitem{Basak:2008na}%
  \BibitemOpen
  \bibfield{author}{%
  \bibinfo {author} {\bibfnamefont{S.}~\bibnamefont{Basak}} \emph{et~al.}
  (\bibinfo {collaboration} {MILC}),\ }%
  \bibfield{journal}{%
  \bibinfo {journal} {PoS}\ }%
  \textbf{\bibinfo {volume} {LATTICE2008}},\ \bibinfo {pages} {127} (\bibinfo
  {year} {2008}),\ \Eprint{http://arxiv.org/abs/0812.4486}{arXiv:0812.4486
  [hep-lat]}%
  \bibAnnoteFile{NoStop}{Basak:2008na}%
\bibitem{Jenkins:1990jv}%
  \BibitemOpen
  \bibfield{author}{%
  \bibinfo {author} {\bibfnamefont{E.~E.}\ \bibnamefont{Jenkins}}\ and\
  \bibinfo {author} {\bibfnamefont{A.~V.}\ \bibnamefont{Manohar}},\ }%
  \bibfield{journal}{%
  \Doi{10.1016/0370-2693(91)90266-S}{\bibinfo {journal} {Phys. Lett.}}\ }%
  \textbf{\bibinfo {volume} {B255}},\ \bibinfo {pages} {558} (\bibinfo {year}
  {1991})%
  \bibAnnoteFile{NoStop}{Jenkins:1990jv}%
\bibitem{Duncan:2004ys}%
  \BibitemOpen
  \bibfield{author}{%
  \bibinfo {author} {\bibfnamefont{A.}~\bibnamefont{Duncan}}, \bibinfo {author}
  {\bibfnamefont{E.}~\bibnamefont{Eichten}},\ and\ \bibinfo {author}
  {\bibfnamefont{R.}~\bibnamefont{Sedgewick}},\ }%
  \bibfield{journal}{%
  \Doi{10.1103/PhysRevD.71.094509}{\bibinfo {journal} {Phys. Rev.}}\ }%
  \textbf{\bibinfo {volume} {D71}},\ \bibinfo {pages} {094509} (\bibinfo {year}
  {2005}),\
  \Eprint{http://arxiv.org/abs/hep-lat/0405014}{arXiv:hep-lat/0405014}%
  \bibAnnoteFile{NoStop}{Duncan:2004ys}%
\bibitem{Hasenfratz:2008fg}%
  \BibitemOpen
  \bibfield{author}{%
  \bibinfo {author} {\bibfnamefont{A.}~\bibnamefont{Hasenfratz}}, \bibinfo
  {author} {\bibfnamefont{R.}~\bibnamefont{Hoffmann}},\ and\ \bibinfo {author}
  {\bibfnamefont{S.}~\bibnamefont{Schaefer}},\ }%
  \bibfield{journal}{%
  \Doi{10.1103/PhysRevD.78.014515}{\bibinfo {journal} {Phys. Rev.}}\ }%
  \textbf{\bibinfo {volume} {D78}},\ \bibinfo {pages} {014515} (\bibinfo {year}
  {2008}),\ \Eprint{http://arxiv.org/abs/0805.2369}{arXiv:0805.2369 [hep-lat]}%
  \bibAnnoteFile{NoStop}{Hasenfratz:2008fg}%
\bibitem{Jung:2010jt}%
  \BibitemOpen
  \bibfield{author}{%
  \bibinfo {author} {\bibfnamefont{C.}~\bibnamefont{Jung}}}%
   (\bibinfo {year} {2010}),\
  \Eprint{http://arxiv.org/abs/1001.0941}{arXiv:1001.0941 [hep-lat]}%
  \bibAnnoteFile{NoStop}{Jung:2010jt}%
\bibitem{Aoki:2009ix}%
  \BibitemOpen
  \bibfield{author}{%
  \bibinfo {author} {\bibfnamefont{S.}~\bibnamefont{Aoki}} \emph{et~al.}
  (\bibinfo {collaboration} {PACS-CS}),\ }%
  \bibfield{journal}{%
  \Doi{10.1103/PhysRevD.81.074503}{\bibinfo {journal} {Phys. Rev.}}\ }%
  \textbf{\bibinfo {volume} {D81}},\ \bibinfo {pages} {074503} (\bibinfo {year}
  {2010}),\ \Eprint{http://arxiv.org/abs/0911.2561}{arXiv:0911.2561 [hep-lat]}%
  \bibAnnoteFile{NoStop}{Aoki:2009ix}%
\bibitem{Ishikawa:2010tq}%
  \BibitemOpen
  \bibfield{author}{%
  \bibinfo {author} {\bibfnamefont{T.}~\bibnamefont{Ishikawa}}, \bibinfo
  {author} {\bibfnamefont{Y.}~\bibnamefont{Aoki}},\ and\ \bibinfo {author}
  {\bibfnamefont{T.}~\bibnamefont{Izubuchi}},\ }%
  \bibfield{journal}{%
  \bibinfo {journal} {PoS}\ }%
  \textbf{\bibinfo {volume} {LAT2009}},\ \bibinfo {pages} {035} (\bibinfo
  {year} {2009}),\ \Eprint{http://arxiv.org/abs/1003.2182}{arXiv:1003.2182
  [hep-lat]}%
  \bibAnnoteFile{NoStop}{Ishikawa:2010tq}%
\bibitem{Izubuchi:2010}%
  \BibitemOpen
  \bibfield{author}{%
  \bibinfo {author} {\bibfnamefont{T.}~\bibnamefont{Izubuchi}}, \bibinfo
  {author} {\bibfnamefont{T.}~\bibnamefont{Blum}}, \bibinfo {author}
  {\bibfnamefont{T.}~\bibnamefont{Doi}}, \bibinfo {author}
  {\bibfnamefont{M.}~\bibnamefont{Hayakawa}}, \bibinfo {author}
  {\bibfnamefont{T.}~\bibnamefont{Ishikawa}}, \bibinfo {author}
  {\bibfnamefont{S.}~\bibnamefont{Uno}}, \bibinfo {author}
  {\bibfnamefont{N.}~\bibnamefont{Yamada}},\ and\ \bibinfo {author}
  {\bibfnamefont{R.}~\bibnamefont{Zhou}},\ }%
  \bibfield{journal}{%
  \bibinfo {journal} {PoS}\ }%
  \textbf{\bibinfo {volume} {LATTICE2010}} (\bibinfo {year} {2010})%
  \bibAnnoteFile{NoStop}{Izubuchi:2010}%
\bibitem{Abramczyk:2009gb}%
  \BibitemOpen
  \bibfield{author}{%
  \bibinfo {author} {\bibfnamefont{M.}~\bibnamefont{Abramczyk}}, \bibinfo
  {author} {\bibfnamefont{T.}~\bibnamefont{Blum}}, \bibinfo {author}
  {\bibfnamefont{G.}~\bibnamefont{Petropoulos}},\ and\ \bibinfo {author}
  {\bibfnamefont{R.}~\bibnamefont{Zhou}}}%
   (\bibinfo {year} {2009}),\
  \Eprint{http://arxiv.org/abs/0911.1348}{arXiv:0911.1348 [hep-lat]}%
  \bibAnnoteFile{NoStop}{Abramczyk:2009gb}%
\bibitem{Allton:2007hx}%
  \BibitemOpen
  \bibfield{author}{%
  \bibinfo {author} {\bibfnamefont{C.}~\bibnamefont{Allton}} \emph{et~al.}
  (\bibinfo {collaboration} {RBC and UKQCD}),\ }%
  \bibfield{journal}{%
  \bibinfo {journal} {Phys. Rev.}\ }%
  \textbf{\bibinfo {volume} {D76}},\ \bibinfo {pages} {014504} (\bibinfo {year}
  {2007}),\ \Eprint{http://arxiv.org/abs/hep-lat/0701013}{hep-lat/0701013}%
  \bibAnnoteFile{NoStop}{Allton:2007hx}%
\bibitem{Furman:1995ky}%
  \BibitemOpen
  \bibfield{author}{%
  \bibinfo {author} {\bibfnamefont{V.}~\bibnamefont{Furman}}\ and\ \bibinfo
  {author} {\bibfnamefont{Y.}~\bibnamefont{Shamir}},\ }%
  \bibfield{journal}{%
  \bibinfo {journal} {Nucl. Phys.}\ }%
  \textbf{\bibinfo {volume} {B439}},\ \bibinfo {pages} {54} (\bibinfo {year}
  {1995}),\ \Eprint{http://arxiv.org/abs/hep-lat/9405004}{hep-lat/9405004}%
  \bibAnnoteFile{NoStop}{Furman:1995ky}%
\bibitem{Blum:1998ud}%
  \BibitemOpen
  \bibfield{author}{%
  \bibinfo {author} {\bibfnamefont{T.}~\bibnamefont{Blum}},\ }%
  \bibfield{journal}{%
  \bibinfo {journal} {Nucl. Phys. Proc. Suppl.}\ }%
  \textbf{\bibinfo {volume} {73}},\ \bibinfo {pages} {167} (\bibinfo {year}
  {1999}),\ \Eprint{http://arxiv.org/abs/hep-lat/9810017}{hep-lat/9810017}%
  \bibAnnoteFile{NoStop}{Blum:1998ud}%
\bibitem{Antonio:2008zz}%
  \BibitemOpen
  \bibfield{author}{%
  \bibinfo {author} {\bibfnamefont{D.~J.}\ \bibnamefont{Antonio}} \emph{et~al.}
  (\bibinfo {collaboration} {RBC}),\ }%
  \bibfield{journal}{%
  \Doi{10.1103/PhysRevD.77.014509}{\bibinfo {journal} {Phys. Rev.}}\ }%
  \textbf{\bibinfo {volume} {D77}},\ \bibinfo {pages} {014509} (\bibinfo {year}
  {2008}),\ \Eprint{http://arxiv.org/abs/0705.2340}{arXiv:0705.2340 [hep-lat]}%
  \bibAnnoteFile{NoStop}{Antonio:2008zz}%
\bibitem{Doi:2006xh}%
  \BibitemOpen
  \bibfield{author}{%
  \bibinfo {author} {\bibfnamefont{T.}~\bibnamefont{Doi}}, \bibinfo {author}
  {\bibfnamefont{T.}~\bibnamefont{Blum}}, \bibinfo {author}
  {\bibfnamefont{M.}~\bibnamefont{Hayakawa}}, \bibinfo {author}
  {\bibfnamefont{T.}~\bibnamefont{Izubuchi}},\ and\ \bibinfo {author}
  {\bibfnamefont{N.}~\bibnamefont{Yamada}},\ }%
  \bibfield{journal}{%
  \bibinfo {journal} {PoS}\ }%
  \textbf{\bibinfo {volume} {LAT2006}},\ \bibinfo {pages} {174} (\bibinfo
  {year} {2006}),\
  \Eprint{http://arxiv.org/abs/hep-lat/0610095}{arXiv:hep-lat/0610095}%
  \bibAnnoteFile{NoStop}{Doi:2006xh}%
\bibitem{press}%
  \BibitemOpen
  \bibfield{author}{%
  \bibinfo {author} {\bibfnamefont{W.}~\bibnamefont{Press}}, \bibinfo {author}
  {\bibfnamefont{S.}~\bibnamefont{Teukolsky}}, \bibinfo {author}
  {\bibfnamefont{W.}~\bibnamefont{Vetterling}},\ and\ \bibinfo {author}
  {\bibfnamefont{B.}~\bibnamefont{Flannery}},\ }%
  \emph{\bibinfo {title} {Numerical Recipes in C}},\ \bibinfo {edition} {2nd}\
  ed.\ (\bibinfo {publisher} {Cambridge University Press},\ \bibinfo {address}
  {Cambridge, UK},\ \bibinfo {year} {1992})%
  \bibAnnoteFile{NoStop}{press}%
\bibitem{Lellouch:2009fg}%
  \BibitemOpen
  \bibfield{author}{%
  \bibinfo {author} {\bibfnamefont{L.}~\bibnamefont{Lellouch}},\ }%
  \bibfield{journal}{%
  \bibinfo {journal} {PoS}\ }%
  \textbf{\bibinfo {volume} {LATTICE2008}},\ \bibinfo {pages} {015} (\bibinfo
  {year} {2009}),\ \Eprint{http://arxiv.org/abs/0902.4545}{arXiv:0902.4545
  [hep-lat]}%
  \bibAnnoteFile{NoStop}{Lellouch:2009fg}%
\bibitem{Durr:2008zz}%
  \BibitemOpen
  \bibfield{author}{%
  \bibinfo {author} {\bibfnamefont{S.}~\bibnamefont{Durr}} \emph{et~al.},\ }%
  \bibfield{journal}{%
  \Doi{10.1126/science.1163233}{\bibinfo {journal} {Science}}\ }%
  \textbf{\bibinfo {volume} {322}},\ \bibinfo {pages} {1224} (\bibinfo {year}
  {2008}),\ \Eprint{http://arxiv.org/abs/0906.3599}{arXiv:0906.3599 [hep-lat]}%
  \bibAnnoteFile{NoStop}{Durr:2008zz}%
\bibitem{Dawson:2009}%
  \BibitemOpen
  \bibfield{author}{%
  \bibinfo {author} {\bibfnamefont{C.}~\bibnamefont{Dawson}},\ }%
  \bibfield{journal}{%
  \bibinfo {journal} {Talk given at Lattice 2009 (Beijing)}}%
   (\bibinfo {year} {2009})%
  \bibAnnoteFile{NoStop}{Dawson:2009}%
\bibitem{Luscher:2008tw}%
  \BibitemOpen
  \bibfield{author}{%
  \bibinfo {author} {\bibfnamefont{M.}~\bibnamefont{Luscher}}\ and\ \bibinfo
  {author} {\bibfnamefont{F.}~\bibnamefont{Palombi}},\ }%
  \bibfield{journal}{%
  \bibinfo {journal} {PoS}\ }%
  \textbf{\bibinfo {volume} {LATTICE2008}},\ \bibinfo {pages} {049} (\bibinfo
  {year} {2008}),\ \Eprint{http://arxiv.org/abs/0810.0946}{arXiv:0810.0946
  [hep-lat]}%
  \bibAnnoteFile{NoStop}{Luscher:2008tw}%
\bibitem{Sturm:2009kb}%
  \BibitemOpen
  \bibfield{author}{%
  \bibinfo {author} {\bibfnamefont{C.}~\bibnamefont{Sturm}} \emph{et~al.},\ }%
  \bibfield{journal}{%
  \Doi{10.1103/PhysRevD.80.014501}{\bibinfo {journal} {Phys. Rev.}}\ }%
  \textbf{\bibinfo {volume} {D80}},\ \bibinfo {pages} {014501} (\bibinfo {year}
  {2009}),\ \Eprint{http://arxiv.org/abs/0901.2599}{arXiv:0901.2599 [hep-ph]}%
  \bibAnnoteFile{NoStop}{Sturm:2009kb}%
\bibitem{Gorbahn:2010bf}%
  \BibitemOpen
  \bibfield{author}{%
  \bibinfo {author} {\bibfnamefont{M.}~\bibnamefont{Gorbahn}}\ and\ \bibinfo
  {author} {\bibfnamefont{S.}~\bibnamefont{Jager}}}%
   (\bibinfo {year} {2010}),\
  \Eprint{http://arxiv.org/abs/1004.3997}{arXiv:1004.3997 [hep-ph]}%
  \bibAnnoteFile{NoStop}{Gorbahn:2010bf}%
\bibitem{Aoki:2010yq}%
  \BibitemOpen
  \bibfield{author}{%
  \bibinfo {author} {\bibfnamefont{Y.}~\bibnamefont{Aoki}},\ }%
  \bibfield{journal}{%
  \bibinfo {journal} {PoS}\ }%
  \textbf{\bibinfo {volume} {LAT2009}},\ \bibinfo {pages} {012} (\bibinfo
  {year} {2009}),\ \Eprint{http://arxiv.org/abs/1005.2339}{arXiv:1005.2339
  [hep-lat]}%
  \bibAnnoteFile{NoStop}{Aoki:2010yq}%
\bibitem{Almeida:2010ns}%
  \BibitemOpen
  \bibfield{author}{%
  \bibinfo {author} {\bibfnamefont{L.~G.}\ \bibnamefont{Almeida}}\ and\
  \bibinfo {author} {\bibfnamefont{C.}~\bibnamefont{Sturm}}}%
   (\bibinfo {year} {2010}),\
  \Eprint{http://arxiv.org/abs/1004.4613}{arXiv:1004.4613 [hep-ph]}%
  \bibAnnoteFile{NoStop}{Almeida:2010ns}%
\bibitem{Arthur:2010ht}%
  \BibitemOpen
  \bibfield{author}{%
  \bibinfo {author} {\bibfnamefont{R.}~\bibnamefont{Arthur}}\ and\ \bibinfo
  {author} {\bibfnamefont{P.~A.}\ \bibnamefont{Boyle}}}%
   (\bibinfo {year} {2010}),\
  \Eprint{http://arxiv.org/abs/1006.0422}{arXiv:1006.0422 [hep-lat]}%
  \bibAnnoteFile{NoStop}{Arthur:2010ht}%
\bibitem{Kaplan:1986ru}%
  \BibitemOpen
  \bibfield{author}{%
  \bibinfo {author} {\bibfnamefont{D.~B.}\ \bibnamefont{Kaplan}}\ and\ \bibinfo
  {author} {\bibfnamefont{A.~V.}\ \bibnamefont{Manohar}},\ }%
  \bibfield{journal}{%
  \Doi{10.1103/PhysRevLett.56.2004}{\bibinfo {journal} {Phys. Rev. Lett.}}\ }%
  \textbf{\bibinfo {volume} {56}},\ \bibinfo {pages} {2004} (\bibinfo {year}
  {1986})%
  \bibAnnoteFile{NoStop}{Kaplan:1986ru}%
\bibitem{Leutwyler:1996qg}%
  \BibitemOpen
  \bibfield{author}{%
  \bibinfo {author} {\bibfnamefont{H.}~\bibnamefont{Leutwyler}},\ }%
  \bibfield{journal}{%
  \Doi{10.1016/0370-2693(96)00386-3}{\bibinfo {journal} {Phys. Lett.}}\ }%
  \textbf{\bibinfo {volume} {B378}},\ \bibinfo {pages} {313} (\bibinfo {year}
  {1996}),\ \Eprint{http://arxiv.org/abs/hep-ph/9602366}{arXiv:hep-ph/9602366}%
  \bibAnnoteFile{NoStop}{Leutwyler:1996qg}%
\bibitem{Nelson:2003tb}%
  \BibitemOpen
  \bibfield{author}{%
  \bibinfo {author} {\bibfnamefont{D.~R.}\ \bibnamefont{Nelson}}, \bibinfo
  {author} {\bibfnamefont{G.~T.}\ \bibnamefont{Fleming}},\ and\ \bibinfo
  {author} {\bibfnamefont{G.~W.}\ \bibnamefont{Kilcup}},\ }%
  \bibfield{journal}{%
  \Doi{10.1103/PhysRevLett.90.021601}{\bibinfo {journal} {Phys. Rev. Lett.}}\
  }%
  \textbf{\bibinfo {volume} {90}},\ \bibinfo {pages} {021601} (\bibinfo {year}
  {2003}),\
  \Eprint{http://arxiv.org/abs/hep-lat/0112029}{arXiv:hep-lat/0112029}%
  \bibAnnoteFile{NoStop}{Nelson:2003tb}%
\bibitem{'tHooft:1986nc}%
  \BibitemOpen
  \bibfield{author}{%
  \bibinfo {author} {\bibfnamefont{G.}~\bibnamefont{'t~Hooft}},\ }%
  \bibfield{journal}{%
  \Doi{10.1016/0370-1573(86)90117-1}{\bibinfo {journal} {Phys. Rept.}}\ }%
  \textbf{\bibinfo {volume} {142}},\ \bibinfo {pages} {357} (\bibinfo {year}
  {1986})%
  \bibAnnoteFile{NoStop}{'tHooft:1986nc}%
\bibitem{Choi:1988sy}%
  \BibitemOpen
  \bibfield{author}{%
  \bibinfo {author} {\bibfnamefont{K.}~\bibnamefont{Choi}}, \bibinfo {author}
  {\bibfnamefont{C.~W.}\ \bibnamefont{Kim}},\ and\ \bibinfo {author}
  {\bibfnamefont{W.~K.}\ \bibnamefont{Sze}},\ }%
  \bibfield{journal}{%
  \Doi{10.1103/PhysRevLett.61.794}{\bibinfo {journal} {Phys. Rev. Lett.}}\ }%
  \textbf{\bibinfo {volume} {61}},\ \bibinfo {pages} {794} (\bibinfo {year}
  {1988})%
  \bibAnnoteFile{NoStop}{Choi:1988sy}%
\bibitem{Banks:1994yg}%
  \BibitemOpen
  \bibfield{author}{%
  \bibinfo {author} {\bibfnamefont{T.}~\bibnamefont{Banks}}, \bibinfo {author}
  {\bibfnamefont{Y.}~\bibnamefont{Nir}},\ and\ \bibinfo {author}
  {\bibfnamefont{N.}~\bibnamefont{Seiberg}}}%
   (\bibinfo {year} {1994}),\
  \Eprint{http://arxiv.org/abs/hep-ph/9403203}{arXiv:hep-ph/9403203}%
  \bibAnnoteFile{NoStop}{Banks:1994yg}%
\bibitem{Davoudiasl:2007zx}%
  \BibitemOpen
  \bibfield{author}{%
  \bibinfo {author} {\bibfnamefont{H.}~\bibnamefont{Davoudiasl}}\ and\ \bibinfo
  {author} {\bibfnamefont{A.}~\bibnamefont{Soni}},\ }%
  \bibfield{journal}{%
  \Doi{10.1103/PhysRevD.76.095015}{\bibinfo {journal} {Phys. Rev.}}\ }%
  \textbf{\bibinfo {volume} {D76}},\ \bibinfo {pages} {095015} (\bibinfo {year}
  {2007}),\ \Eprint{http://arxiv.org/abs/0705.0151}{arXiv:0705.0151 [hep-ph]}%
  \bibAnnoteFile{NoStop}{Davoudiasl:2007zx}%
\bibitem{Creutz:2003xc}%
  \BibitemOpen
  \bibfield{author}{%
  \bibinfo {author} {\bibfnamefont{M.}~\bibnamefont{Creutz}},\ }%
  \bibfield{journal}{%
  \Doi{10.1103/PhysRevLett.92.162003}{\bibinfo {journal} {Phys. Rev. Lett.}}\
  }%
  \textbf{\bibinfo {volume} {92}},\ \bibinfo {pages} {162003} (\bibinfo {year}
  {2004}),\ \Eprint{http://arxiv.org/abs/hep-ph/0312225}{arXiv:hep-ph/0312225}%
  \bibAnnoteFile{NoStop}{Creutz:2003xc}%
\bibitem{Shintani:2008qe}%
  \BibitemOpen
  \bibfield{author}{%
  \bibinfo {author} {\bibfnamefont{E.}~\bibnamefont{Shintani}} \emph{et~al.}
  (\bibinfo {collaboration} {JLQCD}),\ }%
  \bibfield{journal}{%
  \Doi{10.1103/PhysRevLett.101.242001}{\bibinfo {journal} {Phys. Rev. Lett.}}\
  }%
  \textbf{\bibinfo {volume} {101}},\ \bibinfo {pages} {242001} (\bibinfo {year}
  {2008}),\ \Eprint{http://arxiv.org/abs/0806.4222}{arXiv:0806.4222 [hep-lat]}%
  \bibAnnoteFile{NoStop}{Shintani:2008qe}%
\bibitem{Boyle:2009xi}%
  \BibitemOpen
  \bibfield{author}{%
  \bibinfo {author} {\bibfnamefont{P.~A.}\ \bibnamefont{Boyle}}, \bibinfo
  {author} {\bibfnamefont{L.}~\bibnamefont{Del~Debbio}}, \bibinfo {author}
  {\bibfnamefont{J.}~\bibnamefont{Wennekers}},\ and\ \bibinfo {author}
  {\bibfnamefont{J.~M.}\ \bibnamefont{Zanotti}} (\bibinfo {collaboration}
  {RBC}),\ }%
  \bibfield{journal}{%
  \Doi{10.1103/PhysRevD.81.014504}{\bibinfo {journal} {Phys. Rev.}}\ }%
  \textbf{\bibinfo {volume} {D81}},\ \bibinfo {pages} {014504} (\bibinfo {year}
  {2010}),\ \Eprint{http://arxiv.org/abs/0909.4931}{arXiv:0909.4931 [hep-lat]}%
  \bibAnnoteFile{NoStop}{Boyle:2009xi}%
\bibitem{Gasser:1984gg}%
  \BibitemOpen
  \bibfield{author}{%
  \bibinfo {author} {\bibfnamefont{J.}~\bibnamefont{Gasser}}\ and\ \bibinfo
  {author} {\bibfnamefont{H.}~\bibnamefont{Leutwyler}},\ }%
  \bibfield{journal}{%
  \bibinfo {journal} {Nucl. Phys.}\ }%
  \textbf{\bibinfo {volume} {B250}},\ \bibinfo {pages} {465} (\bibinfo {year}
  {1985})%
  \bibAnnoteFile{NoStop}{Gasser:1984gg}%
\bibitem{Amoros:2001cp}%
  \BibitemOpen
  \bibfield{author}{%
  \bibinfo {author} {\bibfnamefont{G.}~\bibnamefont{Amoros}}, \bibinfo {author}
  {\bibfnamefont{J.}~\bibnamefont{Bijnens}},\ and\ \bibinfo {author}
  {\bibfnamefont{P.}~\bibnamefont{Talavera}},\ }%
  \bibfield{journal}{%
  \Doi{10.1016/S0550-3213(01)00121-3}{\bibinfo {journal} {Nucl. Phys.}}\ }%
  \textbf{\bibinfo {volume} {B602}},\ \bibinfo {pages} {87} (\bibinfo {year}
  {2001}),\ \Eprint{http://arxiv.org/abs/hep-ph/0101127}{arXiv:hep-ph/0101127}%
  \bibAnnoteFile{NoStop}{Amoros:2001cp}%
\bibitem{Aubin:2004fs}%
  \BibitemOpen
  \bibfield{author}{%
  \bibinfo {author} {\bibfnamefont{C.}~\bibnamefont{Aubin}} \emph{et~al.}
  (\bibinfo {collaboration} {MILC}),\ }%
  \bibfield{journal}{%
  \bibinfo {journal} {Phys. Rev.}\ }%
  \textbf{\bibinfo {volume} {D70}},\ \bibinfo {pages} {114501} (\bibinfo {year}
  {2004}),\ \Eprint{http://arxiv.org/abs/hep-lat/0407028}{hep-lat/0407028}%
  \bibAnnoteFile{NoStop}{Aubin:2004fs}%
\bibitem{Gasser:1984pr}%
  \BibitemOpen
  \bibfield{author}{%
  \bibinfo {author} {\bibfnamefont{J.}~\bibnamefont{Gasser}}\ and\ \bibinfo
  {author} {\bibfnamefont{H.}~\bibnamefont{Leutwyler}},\ }%
  \bibfield{journal}{%
  \Doi{10.1016/0550-3213(85)90494-8}{\bibinfo {journal} {Nucl. Phys.}}\ }%
  \textbf{\bibinfo {volume} {B250}},\ \bibinfo {pages} {539} (\bibinfo {year}
  {1985})%
  \bibAnnoteFile{NoStop}{Gasser:1984pr}%
\bibitem{Marciano:2004uf}%
  \BibitemOpen
  \bibfield{author}{%
  \bibinfo {author} {\bibfnamefont{W.~J.}\ \bibnamefont{Marciano}},\ }%
  \bibfield{journal}{%
  \bibinfo {journal} {Phys. Rev. Lett.}\ }%
  \textbf{\bibinfo {volume} {93}},\ \bibinfo {pages} {231803} (\bibinfo {year}
  {2004}),\ \Eprint{http://arxiv.org/abs/hep-ph/0402299}{hep-ph/0402299}%
  \bibAnnoteFile{NoStop}{Marciano:2004uf}%
\bibitem{Antonelli:2010yf}%
  \BibitemOpen
  \bibfield{author}{%
  \bibinfo {author} {\bibfnamefont{M.}~\bibnamefont{Antonelli}} \emph{et~al.}}%
   (\bibinfo {year} {2010}),\
  \Eprint{http://arxiv.org/abs/1005.2323}{arXiv:1005.2323 [hep-ph]}%
  \bibAnnoteFile{NoStop}{Antonelli:2010yf}%
\bibitem{Cirigliano:2007ga}%
  \BibitemOpen
  \bibfield{author}{%
  \bibinfo {author} {\bibfnamefont{V.}~\bibnamefont{Cirigliano}}\ and\ \bibinfo
  {author} {\bibfnamefont{I.}~\bibnamefont{Rosell}},\ }%
  \bibfield{journal}{%
  \Doi{10.1088/1126-6708/2007/10/005}{\bibinfo {journal} {JHEP}}\ }%
  \textbf{\bibinfo {volume} {10}},\ \bibinfo {pages} {005} (\bibinfo {year}
  {2007}),\ \Eprint{http://arxiv.org/abs/0707.4464}{arXiv:0707.4464 [hep-ph]}%
  \bibAnnoteFile{NoStop}{Cirigliano:2007ga}%
\bibitem{Beane:2006fk}%
  \BibitemOpen
  \bibfield{author}{%
  \bibinfo {author} {\bibfnamefont{S.~R.}\ \bibnamefont{Beane}}, \bibinfo
  {author} {\bibfnamefont{K.}~\bibnamefont{Orginos}},\ and\ \bibinfo {author}
  {\bibfnamefont{M.~J.}\ \bibnamefont{Savage}},\ }%
  \bibfield{journal}{%
  \Doi{10.1016/j.nuclphysb.2006.12.023}{\bibinfo {journal} {Nucl. Phys.}}\ }%
  \textbf{\bibinfo {volume} {B768}},\ \bibinfo {pages} {38} (\bibinfo {year}
  {2007}),\
  \Eprint{http://arxiv.org/abs/hep-lat/0605014}{arXiv:hep-lat/0605014}%
  \bibAnnoteFile{NoStop}{Beane:2006fk}%
\bibitem{Sasaki:2001nf}%
  \BibitemOpen
  \bibfield{author}{%
  \bibinfo {author} {\bibfnamefont{S.}~\bibnamefont{Sasaki}}, \bibinfo {author}
  {\bibfnamefont{T.}~\bibnamefont{Blum}},\ and\ \bibinfo {author}
  {\bibfnamefont{S.}~\bibnamefont{Ohta}},\ }%
  \bibfield{journal}{%
  \Doi{10.1103/PhysRevD.65.074503}{\bibinfo {journal} {Phys. Rev.}}\ }%
  \textbf{\bibinfo {volume} {D65}},\ \bibinfo {pages} {074503} (\bibinfo {year}
  {2002}),\
  \Eprint{http://arxiv.org/abs/hep-lat/0102010}{arXiv:hep-lat/0102010}%
  \bibAnnoteFile{NoStop}{Sasaki:2001nf}%
\bibitem{Cottingham:1963zz}%
  \BibitemOpen
  \bibfield{author}{%
  \bibinfo {author} {\bibfnamefont{W.~N.}\ \bibnamefont{Cottingham}},\ }%
  \bibfield{journal}{%
  \bibinfo {journal} {Annals Phys.}\ }%
  \textbf{\bibinfo {volume} {25}},\ \bibinfo {pages} {424} (\bibinfo {year}
  {1963})%
  \bibAnnoteFile{NoStop}{Cottingham:1963zz}%
\bibitem{Kelly:2004hm}%
  \BibitemOpen
  \bibfield{author}{%
  \bibinfo {author} {\bibfnamefont{J.~J.}\ \bibnamefont{Kelly}},\ }%
  \bibfield{journal}{%
  \Doi{10.1103/PhysRevC.70.068202}{\bibinfo {journal} {Phys. Rev.}}\ }%
  \textbf{\bibinfo {volume} {C70}},\ \bibinfo {pages} {068202} (\bibinfo {year}
  {2004})%
  \bibAnnoteFile{NoStop}{Kelly:2004hm}%
\bibitem{Renfrew:2009wu}%
  \BibitemOpen
  \bibfield{author}{%
  \bibinfo {author} {\bibfnamefont{D.}~\bibnamefont{Renfrew}}, \bibinfo
  {author} {\bibfnamefont{T.}~\bibnamefont{Blum}}, \bibinfo {author}
  {\bibfnamefont{N.}~\bibnamefont{Christ}}, \bibinfo {author}
  {\bibfnamefont{R.}~\bibnamefont{Mawhinney}},\ and\ \bibinfo {author}
  {\bibfnamefont{P.}~\bibnamefont{Vranas}},\ }%
  \bibfield{journal}{%
  \bibinfo {journal} {PoS}\ }%
  \textbf{\bibinfo {volume} {LATTICE2008}},\ \bibinfo {pages} {048} (\bibinfo
  {year} {2008}),\ \Eprint{http://arxiv.org/abs/0902.2587}{arXiv:0902.2587
  [hep-lat]}%
  \bibAnnoteFile{NoStop}{Renfrew:2009wu}%
\end{thebibliography}%

\end{document}